 \pdfoutput=1

\documentclass[final,1p,times]{elsarticle}%

\usepackage{graphicx}

\usepackage{appendix}
\usepackage{hyperref} 

\usepackage{amssymb}
\usepackage{amsmath}

\usepackage{nonfloat}

\usepackage{bigstrut}

\usepackage{array}			
\newcommand{\unit}[2]{\ensuremath{#1\,}#2}
\newcommand{\un}[1]{\,{\rm #1}}
\newcommand{\dd}{\mathrm{d}}

\newcommand{\degree}{$^{\circ}$}
\newcommand{\degC}{$^{\circ}$C}
\newcommand{\dg}{^{\circ}}

\journal{NIM}

\begin{document}

\begin{frontmatter}


\title{IceTop: The surface component of IceCube\\[5mm]
{\normalsize The IceCube Collaboration}\\[-2mm]}

%

\author[MadisonPAC]{R.~Abbasi}
\author[Gent]{Y.~Abdou}
\author[Zeuthen]{M.~Ackermann}
\author[Christchurch]{J.~Adams}
\author[Geneva]{J.~A.~Aguilar}
\author[MadisonPAC]{M.~Ahlers}
\author[Berlin]{D.~Altmann}
\author[MadisonPAC]{K.~Andeen}
\author[MadisonPAC]{J.~Auffenberg}
\author[Bartol]{X.~Bai\fnref{SouthDakota}}
% ...
\address[Berlin]{Institut f\"ur Physik, Humboldt-Universit\"at zu Berlin, D-12489 Berlin, Germany}
% ...
\address[Christchurch]{Dept.~of Physics and Astronomy, University of Canterbury, Private Bag 4800, Christchurch, New Zealand}
% ...
\address[Geneva]{D\'epartement de physique nucl\'eaire et corpusculaire, Universit\'e de Gen\`eve, CH-1211 Gen\`eve, Switzerland}
\address[Gent]{Dept.~of Physics and Astronomy, University of Gent, B-9000 Gent, Belgium}
% ...
\address[MadisonPAC]{Dept.~of Physics and Wisconsin IceCube Particle Astrophysics Center, University of Wisconsin, Madison, WI 53706, USA}
% ...
\address[Bartol]{Bartol Research Institute and Department of Physics and Astronomy, University of Delaware, Newark, DE 19716, USA}
% ...
\address[Zeuthen]{DESY, D-15735 Zeuthen, Germany}
\fntext[SouthDakota]{Physics Department, South Dakota School of Mines and Technology, Rapid City, SD 57701, USA}
\fntext[LosAlamos]{Los Alamos National Laboratory, Los Alamos, NM 87545, USA}
\fntext[Bari]{also Sezione INFN, Dipartimento di Fisica, I-70126, Bari, Italy}
\fntext[Goddard]{NASA Goddard Space Flight Center, Greenbelt, MD 20771, USA}

\cortext[cor]{Corresponding author}
\begin{abstract}
IceTop, the surface component of the IceCube Neutrino Observatory at the South Pole, is an air shower array with
an area of 1\,km$^2$. The detector allows a detailed exploration of the mass  composition of primary cosmic rays in the energy range from about \unit{100}{TeV} to \unit{1}{EeV} by exploiting the correlation between the shower energy measured in IceTop and the energy deposited by muons in the deep ice. In this paper we report on the technical design, construction and installation, the trigger and data acquisition systems as well as the software framework for calibration, reconstruction and simulation. Finally the first experience from commissioning and operating the detector and the performance as an air shower detector will be discussed.
\end{abstract}


\end{frontmatter}

\setcounter{tocdepth}{2}
\tableofcontents

\section[Introduction]{Introduction}
\label{sec:introduction}
%
%
%

\begin{figure}
	\centering	\includegraphics[width=0.6\textwidth]{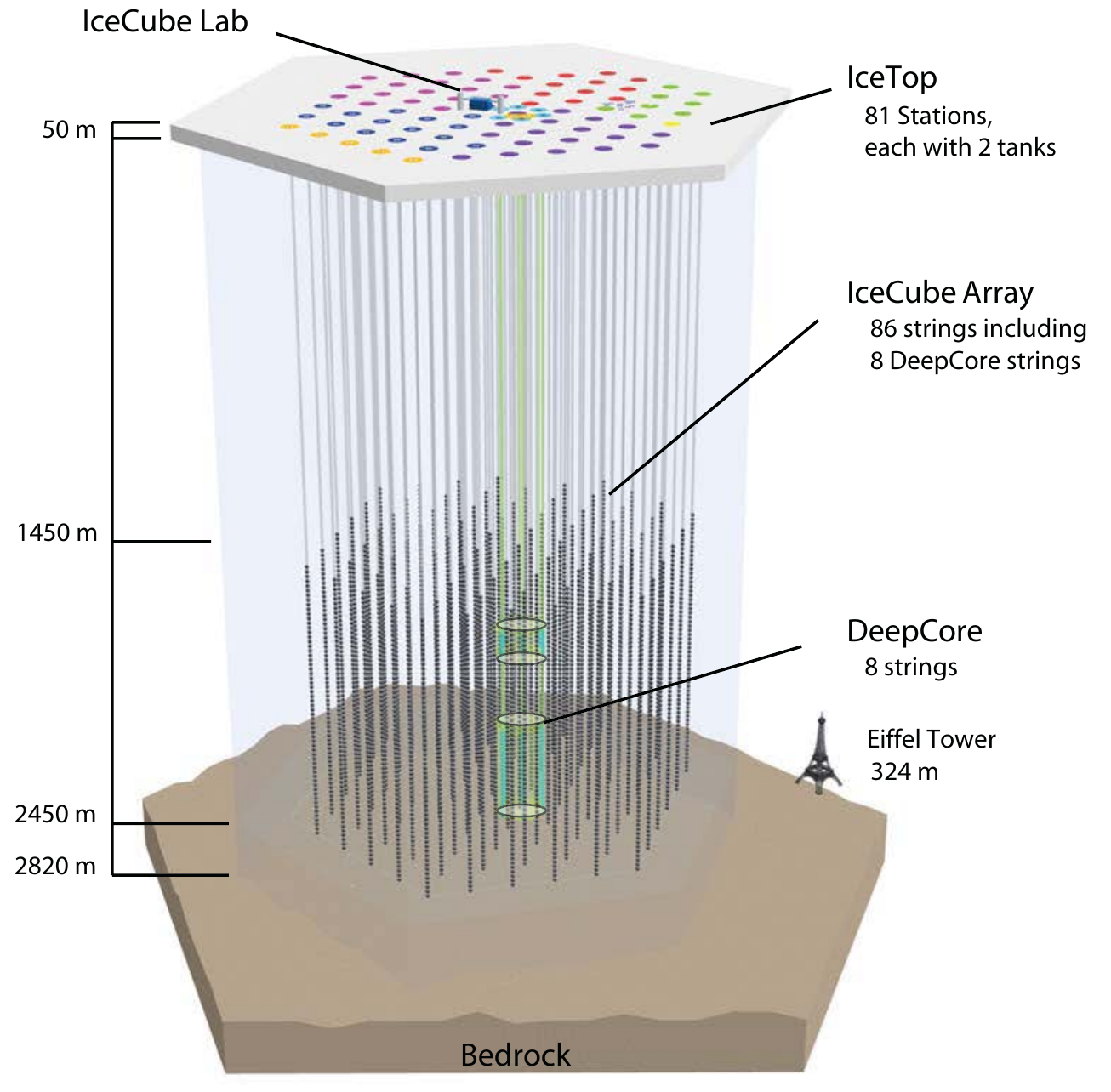}
	\caption{The IceCube detector with its components DeepCore and IceTop in the final configuration (December 2010).}
	\label{fig:I3Array}
\end{figure}

The IceCube Neutrino Observatory at the geographic South Pole consists of a cubic-kilometer detector situated in the ice at a depth between \unit{1450}{m} and \unit{2450}{m} and a square-kilometer detector array at the surface. IceCube is primarily designed to measure neutrinos from below, thus using the Earth as a filter to discriminate against background induced by cosmic rays  \cite{achterberg06,Kolanoski_HLT_icrc2011,Gaisser:2011iz}. The detector employs optical sensors to detect light of charged particles generated by neutrinos in the ice or the Earth's crust.
In addition, IceCube also includes a more densely instrumented part called DeepCore and an extensive air shower array on the surface called IceTop, both 
fully integrated into the IceCube data acquisition system. The IceTop array extends IceCube's capabilities for cosmic ray physics allowing the use of
the full IceCube Observatory (Fig.\ \ref{fig:I3Array}) as a 3-dimensional array for the study of high-energy cosmic rays. Construction of IceCube, including the IceTop component, was completed in December 2010. 

IceTop consists of Cherenkov tanks filled with clear ice that operate on the same principle as the water tanks of the Haverah Park experiment \cite{Haverah_array_Lawrence:1991cc} and the Pierre Auger Observatory \cite{AUGER-detector-2004}.
The tanks are arranged in pairs on the same, approximately \unit{125}{m}, triangular grid as the vertical cables that carry the deep sensors of IceCube. The two tanks at each surface station are separated from each other by \unit{10}{m}. Each tank contains two standard IceCube digital optical modules (DOMs, see Section \ref{subsec:DOM}). Air showers initiated in the atmosphere by cosmic rays are typically spread over a number of stations. The light generated in the tanks by the shower particles (electrons, photons, muons and hadrons) is a measure of the energy deposit of these particles in the tanks. The information from multiple stations is fitted to a model of the overall shower shape and intensity, called `shower size' $S$, and the direction described by the zenith and azimuth angles $\theta, \phi$.

With the IceTop detector configuration at the \unit{2835}{m} altitude of the South Pole surface, the threshold for efficient reconstruction of air showers that trigger 3 or more stations is approximately \unit{300}{TeV} in most of the detector and approximately \unit{100}{TeV} in a denser in-fill region. The geometrical acceptance of the combined surface and in-ice detectors is approximately \unit{0.3}{km$^2$ sr}. This aperture will provide a useful rate of showers with energies up to few \unit{}{EeV}.  Thus, the energy range of  IceCube as a cosmic-ray detector fully covers the knee region of the spectrum and extends to the energies where a transition from galactic cosmic rays to a population of extra-galactic particles may occur.  The key to identifying a transition from one population to another is to know the nuclear composition of the primary cosmic radiation. General reviews of the status of cosmic ray physics in the region covered by IceTop can be found, for example, in \cite{Bluemer:2009zf} or in the cosmic ray review in \cite{PDG2010}. A recent summary of composition measurements is given in \cite{Kampert:2012mx}.

The primary mass determination from extended air showers (EAS) is notoriously difficult because the measurements are indirect and have to rely on models for the hadronization processes. Observables sensitive to the primary mass composition are mainly the height of the shower maximum (measured through fluorescence or Cherenkov emission) and the number of muons in a shower. The highest energy muons stemming from the first interactions in the higher atmosphere are most closely correlated to the mass of the primary nucleus. IceCube, in combination with IceTop, offers the unique possibility to observe these muons, typically with initial energies above 500\,GeV, in the deep ice in coincidence with the mostly electromagnetically deposited shower energy measured at the surface. This provides a particularly powerful method for the determination of the mass composition.

To mitigate the dependence on hadronization models, several alternative methods for studying mass composition with IceTop are available. Other mass sensitive observables are for example: the shower absorption in the atmosphere at different zenith angles, the number of dominantly low-energy muons in the surface detector, and other shower properties such as shower age (state of shower development) and shower front curvature.

The IceTop array has additionally been used to study high-p$_{\rm T}$ muons, PeV gamma rays and transient events, such as the radiation effects of solar flares. It also serves as a partial veto for the detection of downward-going neutrinos with IceCube and for direction calibration. 

This paper is organized as follows: First, we describe the technical design, deployment and commissioning of IceTop in Section \ref{sec:design}, followed by the description of the front-end electronics in Section \ref{sec:icetop_fee} and the trigger and data acquisition systems in Section \ref{sec:trigger_daq}.  Next, the calibration procedures and the environmental conditions which influence the shower detection are discussed in Section \ref{sec:calibration} and \ref{sec:icecube:environment}, respectively. The next part deals with the data handling and analysis, beginning with the signal processing and data preparation in Section \ref{sec:signal_proc} and continuing with the description of the air shower reconstruction in Section \ref{sec:reconstruction} and the simulation of air shower events in Section \ref{sec:simulation}. Finally the performance of the detector is demonstrated in Section \ref{sec:performance} with examples from finished and ongoing analyses.  The paper ends with a summary and outlook. 
\begin{table}
	\centering
		\caption{Acronyms used in this article.}
	\label{tab:AcronymsUsedInThisArticle}
		\begin{tabular}{ p{.2\textwidth}  p{.74\textwidth} }
		\hline
 ADC  & Analog-to-Digital Converter\\
 ATWD  & Analog Transient Waveform Digitizer (front-end chip for signal recording)\\
 DOM & Digital Optical Module  (light sensor and readout electronics )\\
 DOMHub & Computer in the IceCube Lab which receives the signals from the DOMs\\
 fADC  & fast Analog-to-Digital Converter (used in the DOM to record waveforms at a sampling rate of 40 MHz)\\
 HG & high gain (`HG DOM': one of the two DOMs in a tank)\\
 HLC & `Hard Local Coincidence' (in combination `HLC hit' or `HLC mode', describes an operation mode with full waveform readout)\\
 HV & high voltage \\
 IC  & IceCube (in-ice part)\\
 IT  & IceTop \\
 LC & local coincidence  (of the two high-gain DOMs of a station)\\
 LG & low gain  (`LG DOM': one of the two DOMs in a tank)\\
 MPE  & multiple photoelectrons (used in `MPE discriminator', see also SPE)\\
 PE & photoelectron  (charge which a single photoelectron generates and is seen by the electronics, including PMT gain, used as charge unit)\\
 PMT  & photomultiplier tube\\
 rms &root-mean-square \\
 SLC & `soft local coincidence'  (in combination `SLC hit' or `SLC mode', analogous to `HLC mode' to describe the operation mode; despite the name, `SLC' is not a `coincidence')\\  
 SPE  & single photoelectrons (used in `SPE discriminator', see also MPE)\\
 UTC & Coordinated Universal Time (reference time for all IceCube signals)\\
 VEM & vertical equivalent muon  (charge which a vertical muon generates in the  IceTop DOM electronics of a tank, used as charge unit for IceTop signals)\\    
    \hline
		\end{tabular}
\end{table}
In all descriptions we concentrate on the current status of the detector and the software (spring 2012) and refer to previous configurations only if there were significant changes  which matter for analyses using those data. Acronyms used in the following text are explained in Table \ref{tab:AcronymsUsedInThisArticle}.

\section[Technical design and deployment]{Technical design and deployment}
\label{sec:design}
%
%
%

\subsection{Design considerations}
\label{subsec:design_considerations}

The IceTop air shower array was proposed as an addition to the IceCube neutrino telescope to extend the telescope's capabilities for cosmic ray physics and to partially veto the background of down-going muons. In order to realize this addition in the most cost effective way, technical developments and infrastructure of IceCube were employed as much as possible. These considerations defined the following guidelines for the design of IceTop: 
\begin{itemize}
	\item IceTop uses the same cables as laid out in trenches for the IceCube holes. That constrains IceTop's array size to \unit{1}{km$^2$} and the grid spacing to 125 m. Both together define the energy range from some \unit{100}{TeV} to about \unit{1}{EeV}.
	\item Whenever possible IceTop uses the same detector hardware, electronic readout, triggering scheme and data acquisition as in the deep-ice part of IceCube (henceforth referred to as `in-ice IceCube'). 

\end{itemize}
Using the same Digital Optical Modules (DOMs) as in the deep ice for detection of Cherenkov light with the same readout scheme leads to the choice of ice as detector medium.  The ice is produced in tanks similar to the water-filled tanks used in the Haverah Park experiment \cite{Haverah_array_Lawrence:1991cc} and the Pierre Auger Observatory \cite{AUGER-detector-2004}. Such ice tanks are well suited for the detection of shower particles because a particle produces enough light so that signal fluctuations are not dominated by photon statistics. For example, in the finally produced tanks a vertically through-going muon produces approximately 125 photoelectrons.  The choice of ice Cherenkov detectors
saved design and  development costs and also simplified integrating the IceTop signals into the in-ice data acquisition system. 

Given the constraints on size, spacing, detector technology and basic components the following further considerations aim at optimization for air shower physics: 
\begin{itemize}
	\item Air showers have a huge range of different energy depositions in the tanks. To increase the dynamic range offered by the DOM electronics (see Section \ref{subsec:DOM}) each tank has two light detectors running at different gains. Having two DOMs in each tank also allows for an experimental determination of internal signal fluctuations by setting both DOMs to the same gain during special calibration runs.
	\item Each detector station consists of two tanks for the following reasons:
\begin{itemize}
	\item By requiring events to trigger both tanks of a station (`local coincidence')  signals from extensive air showers can be distinguished from the high (typically \unit{2}{kHz}) individual event rate generated mostly by low energy showers.  
\item Signals seen in both tanks of a station can be compared on an event-by-event basis to give a measure of the intrinsic physical fluctuations in the shower front, both in timing and in amplitude.
\item IceTop can be divided into two very similar sub-arrays, each comprising one of the two tanks of each station, to measure the fluctuations of reconstruction parameters, such as shower front curvature, lateral distribution, core location accuracy, angular resolution.
\item Single-station events that trigger both tanks of a single, interior station without triggering adjacent stations could be exploited to extend the IceTop reach to lower energies. Such events have primary energies around 30 TeV and have  typically at most one muon with sufficient energy to reach the deep detector. 
\end{itemize}
 \item The size of the tanks and their wall reflectivity were optimized for large light output and short signal pulses. Both, large tank size and high reflectivity, yield high signals but long pulse lengths (in the  final tank design the decay constant is about 30 ns).
\end{itemize}

\begin{table}%
\centering
\tabcaption{List of the years when a certain configuration of IceCube (IC), IceTop (IT) became operational together with the numbers of the stations which were added in that year (fourth column). The numbers for IC include also the DeepCore strings. We will use abbreviations like IC79/IT73 for the constellation in 2010, for example. The fourth and following columns list stations  using the numbers given in Fig.\ \ref{fig:map2011}.  The fifth column lists the tanks which have 
 Tyvek liners (the others are coated with zirconia liner). However only the tanks installed in 2005 have higher reflectivities than the other tanks.
The sixth column reports the DOMs which have the `old' transformers with \unit{43}{$\Omega$} and a short time constant, see Sections \ref{subsec:signal_capture}, \ref{subsec:droop}  and Table \ref{tab:FrontEndComponents}. The last two digits in this column (62 or 64) are the DOM numbers in a station according to Fig.\ \ref{fig:Tank-DOM-positions}.} \label{tab:ICIT_configs}
\vspace{3mm}
\begin{tabular}{cccccc}
Year & IC  & IT  & new IT stations & Tyvek liner & old transformer\\
 & strings & stations &  &  & (\unit{43}{$\Omega$})\\
\hline
2005 &  1 &  4 & 21	29	30	39 & all & 	all \\
2006 &  9 &  16 & 38	40	47-50	57-59	66	67	74 & -& all\\
2007 & 22 & 26 & 46	55	56	64	65	71-73	77	78 & -& (77,46,56)-62\\ &&&&&(77,46,71)-64\\
2008 & 40 & 40 &44	45	52-54	60-63	68-70	75	76 & -&  (53,55)-62\\ &&&&&(52,55,62,68-70)-64\\
2009 & 59 & 59  &2-6	9-13	17-20	26-28	36	37 & -&37-64 \\
2010 & 79 & 73 &8	15	16	23-25	32-35	41-43	51 & -& 23-62 \\
2011 & 86 & 81  &1	7	14	22	31	79-81 & 79A, 80A, 81A/B & -\\
\hline

\end{tabular}
\vspace{3mm}
\end{table}
 \begin{figure}
\includegraphics[width=0.8\textwidth]{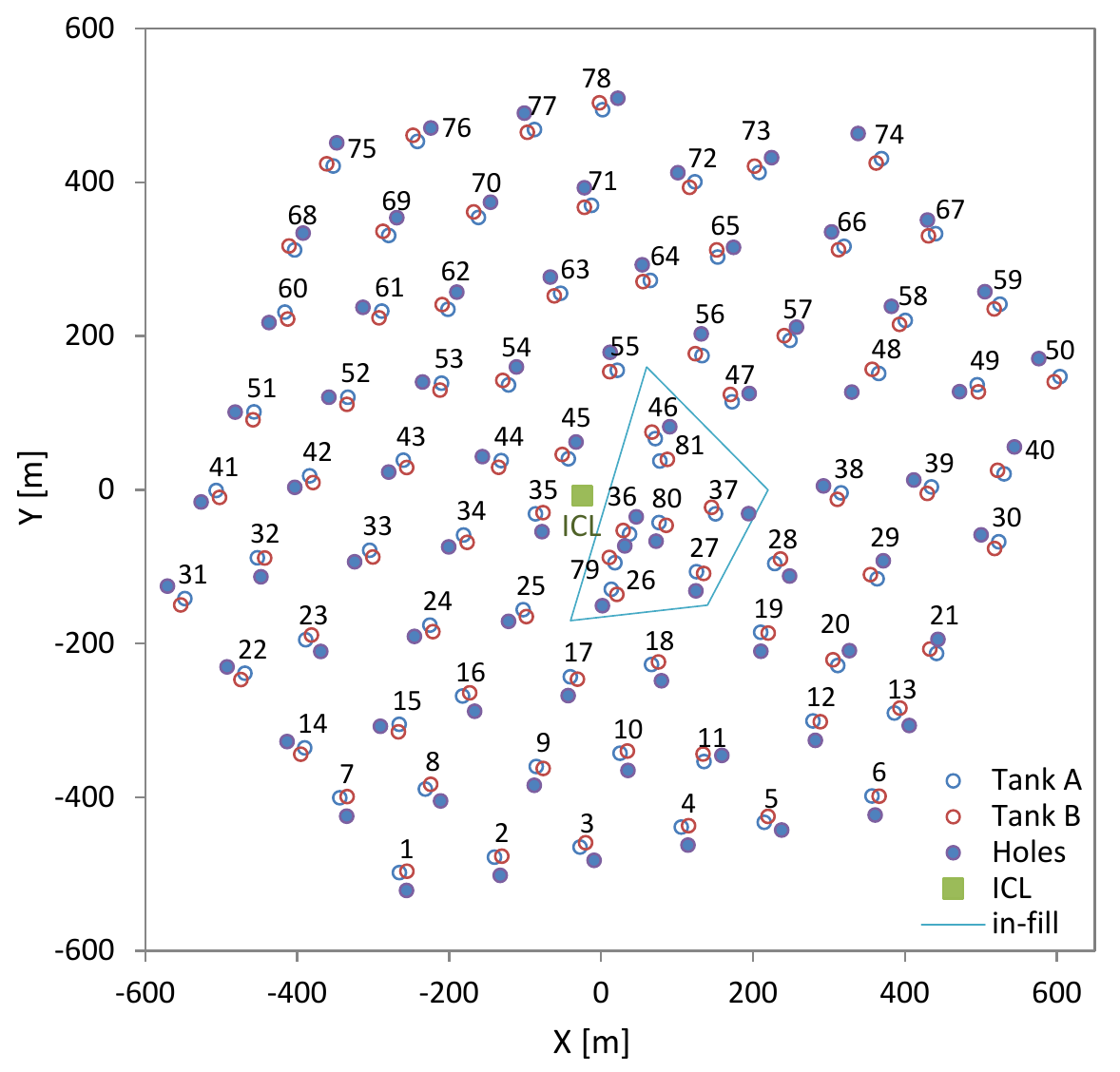}

   \figcaption{Locations of IceCube string holes and IceTop tanks with the IceCube Lab (ICL) in the center in the final configuration after 2010. The holes 81 to 86, belonging to DeepCore and not related to IceTop tanks, are not shown.   IceTop stations are located next to IceCube strings (except for the in-fill station 81) and consist of two tanks, A and B.  The irregularity of the array arises because tank locations were constrained by surface cabling and IceCube drilling operations. A denser in-fill array is formed by the stations 26, 27, 36, 37, 46, 79, 80, 81.}
   \label{fig:map2011}
 \end{figure}

\subsection{General detector layout}
\label{subsec:detector_layout}

The IceTop air shower array is located above the in-ice IceCube detector at a height of
2835\,m above sea level, corresponding to an atmospheric depth of about 680
 g/cm$^2$. It consists of 162
 ice Cherenkov tanks, placed at 81 stations and distributed over an area of 1\,km$^2$ on a grid with mean spacing of 125\,m. Figure \ref{fig:map2011} shows a plan view of the final IceTop array.  
In the center of the array, three stations (numbered 79, 80, 81) are installed at intermediate positions. Together with the neighboring stations (numbered 26, 27, 36, 37, 46) they form an in-fill array for denser shower sampling. Each station comprises two cylindrical tanks, A and B,  \unit{10}{m} apart from each other. The tanks are embedded into the snow so that their top  surface is level with the surrounding snow to minimize temperature variations and accumulation of drifting snow.

IceTop, as well as in-ice IceCube, took data already during the construction phase; the periods during which certain IceCube-IceTop configurations had been commissioned are reported in Table \ref{tab:ICIT_configs}.

\subsection{Tank design}
\label{subsec:tank_design}

\begin{figure}
	\centering
	\includegraphics[width=0.80\textwidth]{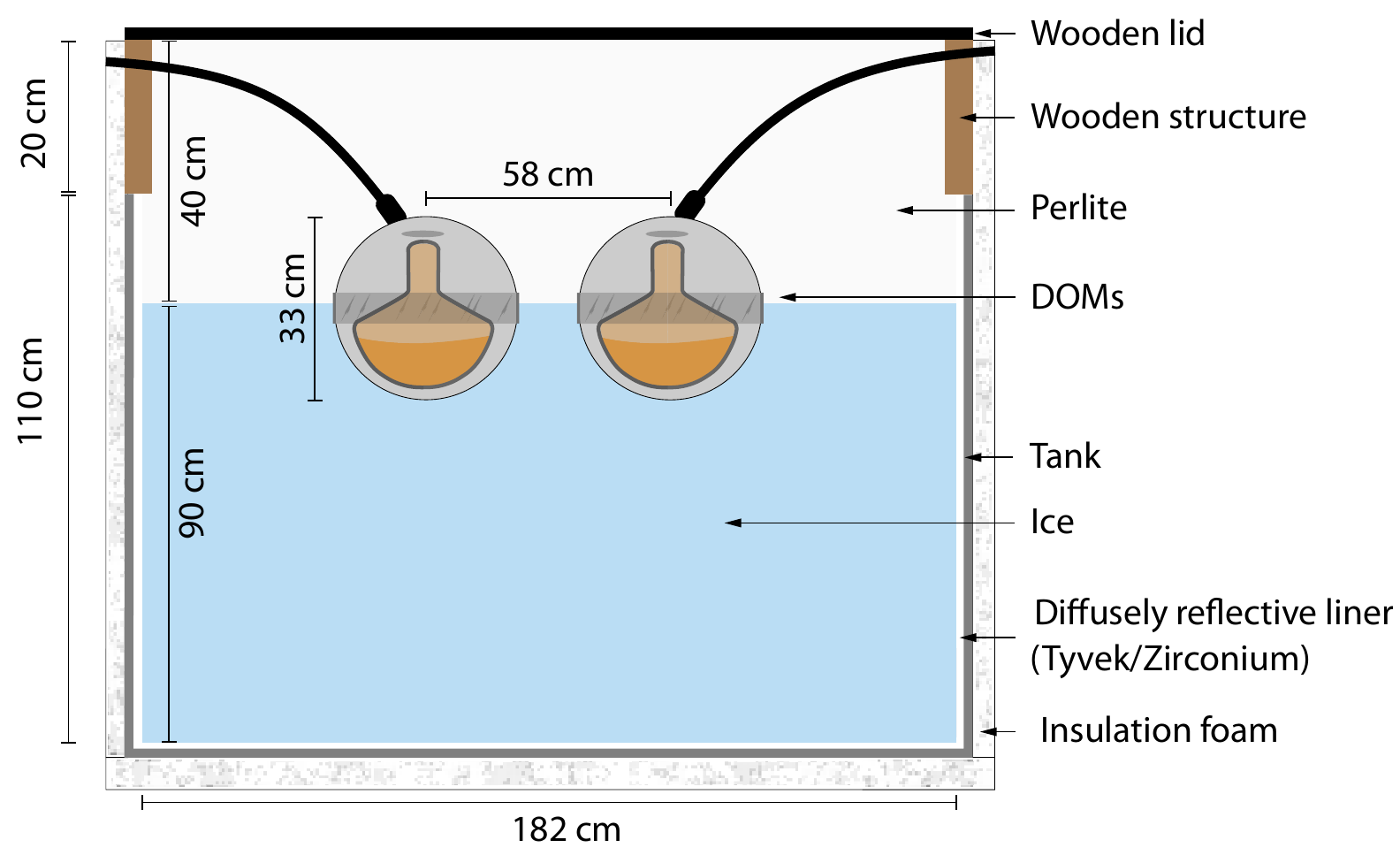}
	\caption{Tank dimensions in a cross-sectional view. Wall thicknesses are not to scale (see Table \ref{tab:IceTopTank}).}
	\label{fig:TankSchematic}
\end{figure}

\begin{table}
	\centering
	\caption{Dimensions of IceTop tanks and assembly. The numbers given in units of m have tolerances of the order of \unit{1}{cm}.}	
		\begin{tabular}{lll}
		component & quantity & value\\
		\hline
		polyethylene tank & height  & \unit{1.10}{m}  \\
				& wooden extension &  \unit{0.20}{m} \\
			 & inner diameter &  \unit{1.82}{m} \\
			 & wall thickness & \unit{6}{mm}\\
			 & zirconia liner & \unit{4}{mm}\\

		ice & height &  \unit{0.90}{m} \\
		DOMs & distance between centers &  \unit{0.58}{m} \\
		perlite & thickness &  \unit{0.40}{m} \\
		outside tank			 &polystyrene below tank &  \unit{(100\pm 2)}{mm}\\ 
		&polyurethane foam around tank&  \unit{(50 - 100)}{mm}\\ 
		\hline
		\end{tabular}

	\label{tab:IceTopTank}
\end{table}

A schematic cross section of the IceTop detectors is shown in Fig.\ \ref{fig:TankSchematic} (dimensions are reported in Table \ref{tab:IceTopTank}).  The tanks are made of black, cross-linked polyethylene, \unit{6}{mm} thick, \unit{1.1}{m} high, with a \unit{1.82}{m} inner diameter and are filled with ice to a height of \unit{0.90}{m}. 
Most of the tanks have an integral diffusely reflective white liner
made by dispersing zirconium dioxide powder\footnote{Supplier: Stanford Materials, Irvine, CA 92618 U.S.A.}, referred to as zirconia, into  High Density Polyethylene (HDPE) by extrusion\footnote{Manufacturer: PlastiScience, LLC, Smyrna, DE 19977}. The polyethylene, containing 6\% zirconia by volume, is milled to 
fine white powder\footnote{Manufacturer: Power King, Texas} with average particle size \unit{45}{$\mu$m}.
The tanks are produced by a rotational molding technique\footnote{Manufacturer: PolyProcessing, Winchester VA}  starting with molding the \unit{6}{mm} thick, black outside layer and then covering it using the white powder to form a \unit{4}{mm} thick layer. A final curing process leads to cross-linked bonds in the HDPE polymer structure which strengthens the tanks and finally determines the diffusive reflectivity of the liner.
\begin{figure}
	\centering
\includegraphics[width=0.60\textwidth]{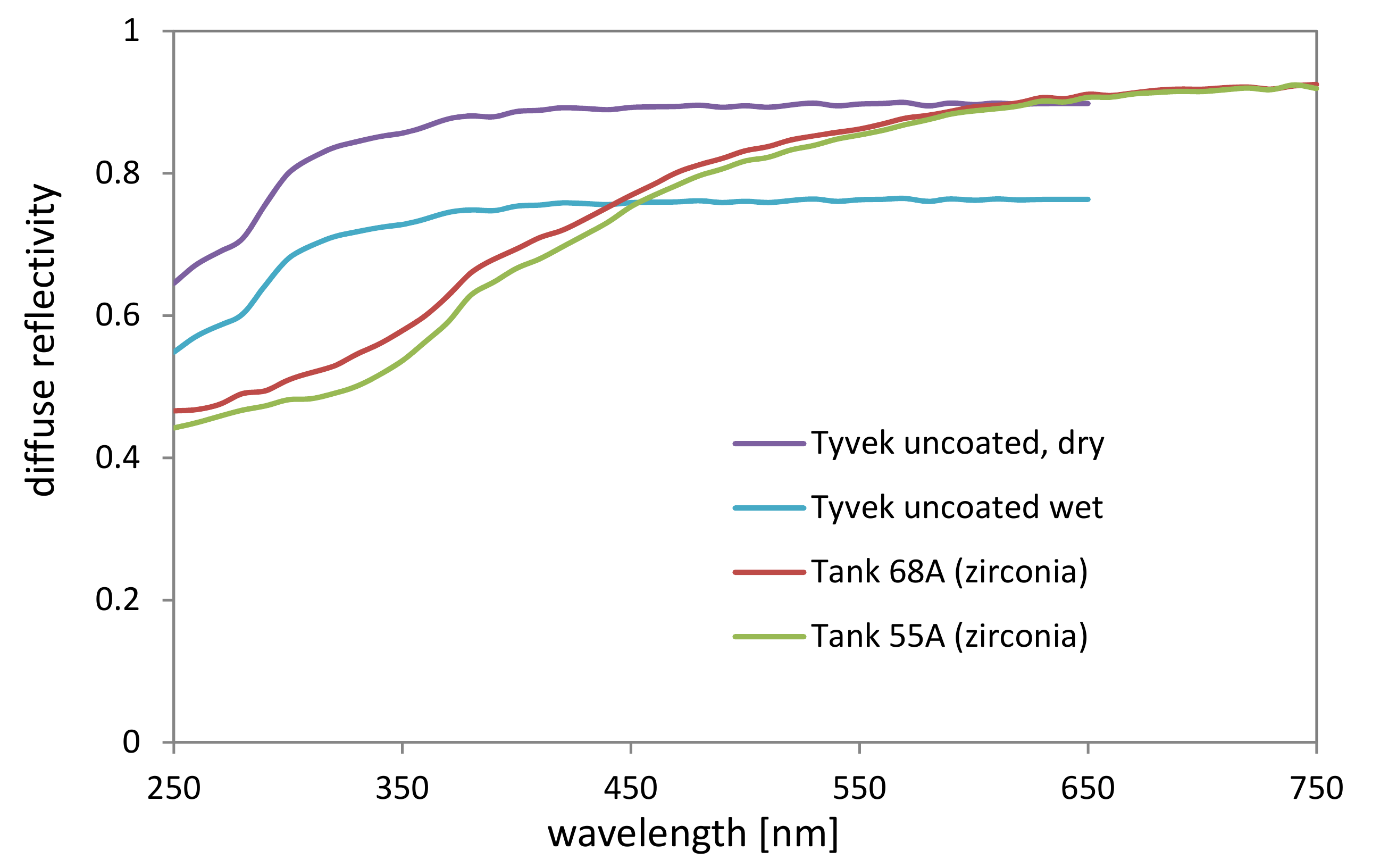}
	\caption{Measurements of the wavelength dependence of diffuse reflectivity on samples from Tyvek bags (`uncoated' refers to the 2005 tanks) from tanks with zirconia liners. For more details see text.}
	\label{fig:DifReflectivity}
\end{figure}
 Eight tanks commissioned in 2005 and four tanks deployed in 2011 have Tyvek\footnote{Dupont brand of a synthetic textile made of high-density polyethylene fibers} linings in form of bags loosely covering the tank walls (Table \ref{tab:ICIT_configs}).

Figure \ref{fig:DifReflectivity} shows measurements of the diffuse reflectivity of tank liners as a function of the wavelength. The measurements were done outside of the tanks and usually in dry condition. For Tyvek also a comparison of reflectivities of dry and wet material from the 2005 tanks is shown in the plot. The difference between dry and wet Tyvek is relatively large because the water fills the Tyvek matrix which is for dry material filled by air. The reflectivity depends on the ratio of the refractive indices of filler and matrix which is difficult to predict for Tyvek under real conditions in ice-filled tanks.  The Tyvek tanks commissioned in 2011 (Table \ref{tab:ICIT_configs}) exhibit in ice 
 optical properties which are resembling much more the properties of zirconia tanks according to pulse decay time and signal size (see Table \ref{tab:IceTopPulseCharacteristics}). The reason for this unexpected behavior could not be fully traced back because several other conditions were different for the first tanks installed in 2005. For example, in 2005 the water came from the station, while later the water was prepared in the drill camp and additionally filtered. For zirconia tanks all reflectivity curves are quite similar as demonstrated by the two examples in  Fig.\ \ref{fig:DifReflectivity}. In contrast to the Tyvek liner the zirconia liner does not absorb water. However, the real conditions for the liner-ice boundary are not known in both cases. In the simulation (described in Section \ref{subsubsec:tanktop}) the measured reflectivity curves are scaled to match the observed optical properties of the tanks, which are the pulse height and decay time (for typical single-particle pulses as measured with muons, see Section \ref{sec:Discriminators}).  

\begin{figure} 
\includegraphics[width=0.7\textwidth]{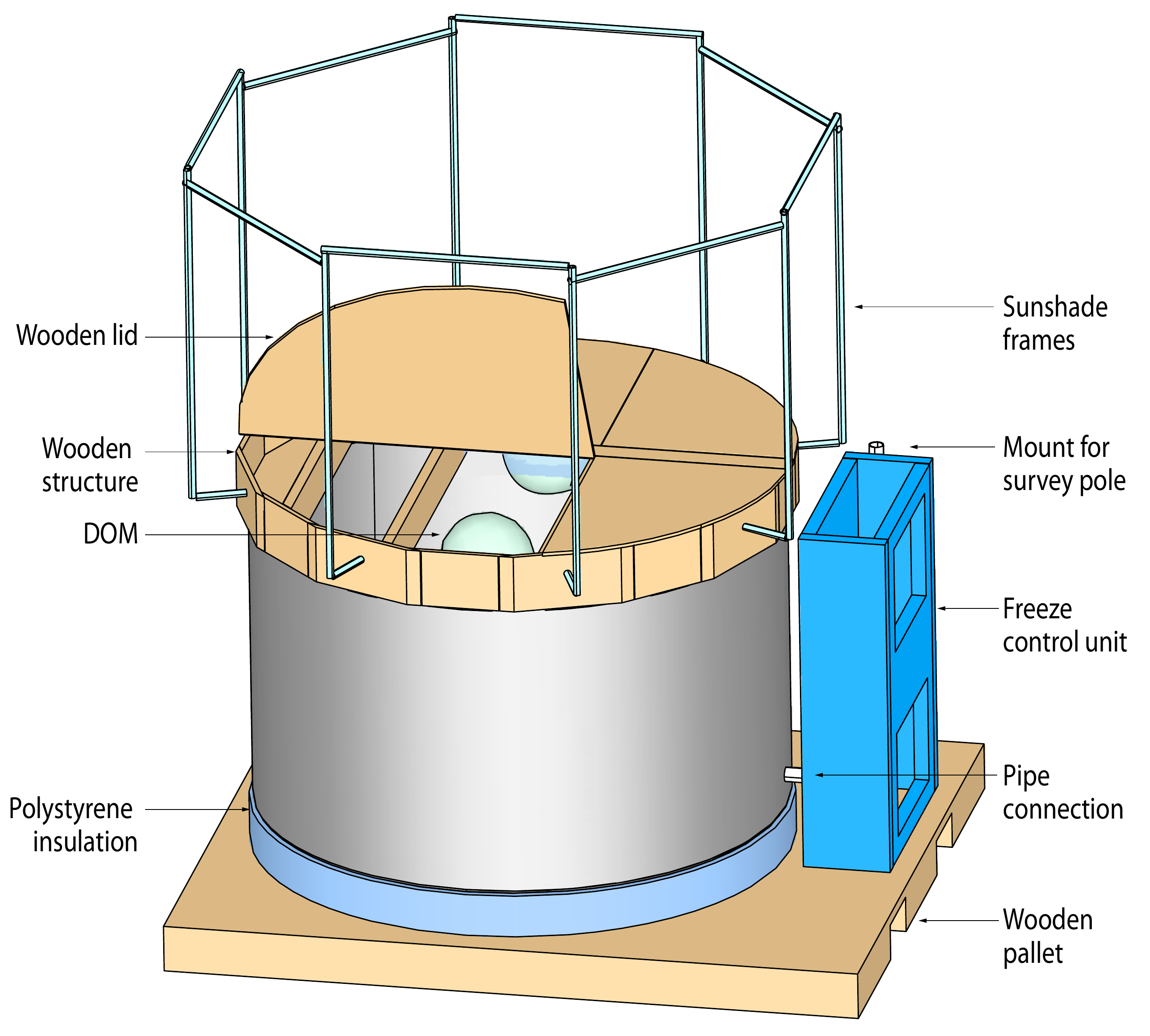} 
   \caption{Design rendering of an IceTop tank assembly before the outer foam insulation is applied.}
   \label{fig:PaulTank}
 \end{figure}

Figure \ref{fig:PaulTank} shows the tank assembly. Attached to the top of the tank is a wooden support structure (shown in brown in Fig.\ \ref{fig:PaulTank}) 
 that consists of $2'' \times 8''$ joists covered by hinged, plywood lids.
The joists support the DOMs before the water freezes and 
the whole top serves as a structural platform to protect the detector after deployment. The hinged panels allow access for installing  DOMs and let heat escape during the freezing process.
The space between the ice surface and the lid is filled with expanded perlite (amorphous volcanic glass, expanded to low density with grain sizes of the order of \unit{1}{mm}  \cite{PerliteManufacturer}) for thermal insulation and light protection (see Table \ref{tab:g4material} for composition).  

The tank rests on a base of \unit{10}{cm} thick rigid polystyrene insulation 
placed on top of a wooden pallet.  The tank is 
mechanically strapped to the pallets with ropes. 
  An enclosure for the Freeze Control Unit, described below, is mounted on the tank.
On the right side of the Freeze Control Unit the drawing shows a tube into which  poles carrying survey markers are inserted. Two vertical lines of temperature sensors are
 installed on the outer wall of the tank (not shown in Fig.~\ref{fig:PaulTank}) to monitor the freezing progress, one next to the Freeze Control Unit and one on the opposite side. There are 8 sensors per line in distances of about \unit{15}{cm}, starting very near the tank bottom and ending about \unit{15}{cm} above the water/ice level (measurements with these sensors are displayed in Fig.\ \ref{fig:freeze-history}). The whole tank structure is integrated by applying expanding polyurethane foam around the tank from the pallet to the lid.  The frame structure above the tank is only attached during the freezing process and carries sunshades to keep direct sunshine off the ice surface. 
 
 The assembly took place in the US and the completed units were transported to the South Pole. The sunshade frames were installed before deployment and removed after freezing was complete.

\subsection{Deployment and commissioning}
\label{subsec:tank_deployment}

\begin{figure}
	\centering
		\includegraphics[width=0.60\textwidth]{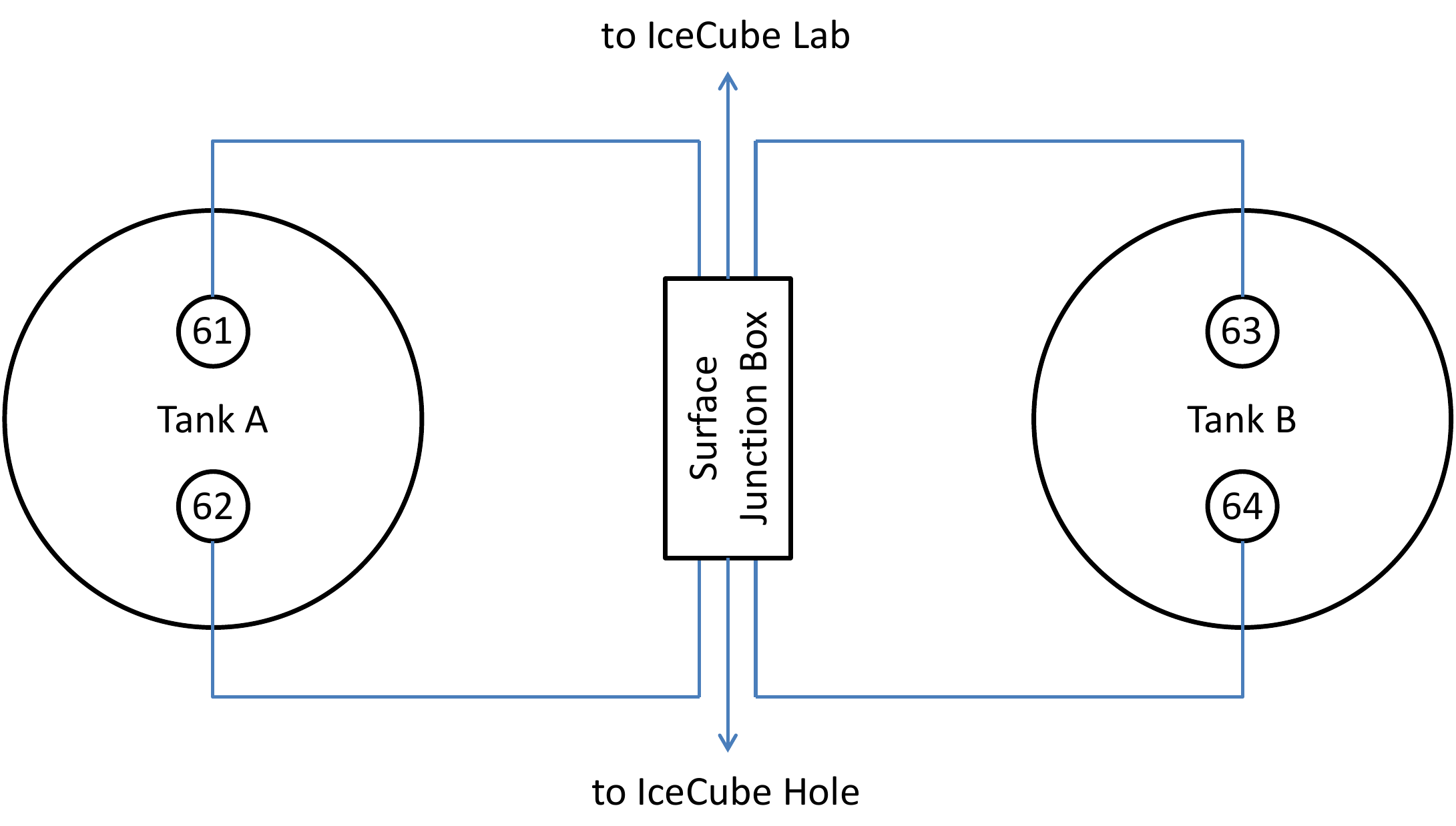}
	\caption{Sketch of the orientation of tanks and DOMs. Indicated is the position of the tanks relative to the nearest IceCube hole. Between both tanks is the `surface junction box'  which connects the 64 DOMs of the IceCube string and the IceTop tanks to the cable running to the IceCube Lab. The string DOMs have the numbers 1 to 60, the IceTop DMs from 61 to 64. The IceTop DOM numbering scheme is shown in the sketch. The DOMs 61 and 63 have high gains and  62 and 64 low gains (see also Fig.\ \ref{fig:icetop_lc}).}
	\label{fig:Tank-DOM-positions}
\end{figure}

\begin{figure}
  \includegraphics[width=0.5\textwidth]{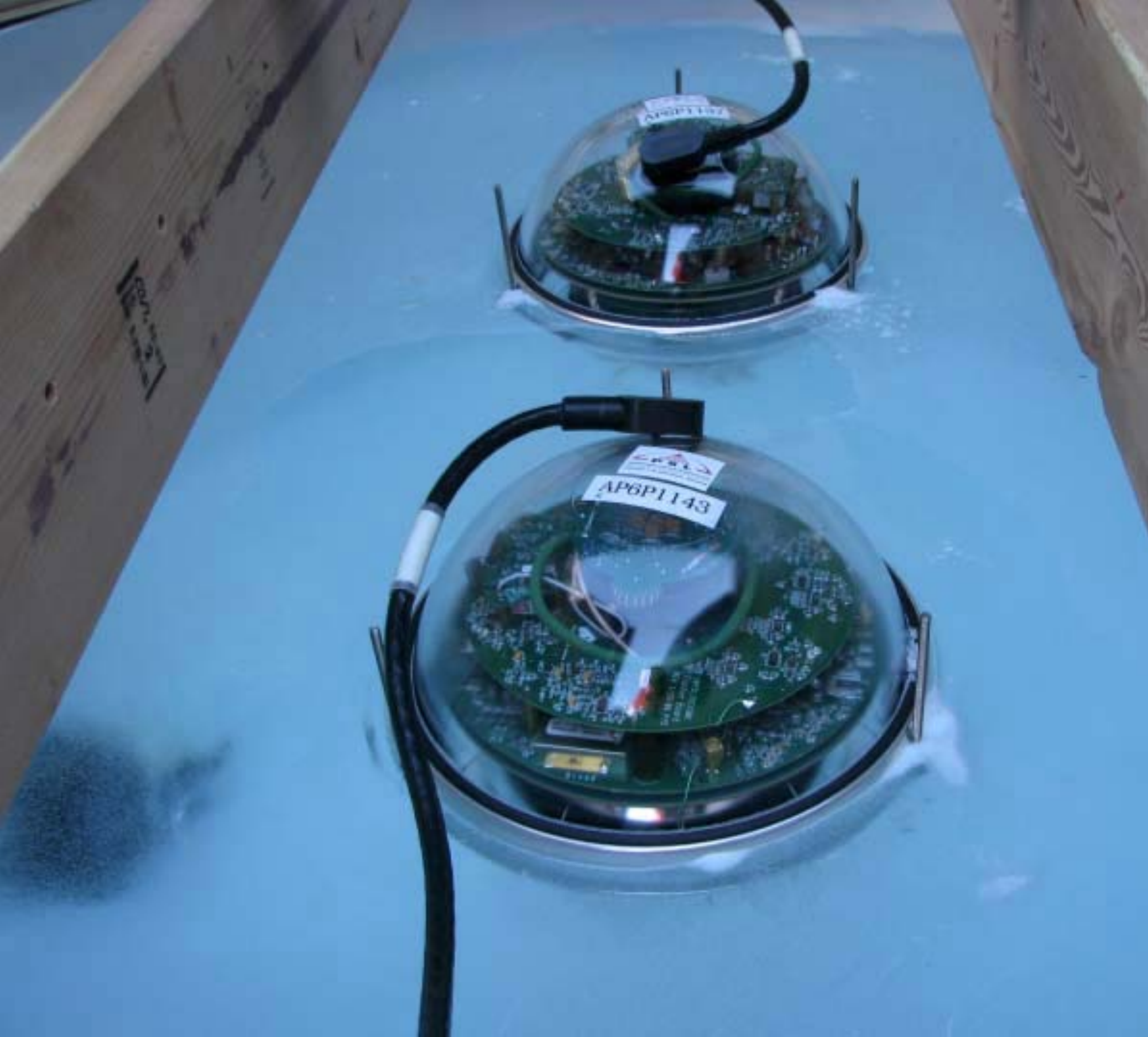}  
   \caption{Photograph of a frozen tank ready to be closed.  The degasser unit, which remains 
frozen at the bottom, is visible under \unit{90}{cm} of ice in the lower left.  The fuzziness in the outline of the degasser is mainly the result of imperfections in the surface
of the ice.}
   \label{fig:54Brev}
 \end{figure} 

IceTop deployment was integrated with the in-ice drilling and deployment.  Tank locations were laid out during the survey for the surface cables of IceCube.   Stations consisting of two tanks, \unit{10}{m} apart, were placed nominally \unit{25}{m} from the top of the corresponding in-ice hole.
Looking from the hole towards the tanks the left tank was labeled `A' and the right one `B' (Fig.\ \ref{fig:Tank-DOM-positions}), with the exception of station 48 where it was accidentally reversed. 
 Tanks were set into trenches so the tops of the tanks were even with the snow surface to minimize drifting.  After the tanks were in position, the DOMs were mounted on hangers attached to the structural joists of the tanks (Fig.\ \ref{fig:PaulTank}) with
the PMT half facing down. The equator of the DOM was set at \unit{90}{cm} above the
bottom of the tank so that the photocathode portion of the DOM was submerged.
The two DOMs in each tank were placed along a diameter of the tank
symmetrically about its center with a separation of \unit{58}{cm} 
between the centers of the two DOMs.  
  Figure~\ref{fig:54Brev} shows a tank with its two DOMs installed in the ice just before the tank was filled with perlite insulation and closed. 

A `surface junction box' was placed between the two tanks of each station  (Fig.\ \ref{fig:Tank-DOM-positions}) and the four IceTop DOMs as well as the corresponding in-ice cable with wires for its 60 in-ice DOMs were connected to it. The connection to the IceCube Lab (ICL) at the center of the array was made by a surface cable carrying the wires for the 64 DOMs at each hole/station.
The order of deployment of the IceTop stations followed the trenching and laying of surface cables.  
 The surface cable also carries wires to provide power and housekeeping signals for the Freeze Control Unit.  At the end of the season, these wires were secured to the survey marker of each tank and are available for possible future use.

Since the Freeze Control Units had to be kept warm, the tanks could not be filled until the surface cables were connected in the IceCube Lab.  When the surface cables were laid and connected to power, the Freeze Control Units were installed and immediately powered up.  Soon after,
the tanks were filled with water from the in-ice hot water drilling system (except for the tanks commissioned in 2005, see remark in Section \ref{subsec:tank_design}). 
 At the beginning of the drilling season the water contained as much as 4000 ppm of propylene glycol, which remained from the system protection for the winter. A filter assembly reduced the glycol content to a level of less than 10 ppm in the IceTop tanks.
The filtered water was transported to each station in an insulated water reservoir mounted on a dedicated sled.  
The water was filled to a depth of \unit{90}{cm} and the surface was maintained at that level until the freezing was complete.

Immediately after the tanks were filled, the Freeze Control Units were turned on with the tank lids closed for an initial two to three day period of cooling and degassing.  
During this period, sunshades were mounted on pre-installed frames that extended above the tanks (Fig.\ \ref{fig:PaulTank}).  The tanks were then opened and the freeze-in process lasting about \unit{50}{days} began.  The surface of the ice in the tanks had to be kept clear of snow to prevent an accumulation that would have insulated the tank and slowed or stopped the freeze process.  After an initial period of operating the degasser nearly continuously, the duty cycle was
decreased in steps to the minimal level needed to insure clear ice. After the freeze was complete 
the top of each tank (\unit{40}{cm} between the ice surface and the lid) was filled with expanded perlite insulation and the top closed. The sunshades  were removed, and the DOMs could be turned on and commissioned. The snow was backfilled and groomed around the tanks. The Freeze Control Units were retrieved from the tanks and stored for future use.

\subsection{Water freezing} \label{subsec:freezing}

A natural process that allows the water to freeze from the top to the bottom 
was adapted to make clear, crack-free ice.  When ice freezes in a confined volume
from the top down, two problems must be managed.
As the ice front advances, air is trapped in the remaining water and would soon rise to saturation level
and make bubbles if it were not removed. The expansion at the phase transition from water to ice also has to be dealt with.

\begin{figure}
	\centering
		\includegraphics[width=0.60\textwidth]{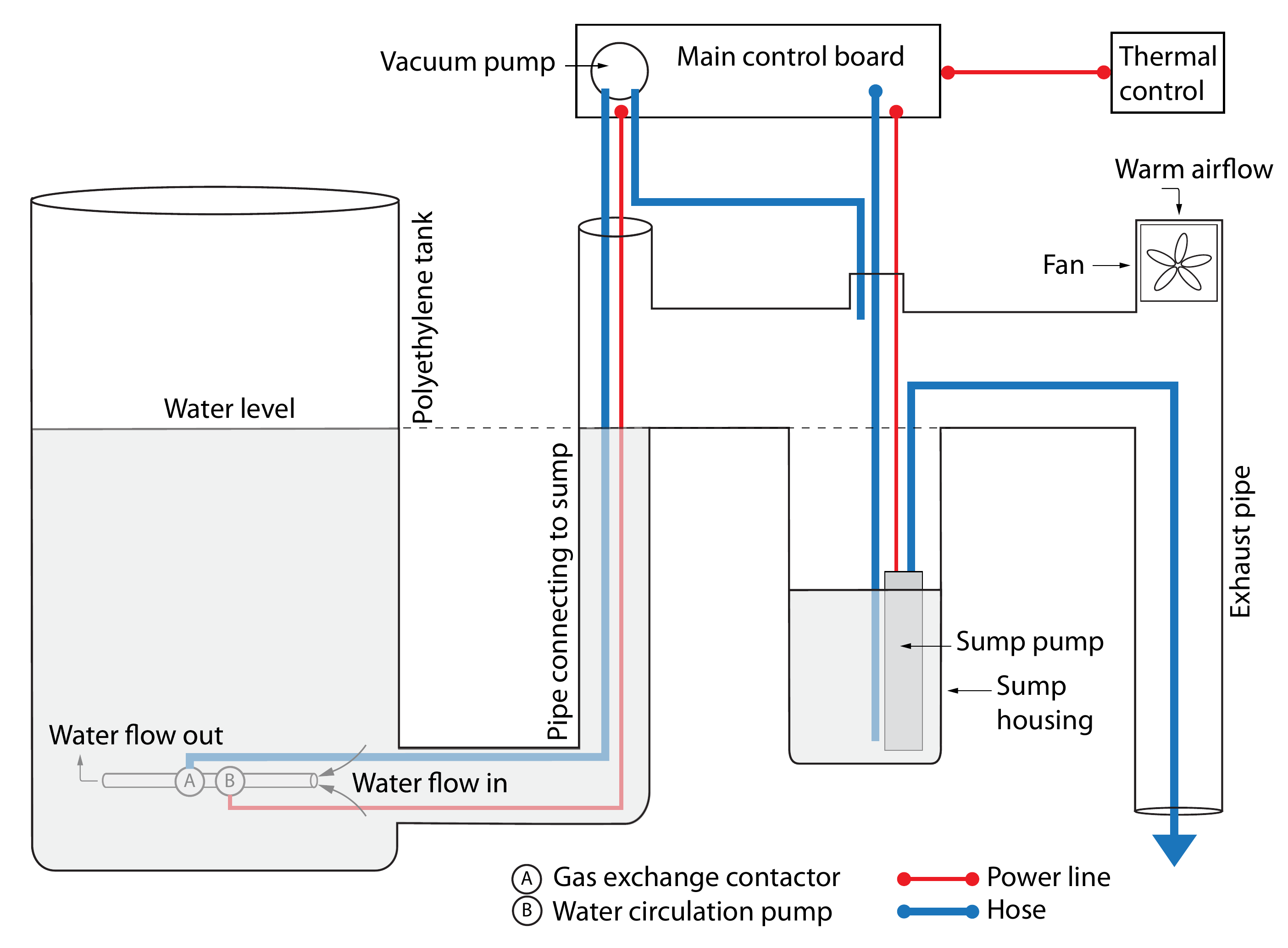}
	\caption{Schematic of the operation principle of the Freeze Control Unit: Initially the tank is filled until the water level spills over to the second container with the sump pump which pumps overflow water to the outside (third pipe). At the bottom of the tank, a pump pushes water through the degasser unit, keeping the water circulating until all the water is frozen.} 
	\label{fig:IceTopTankSketch}
\end{figure}

The principle of the freezing operation is sketched in Fig.\ \ref{fig:IceTopTankSketch}. The air is removed using a commercially available {\it contactor} (A in Fig.\ \ref{fig:IceTopTankSketch}) consisting of a cluster of tiny tubes made out of a semi-permeable membrane passing through a plastic cylinder. A submersible pond pump (B in Fig.\ \ref{fig:IceTopTankSketch}) with a five-micron filter to keep any sediment from fouling, forces water through these tubes.
The contactor is kept at a modest vacuum (\unit{130}{hPa}).  About half the 
dissolved air is removed in one pass with this setup.  Except for the vacuum line and the power supplying the pump, the unit is self-contained. A negative pressure ensures any leaks are inward, and the resulting water is removed by a water tolerant vacuum pump.
Tests showed that it is sufficient to run this assembly at the bottom of the tank. 
If the tank were perfectly insulated, the densest water (4\,\degC ) would concentrate at the bottom and only the lower portion of the tank would be degassed.
Even though the sides of the tank are insulated, there is enough thermal transfer so that a stable temperature gradient from top to bottom was never set up. The small amount
of heat loss through the sides is sufficient to keep the water in the tank well mixed during the freezing process. The temperature of the entire volume, which remains near 0\,\degC , and the dissolved
 gas content are nearly independent of depth.  With the pump operating at full capacity of 2 liter per minute, 
the entire 2500 liter volume of water in the tank can be processed about once per day.

Removal of the water as the ice front advances is also straightforward. The ice rapidly bonds to the sides of the tank with sufficient strength to form a plug.  The pressure builds only high enough to force the water out 
through a pipe (Fig.\ \ref{fig:IceTopTankSketch}) open at the bottom of the tank that extends into the insulated and heated Freeze Control Unit enclosure. The short horizontal section sets the initial water level in the tank.  Inside the Freeze Control Unit, the water collects in a sump which is pumped out when full.  Pumping water out rapidly when needed prevents the water from freezing and plugging that discharge hose which is exposed to about \unit{-30}{\degC} temperatures. The Freeze Control Unit controls the circulation and vacuum pumps for the degasser, and the sump pump. Resistive heaters are used to keep the temperature inside the Freeze Control Unit at \unit{30}{\degC}.
 
\begin{figure}
	\centering
		\includegraphics[width=0.80\textwidth]{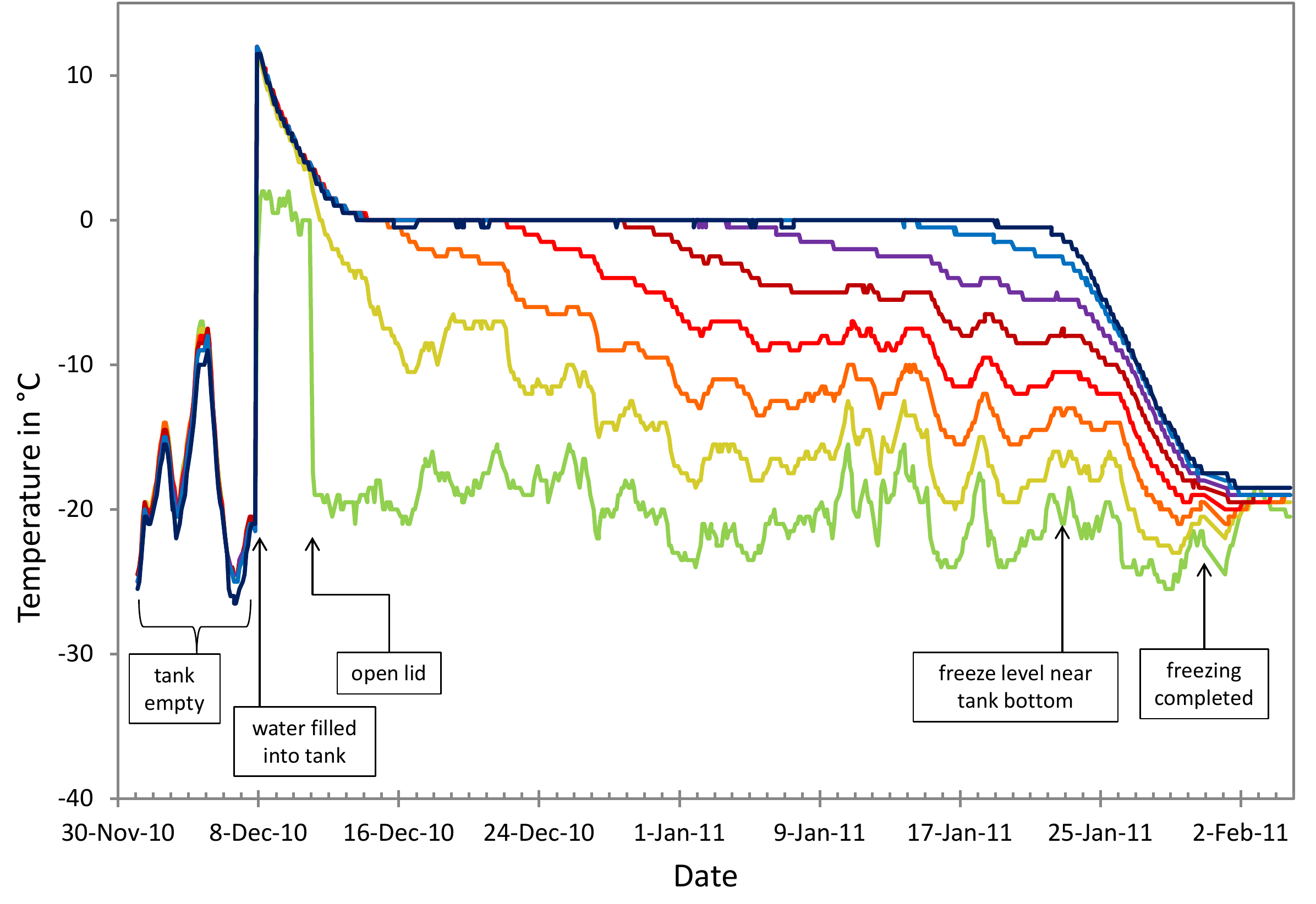}
	\caption{History of the freezing process as recorded by the temperature sensors which are mounted to the wall of the tank at different heights (8 sensors starting from the tank bottom to 15 cm above the water/ice level, see Section \ref{subsec:tank_design}). In the plot the lowest curve, staying at about \unit{-20}{\degC}, is measured by the highest sensor (above the water), the next curves following upwards are measured by the sequentially lower sensors.}
	\label{fig:freeze-history}
\end{figure}

\paragraph{Freeze history} The history of a freezing tank is illustrated in
 Fig.\ \ref{fig:freeze-history} depicting the depth profile of water/ice
  temperatures in the tank during the freezing process. The temperatures are measured through the line of temperature sensors mounted vertically on the
  outside. Initially the tank is empty so all of the sensors measure the ambient air temperature (here around \unit{-20}{\degC}). Water is then added to the tank and all but one of the sensors now follow the water temperature. The exception is one sensor that is located above the surface level of the water. At this stage the tank lid is kept closed so the water can cool uniformly and the degassing unit can reduce the concentration of air in the water (see description above). 
When the water has reached equilibrium and is near the freezing point the lid
is opened, causing an immediate drop in the `free air' sensor (the highest in the sensor line) to near ambient level. The water begins to freeze from the top down. As the freezing front passes each sensor in turn the temperature drops below \unit{0}{\degC}. The thermal profile of the sensors near the surface follows the ambient temperature more closely than that of the deeper sensors as would be expected from the need for the heat to diffuse through the block of ice.
Eventually all of the sensors fall below zero, but the thermal gradient remains high since there still is liquid water in the center of the tank even though the sides are completely frozen. At this stage water is also still being drained into the sump, further indication that the freeze is not complete. Finally, when the last water freezes the temperature gradient collapses and water evolution ceases. The freeze control unit is now disconnected, the perlite insulation placed on top of the tank, and the lid screwed shut. At this point the `free air' sensor now joins the others as it is buried in the insulation and thermally better connected to the ice than it is to the outside air.

\subsection{Survey of the tanks} \label{subsec:survey}
Surveyors determined the location of the center of each IceTop
tank and the location of the survey marker on the tank.  From those two
measurements and the known tank parameters, the location of the center of
each IceTop DOM is determined to an accuracy of 5 cm.  The tops of the
deep holes of IceCube were determined to a similar accuracy in the
same coordinate system.  Initial location of the deep DOMs was determined from the cable payout records and pressure sensor data. 
In a second stage LED light pulses generated in the DOMs (flashers) were used to determine relative vertical offsets of strings.  Reconstructed muon data and flasher data are used to monitor for shearing or other changes over time, which
have so far not been observed.  Locations of deep DOMs are known to an accuracy of better than \unit{1}{m}. 
%

The IceTop tanks lie roughly on a tilted plane with the top of the tanks at \unit{z=1944.54}{m}  for the lowest point (tank 18A in Fig.\ \ref{fig:map2011}) and  at \unit{z=1950.08}{m}  for the highest (tank 74A) where $z=0$ is defined in the center of IceCube.


\section[Front-end electronics]{Front-end electronics}\label{sec:icetop_fee}

The pivotal components of the detector are the Digital Optical Modules (DOMs). They detect the Cherenkov light produced by the shower particles in the ice of the tanks and convert these signals into digital information, which is sent to the central counting house, the IceCube Lab. 

IceTop DOMs have the same hardware as all other DOMs in IceCube, but some details of the data acquisition system are different due the different physics requirements and different environmental conditions. For example, the pulses generated by air showers deliver in general more charge and thus the pulse thresholds are higher; another difference is that the IceTop DOMs are subject to the seasonal temperature changes while the temperature in the deep ice is extremely stable.

\subsection{Digital Optical Module (DOM)}\label{subsec:DOM}

\begin{figure}
  \centering
  
\includegraphics[width=0.49\textwidth]{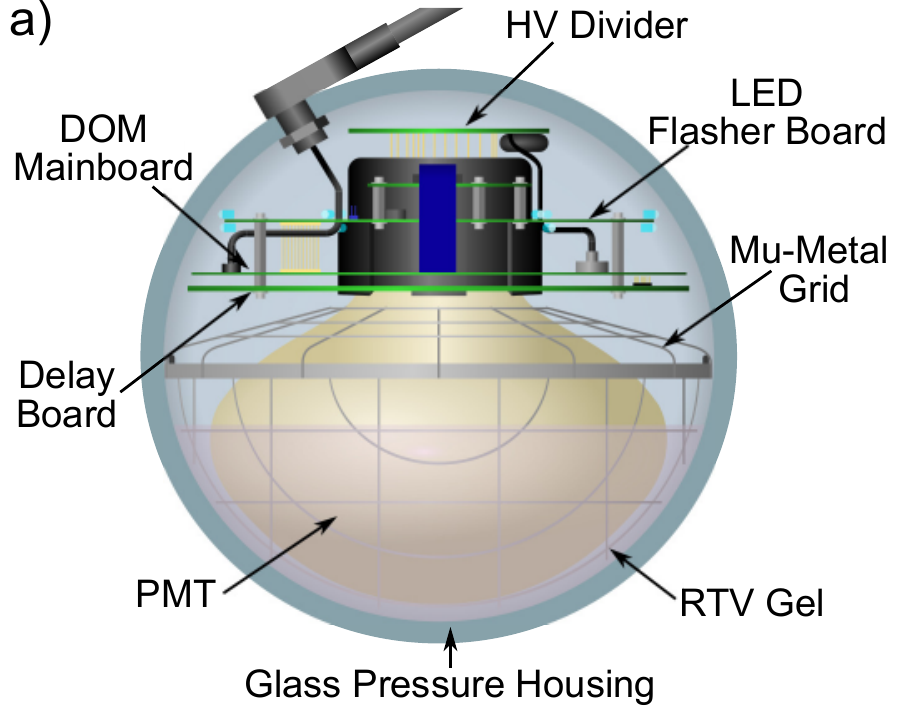}\hfill
\includegraphics[width=0.49\textwidth]{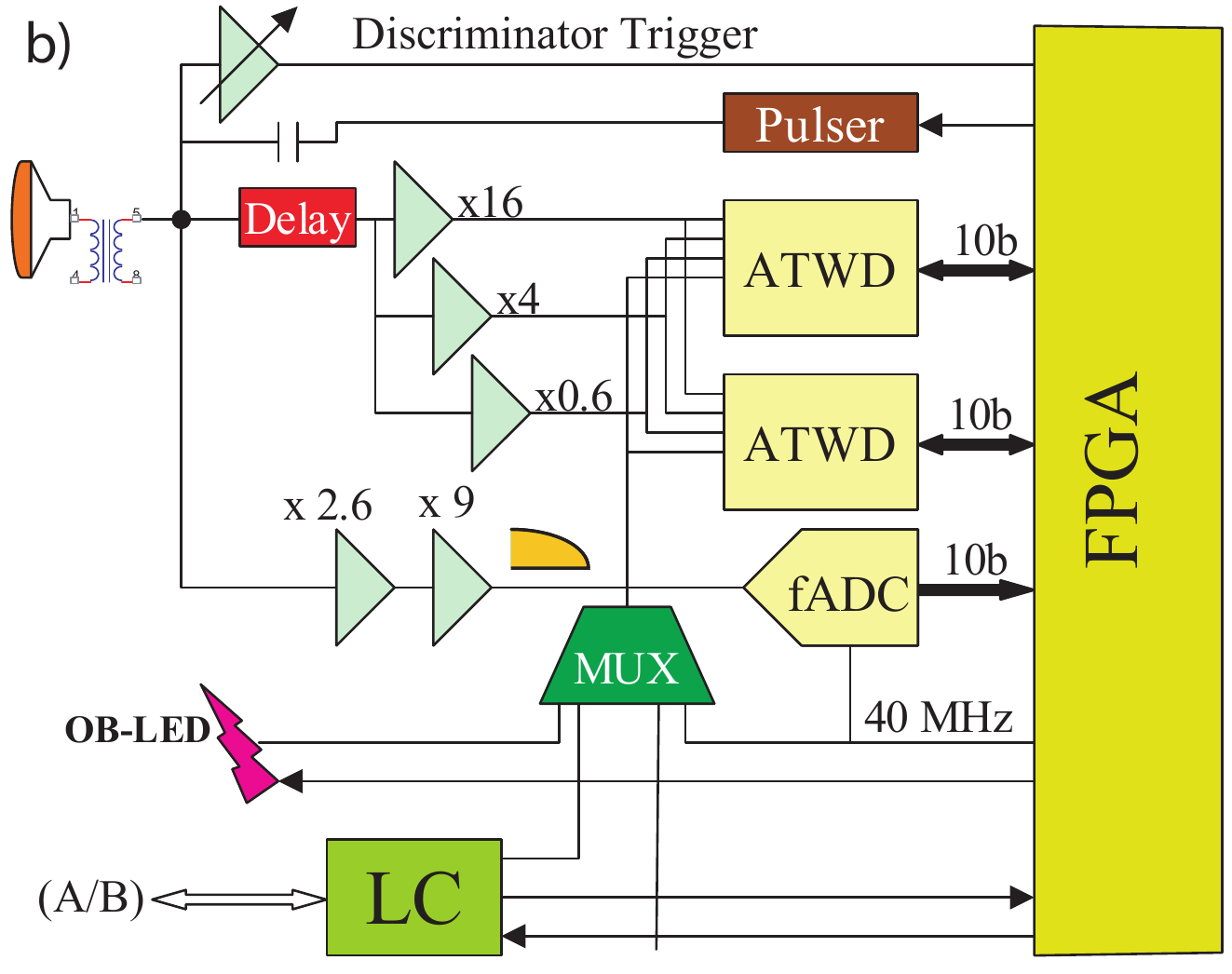}
  \caption[Digital Optical Module]{The IceCube Digital Optical Module (DOM): a) Schematic view of a DOM. b) Analog signal part of the block diagram of the DOM mainboard.}
  \label{fig:icecube:dom}
\end{figure}

Each DOM \cite{DOMPaper} consists of a 10" Hamamatsu photomultiplier tube (PMT) \cite{PMTPaper} and electronic circuitry, contained inside a $33\un{cm}$ glass pressure sphere, as shown in Fig.\ \ref{fig:icecube:dom}\,a. The PMT is shielded from Earth's magnetic field by a mu-metal grid. The two DOMs of a tank are operated with different PMT gains,  high gain (HG) and low gain (LG), to adapt the dynamical range to the extremely different signals in air showers (Table \ref{tab:FrontEndComponents}). 

Data taking, triggering, digitization and communication with the IceCube Lab are controlled by an FPGA (Field Programmable Gate Array) with an on-chip CPU on the DOM mainboard. The  CPU permits remote DOM firmware updates.
Timing of the DOM is controlled by a free-running $20\un{MHz}$ oscillator which is regularly calibrated with the master clock in the IceCube Lab. 
In addition, each DOM mainboard is equipped with an LED for calibration, see Section \ref{subsec:calibration_dom} (there are other LEDs in the DOM which are used to study the optical properties of the deep ice but are not used in IceTop).

A detailed description of the Digital Optical Modules can be found in \cite{DOMPaper}. Some characteristic parameters for the IceTop DOM operation are summarized in Table \ref{tab:FrontEndComponents}.

\begin{table}[pt]
	\centering
	\caption{Important parameters of front-end components.}	
	\label{tab:FrontEndComponents}
		\begin{tabular}{lll}
		component & quantity & value\\
		\hline
			PMT 		& quantum efficiency at \unit{390}{nm} (near maximum) & $\sim 25\%$ \cite{hamamatsu-PMTdata}\\
				 & quantum efficiency, av. for \unit{300-650}{nm} & $\sim 18\%$ \cite{PMTPaper}\\
			 &cathode area & \unit{\sim\,550}{cm$^2$} \cite{hamamatsu-PMTdata}\\ 
				 & high voltage, gain  & HG: $5\times10^6$ at $\sim$1250\,V\\ 
			&  & LG:  $10^5$ at $\sim$750\,V \\
			& saturation current (50\% from linearity)& HG: \unit{140}{mA} 	\\
			& & 	 LG: \unit{50}{mA} \\
			transformer & time constant (at \unit{-30}{\degree C}): &\unit{1-2}{$\mu$ s} (old), \unit{12}{$\mu$s} (new) \\
	& front-end impedance &  \unit{50}{$\Omega$} (old), \unit{43}{$\Omega$} (new)  \\	
	\hline		

		\end{tabular}
\vspace{10mm}
	\centering	
	\caption{Characteristics of the three ATWD readout channels (nominal or approximate values). The smallest digitization value of each channel is given in the third column as LSB = least significant bit.}
	\label{tab:ATWDProperties}
	\begin{tabular}{lccr}
channel & gain & LSB [mV] & saturation [mV]\\
\hline
    0 (high gain)&16 & 0.125 & $\sim$100\\
    1 (medium gain) & 2& 1.0 & $\sim$800 \\
    2 (low gain) & 0.25& 8.0 & $\sim$7500 \\   
    \hline 
%
		\end{tabular}
\vspace{10mm}
	\centering
	\caption{IceTop pulse characteristics. The numbers are average or approximate values, actual values can deviate. The notations `old' and `standard' tanks refer to the tanks commissioned in 2005 and after 2005, respectively.}	
	
		\begin{tabular}{l|cccc}
			tank liner  &   \multicolumn{2}{c}{`standard' tanks} &  \multicolumn{2}{c}{`old' tanks} \\
		PMT gain	 & HG & LG& HG& LG\\
			\hline
			rise time (10-90\%) & ~10 &~10 & ~10 & ~10 \\
			decay time (observed) [ns]  &    31.0 & 31.0  &43.5&43.5\\
		 decay time (input for simulation) [ns]  &  26.5 & 26.5  &41.9&41.9\\
		 VEM charge [PE] &    125& 110 & 200 & 180 \\
		 fraction of charge in peak bin  &   0.07 & 0.07 & 0.05&0.05 \\
		 peak voltage of 1 VEM [mV]  & 110&   2.2  & 100 & 2.0 \\
		 peak voltage of 1 PE [mV]  & 0.88 &0.02 &0.5 & 0.01\\
		 		  total charge of 1 PE [mV] (sum of ATWD bins) & 12.6 & 0.29   & 10.0& 0.2\\%
		 		  saturation in PE & 8\,000 &	125\,000 &	14\,000 &	225\,000 \\
		 		  saturation in VEM & 64 &	1140 &		70 &	1250 \\

		  \hline
		\end{tabular}
	\label{tab:IceTopPulseCharacteristics}
\end{table}

\subsection{Signal capture and digitization}\label{subsec:signal_capture}\noindent
The PMT signals are captured by the DOM electronics when the peak pulse voltage surpasses a programmable discriminator threshold. 
The relation between the pulse voltage, which the discriminator triggers on, and the pulse charge, which is the calibration quantity depends on the pulse shape and thus on the front-end electronics. Because of the importance for IceTop signal and threshold calibration we briefly describe the two electronic paths for pulse discrimination and recording. 

The PMT's cathode is kept grounded 
with the high voltage (HV) applied to the anode from where the signals are read out. The signals are decoupled from the HV by a wide-band inductive transformer (also referred to as `toroid') which is expected to have a higher reliability as compared to the more common capacitive decoupling. The effect of the droop caused by a transformer and its correction will be discussed below (Section \ref{subsec:droop}).

The signal paths described here can be followed in the schematic given in Fig.\,\ref{fig:icecube:dom}\,b. From the PMT's anode the signal passes the transformer and is then split into a path to the discriminator trigger and two paths to the signal recording electronics. A `Pulser', feeding into the same paths, is used in the  calibration procedure (Section \ref{subsec:calibration_dom}).

If the signal passes the discriminator threshold, the PMT output is sampled by a custom-made integrated circuit called `Analog Transient Waveform Digitizer' (ATWD)
\cite{ATWD-paper} in $3.33\un{ns}$ wide bins for a total of 128 bins, corresponding to a total sampling time of about \unit{427}{ns}. The signals on the ATWD path are delayed  by \unit{75}{ns} to allow for the trigger decision. Each ATWD has four channels, labeled ATWD0 to ATWD3, three of which are used for waveform recording. The fourth channel is not used for regular data taking but, depending on the running mode, other information can be recorded using a multiplexer (MUX in Fig.\,\ref{fig:icecube:dom}\,b), see details in \cite{DOMPaper}. To minimize dead-time, each DOM is equipped with two ATWD chips which are used alternately.

The inputs to the channels 0, 1 and 2 are amplified by nominal gain factors 16, 2 and 0.25, respectively (Table \ref{tab:ATWDProperties}). 
In the case of a discriminator trigger, the FPGA opens the input channels of one of the two ATWD chips and the three scaled signals are recorded synchronously. After the analog sampling, 128 Wilkinson ADCs digitize with 10 bit precision each of the 128 ATWD cells of a channel in parallel. The three channels are digitized sequentially, starting with the highest gain channel; channels with lower gain are only digitized if in the higher gain channels any bin is above 768 counts.  This takes about 30 $\mu$s
 per channel and up to 100 $\mu$s for all three channels. 
The three different gain channels of the DOMs make up for an effective dynamic range of 16 bit. Measuring charges in units of PE, the charge yield of a single photoelectron, the effective dynamic range  
of the high-gain DOM in a `standard' tank (Table \ref{tab:IceTopPulseCharacteristics}) spans from about 0.15 PE to about 10\,000 PE.  
The conversion from mV to PE is done by using the numbers in Table  
\ref{tab:IceTopPulseCharacteristics} (in this case for the `standard' tank) which are specific to the multi-PE pulses seen in IceTop tanks and not to the pulse  shape from single photoelectrons. The standard high-gain IceTop pulse has a peak voltage of 0.88 mV/PE corresponding to about 7\% of the total charge in the peak bin (for the single photoelectron pulse shape these numbers are about 40\% or 5 mV/PE in the peak bin).

 In response to a trigger, the ATWDs are launched in discrete time steps of \unit{25}{ns}, phase synchronized to  the frequency-doubled \unit{20}{MHz} DOM clock.
 The final time resolution is obtained from the position of the waveform in the ATWD with respect to the start time. The pulse time determination is described in Section \ref{sec:signal_proc}.

In parallel, the pulses are digitized continuously by a commercial 10-bit fast ADC (fADC in Fig.\ \ref{fig:icecube:dom}\,b) at a sampling rate of \unit{40}{MHz}. The fADC inputs are shaped with a time constant of 180 ns limiting the bandwidth to adjust to the lower sampling speed. In response to a trigger (Section  \ref{sec:trigger_daq}) the downstream DOM electronics records fADC data for \unit{6.4}{$\mu$ s}. The fADC data have not yet been used in IceTop analyses but could be used for studies of signal tails extended in time.

If DOM local coincidence or certain other conditions as descibed in Section  \ref{sec:trigger_daq} are satisfied, the ATWD and fADC digitization streams are losslessly compressed and written to memory. If a discriminator has triggered and an ATWD is available, at least the sum of the ATWD bin contents, called chargestamp, and a timestamp is extracted and written to memory. The ATWD recording is not possible while a waveform acquisition is still in progress or when both ATWDs are busy (see also the discussion on dead time in Section  \ref{sec:deadtime}).

\subsection{Pulse shapes and discriminator thresholds}\label{sec:Discriminators}

As described above, the readout of signals is initiated if a pulse crosses a voltage threshold set by the DOM discriminator (upper path in the schematic in Fig.~\ref{fig:icecube:dom}\,b). Any relation between voltages and charges depends on the pulse shape, which is determined by the tank properties and the electronics. Therefore we first discuss the typical shapes of tank pulses before turning to discriminator thresholds.

\subsubsection{Pulse shapes and VEM units} \label{subsubsec:pulsesVEMunit}
\begin{figure}
	\centering
	
\includegraphics[width=0.48\textwidth]{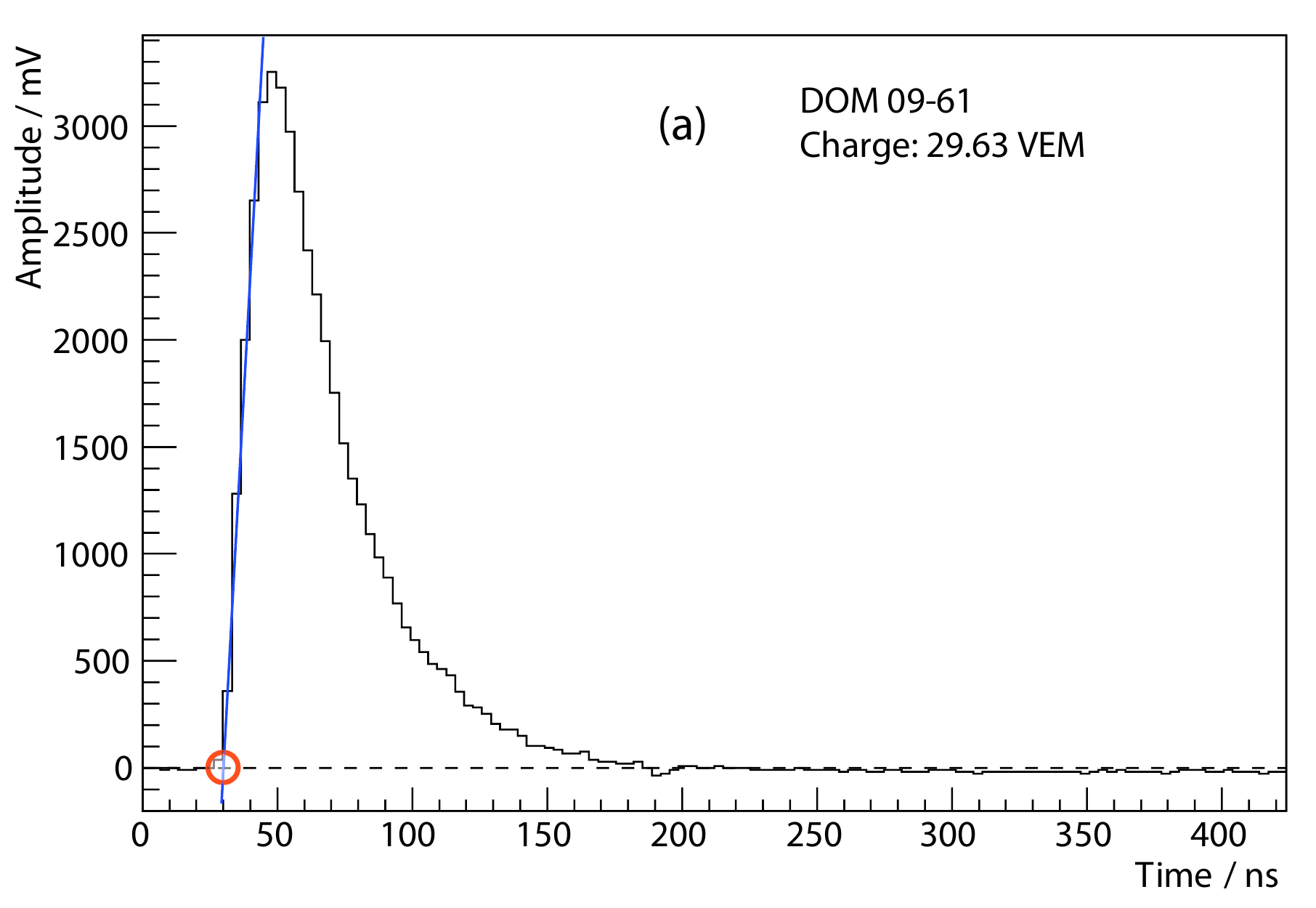}\hfill\includegraphics[width=0.48\textwidth]{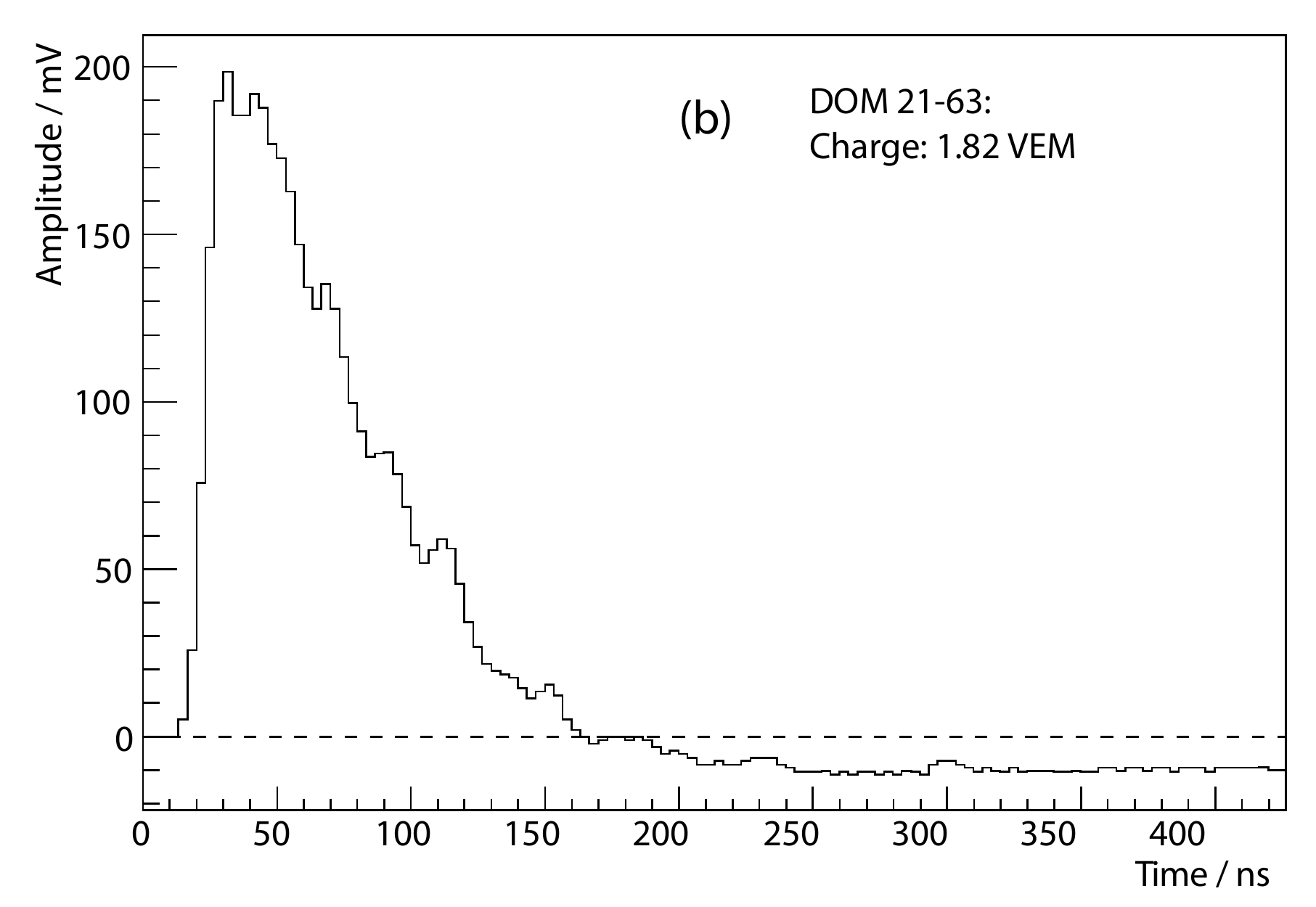}
	\caption{Typical IceTop waveforms. The most frequent shape has a steep rise, a single peak and an exponential tail. The waveform is here not droop corrected, but in panel a) the undershoot below the baseline after the pulse is very small, a feature of the `new' transformers installed in DOM 09-61 (Table \ref{tab:IceTopTank}).  An example for the 'old' transformer characteristics is shown in panel b) with a quite distinct undershoot. This DOM 21-63 belongs to the tanks installed in 2005 (Table \ref{tab:IceTopTank}) which exhibit longer decay times of the waveforms (Table \ref{tab:IceTopPulseCharacteristics}).  
In the left plot the blue line depicts the slope of the leading edge with the extrapolation to the baseline defining the pulse time (red circle), see Section \ref{sec:signal_proc}.}
	\label{fig:waveforms}
\end{figure}

Most of the pulses  have a similar shape given mainly by the reflectivity of the tank walls because the spread of crossing times of the shower particles are short compared to the decay time due to reflections (Fig.~\ref{fig:waveforms}). They can mostly be described by a superposition of standard pulses which is well represented by the signal of a single muon. The signal charge of a single muon is also used as a calibration standard for normalizing the signals of different tanks. The normalized charges are expressed in units of `vertical equivalent muons' (VEM) determined by calibrating each DOM with muons (see Section \ref{sec:calibration}). 

The standard pulses are characterized by a fast rise of the leading edge, about \unit{10}{ns} from 10\% to 90\%, and an exponential decay with a time constant of on average \unit{31}{ns} for most of the tanks (Fig.~\ref{fig:waveforms}\,a) and \unit{43.5}{ns} (Fig.~\ref{fig:waveforms}\,b) for the
8 tanks installed in 2005 (Section \ref{subsec:tank_design}).
If not otherwise mentioned we use as nominal pulse shapes those obtained for the  tanks which were installed after 2005 (Table \ref{tab:IceTopPulseCharacteristics}). In Table \ref{tab:IceTopPulseCharacteristics} two decay times are given: as observed and as put into the simulation. The difference is due to the electronic shaping which is applied also in the simulation as described in the next section.

\subsubsection{Pulse shaping by the electronics}
IceTop analyses are particularly dependent on a good simulation of pulse shapes to correctly  reproduce the experimental thresholds: Unlike the case for in-ice DOMs where the thresholds are set on fractions of a PE, the multi-PE thresholds in IceTop are more affected by the actual pulse shape generated by many PE's. 

\begin{figure}
	\centering
\includegraphics[width=0.40\textwidth]{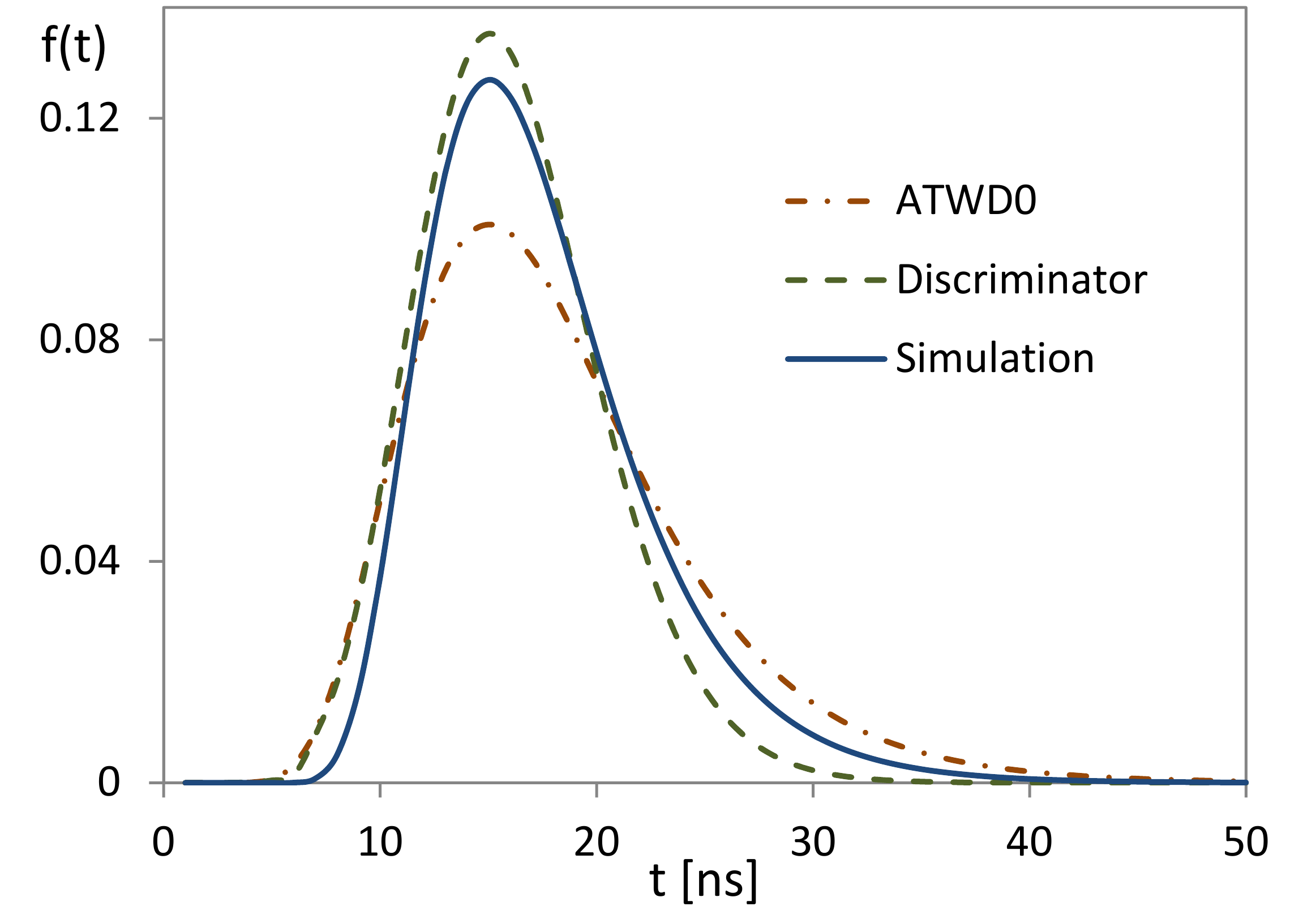}
	\caption{Comparison of the transfer function currently used in the detector simulation, as described by the function \eqref{eq:atwd_tf} with the parameters given in the text, with the measured transfer functions of the ATWD and discriminator paths (for `new' transformers). }
	\label{fig:TranferFunction}
\end{figure}

For the simulation, pulse shaping is described by transfer functions for the different paths with which the PMT output has to be convoluted.
 The software simulating the DOMs \cite{DOMsimulator} currently uses the following transfer function for the three ATWD channels and the discriminator path:
\begin{equation}\label{eq:atwd_tf} 
  f(t) = \frac{1}{t \sqrt{\sigma \pi}} \exp \left ( - \frac{(\ln(t) - \ln(t_{m}))^{2}}{2 \sigma^{2}} \right )
\end{equation}
with the parameters $\sigma$ and $t_{m}$ characterizing the general shape of the signal. The parameter $\sigma$ depends on the transformer version (0.41 for the
`old', 0.35 for the `new' transformers) and $t_{m}$ is set for both to 13.53 ns. The function has a maximum at the time $t_{max}$ which is related to the parameter $t_{m}$ by:
\begin{equation}\label{eq:tm_tmax} 
  t_{max} = t_{m} \: e^{ - \sigma^{2}}\, .
\end{equation}
A detailed study \cite{ClaireReport} found that the ATWD and discriminator paths have similar, but
 not identical, transfer functions. Both can be described by \eqref{eq:atwd_tf}, but the parameters $\sigma$ and $ t_m $ are significantly different resulting in narrower pulses at the discriminator than at the ATWD. The shapes of the transfer functions with the measured parameters for both paths and those currently used in the simulation are displayed in Fig.\ \ref{fig:TranferFunction}.  After convolution of the PMT pulses with the transfer functions, differences between both paths are reduced and a single set of simulation parameters gives a reasonable approximation. 
 
\subsubsection{Discriminator thresholds}  \label{subsubsec:DiscrThresholds}

\begin{figure}
	\centering
		\includegraphics[width=0.80\textwidth]{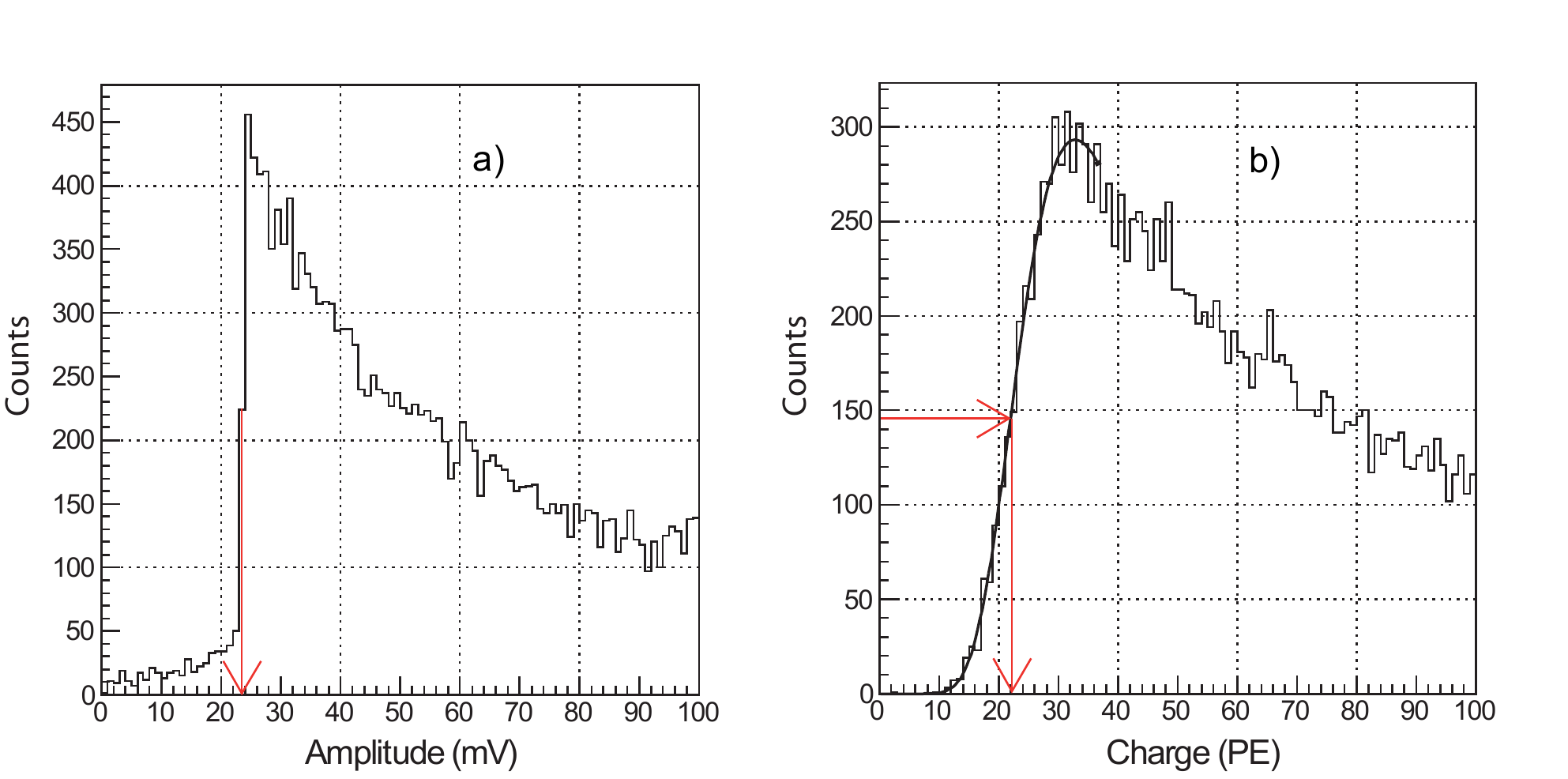}
	\caption{Threshold monitoring plots for the high-gain
DOM in tank 26A.  a) The peak voltage of the first arriving pulse of the waveform shows a sharp rise at a threshold of \unit{24}{mV}. b) The threshold in terms of charge is smeared out. The charge threshold can be defined by 50\% point of the leading slope of the charge distribution as shown on the plot.}
	\label{fig:2662-120014-qmoni-thr_upper}
\end{figure}

Each DOM has two voltage sensitive discriminators, called SPE (single-PE) and MPE (multiple-PE), which are generally used for triggering waveform capture. Both discriminators feature programmable thresholds with ranges differing by a factor 10.  In IceTop, both discriminators are used for multiple-PE thresholds.
In high-gain DOMs, the MPE discriminators are used for triggering on air showers with  thresholds of about \unit{20}{mV} corresponding to a signal charge of about \unit{23}{PE} for a standard pulse (see Fig.\ \ref{fig:2662-120014-qmoni-thr_upper}), while the low-gain DOMs use the SPE discriminators for the same purpose 
 with thresholds of about \unit{4}{mV} corresponding to \unit{270}{PE} \cite{IceTopConfigurations}.
With these thresholds, the single DOM rates are about \unit{1600}{Hz}. 

\begin{figure}
	\centering
		\includegraphics[width=0.50\textwidth]{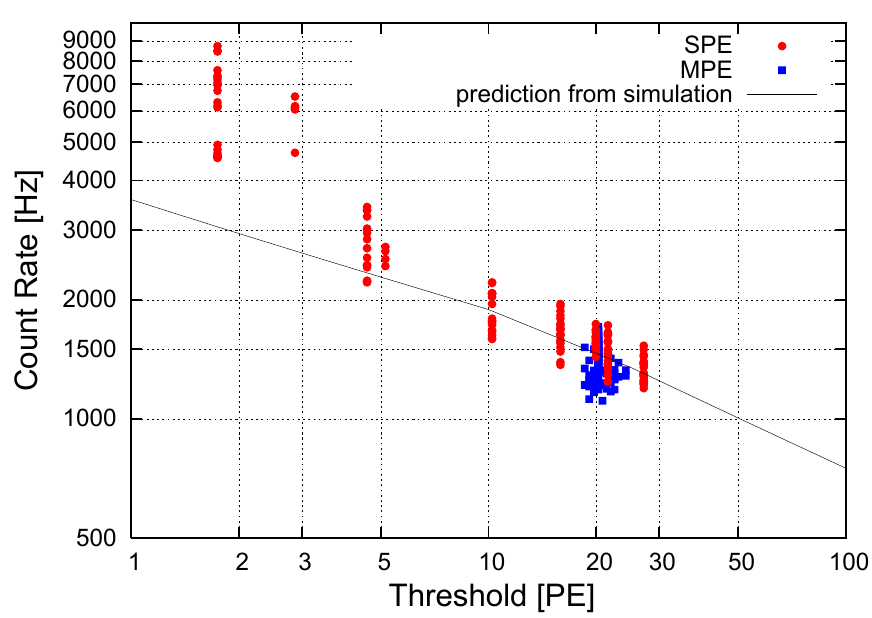}
	\caption{Count rate as a function of discriminator setting
of different DOMs. Rates are averaged over the interval from 17 February 22:00 UT to 18 February 02:00, which defines the ``base count rate'' for a study of a Forbush decrease observed in February, 2011 \cite{Takao_IT_icrc2011}.
The line shows the expected values based on a simulation \cite{Clem:2007zz}.}
	\label{fig:icrc0921_fig02}
\end{figure}

The SPE discriminators of the high-gain DOMs are used to record `scaler rates' at various thresholds (from about \unit{0.5}{PE} to \unit{30}{PE}) without triggering waveform capture. An example for the threshold dependence of rates is shown in Fig.\ \ref{fig:icrc0921_fig02}. The scaler rates are monitored continuously for solar flares, gamma ray bursts and other transient events. 
%

\subsection{PMT gains and saturation}\label{sec:PMT_gain}
The two DOMs in each tank are operated at different PMT gains (Table \ref{tab:FrontEndComponents}): 
The high-gain DOMs are nominally set to a gain of $5\times 10^6$, half the standard gain for in-ice DOMs. At this gain the saturation current (deviation by 50\% from linearity) is about \unit{140}{mA} corresponding to about 8\,000 PE for the standard IceTop pulses (Table \ref{tab:IceTopPulseCharacteristics}). 
The low-gain DOMs are set to a gain of $10^5$ and have a saturation current of \unit{50}{mA} which corresponds to about 125\,000 PE for standard IceTop
 pulses. Although the gain of the low-gain DOMs is 50 times smaller than of the high-gain DOMs, the number of photoelectrons at saturation level increases only by a factor of 15. The available 16 bit dynamic range of the ATWDs is
 about 10\,000 PE for high-gain DOMs and 450\,000 PE for low-gain DOMs (standard pulses), thus covering the non-saturated PMT range. The change from high-gain DOM to low-gain DOM signals is fixed to a much lower PE value, corresponding to less deviation from linearity (Section \ref{subsubsec:LG_cross-calibration}). Since the pulse shape stays very similar in a wide range (except for very low pulses and beyond saturation) the saturation can also be expressed in terms of VEM (for the `standard' pulse, see Table \ref{tab:IceTopPulseCharacteristics}). The low-gain DOM saturation determines the IceTop tank saturation which is typically of the order of 1000 VEM. 

The implementation of saturation effects in the simulation is described in Section \ref{sub:detsim}.


%

\subsection{Droop} \label{subsec:droop}
The transformer described in Section \ref{subsec:signal_capture} leads to a characteristic high-pass frequency response for all signals. While the time constant is large compared to IceTop pulse widths, the later parts of each waveform show a `droop' in amplitude, followed by undershoot below the baseline after the pulse ends. 
The pedestals of the ATWD and fADC are set to about 10\% of the maximum scale, to permit waveforms with below-baseline excursions (that effectively also reduces the available 10 bit range).

Choosing the time constants of the transformer requires a compromise between loss of precision in high-frequency signals, and large droop after large signal amplitudes.
The first DOMs produced had transformers  
with short time constants of about \unit{1}{$\mu$s} at \unit{-30}{\degree C} leading to a strong droop effect (Fig.~\ref{fig:waveforms}\,b),  
whereas DOMs produced later (since 2005) have larger time constants of about \unit{12}{$\mu$s} at \unit{-30}{\degree C} with reduced droop (Fig.~\ref{fig:waveforms}\,a). 
In IceTop, 79 of the 324 DOMs or  27\% of all DOMs  have transformers with large droop, see Table \ref{tab:ICIT_configs}. 

The droop effect can be described empirically by an impulse response with two time constants~$\tau_{1,2}$ \citep{roucelle07},
\begin{equation}\label{eq:icecube:droop}
  \delta(t) \to \delta(t) - N\left((1-f)\, {\rm e}^{-t/\tau_1} + f\, {\rm e}^{-t/\tau_2}\right),
\end{equation} 
where $N$ is a normalization constant.
The formula is used in a discretized form for simulation and corrections of measured pulses (see Section \ref{sec:signal_proc}).
The droop constants $f$, $\tau_1$, and $\tau_2$ have been measured in the laboratory as a function of DOM temperature. The temperature dependence is ($T\ {\rm in\ \dg C}$) 
\begin{equation}\label{eq:droop_temperature}
	\begin{array}{rcl}
\displaystyle		\tau_1(T)& = &\displaystyle	 A + \frac{B}{\,\!1+e^{-T/C}}\\
\displaystyle		\tau_2(T) & = &\displaystyle	 0.75\, \tau_1(T)
	\end{array}
\end{equation}
Using a single-tau approximation by setting $f=0$ the parameters are on average: 
\begin{equation} \label{eq:droop_temp_single-tau}
	\begin{array}{llll}
	A=0.4\, {\rm \mu s}, & B= 5.0 \, {\rm \mu s},& C=16\, {\rm \dg C} & {\rm (old\ transformers)}\\
		A=1.0\, {\rm \mu s}, & B= 50.0 \, {\rm \mu s},& C=25\, {\rm \dg C} & {\rm (new\ transformers)}
	\end{array}	
\end{equation}	

In the IceTop tanks the DOM mainboard temperatures vary between about \unit{-20}{\degree C} and \unit{-40}{\degree C} throughout the year (in the deep ice the temperatures are stable), yielding for the new transformers $\tau$ values between about 9 and 17 ${\rm \mu s}$. This has a major impact on the signal extraction and requires a continuous monitoring of the DOM temperatures. The inversion of the formula \eqref{eq:icecube:droop} is used to remove the droop in the data, see Section \ref{sec:signal_proc}.

\section[IceTop triggers and data acquisition]{IceTop triggers and data acquisition} \label{sec:trigger_daq}

\subsection{General scheme}\label{sec:trigger_daq_scheme}
In normal operation, the DOMs act as autonomous data collectors, initiating the capture and digitization of waveforms in response to one of many triggering conditions (generated either internally to DOM or from within a full IceTop station). A DOM is capable of sending a variety of data records, depending on 
selected operation modes, to the IceCube Lab for consideration in detector-wide triggers and event construction or for calibration and monitoring purposes. IceTop data included in these triggered detector-wide readouts, known as an `event', are available to a suite of filter algorithms, selecting the most interesting events for immediate transmission north via a dedicated satellite link.

\paragraph{Trigger scheme} To select physically interesting events and to save bandwidth for data transfer IceCube has a multi-level
 trigger system including both hardware and software decisions. In the following an overview of the different `triggers' is given, which also serves to define the terminology:
\begin{description}
	\item[Discriminator Trigger:] Normally the capture of the waveform by an ATWD is initiated by the FPGA, when a pulse passes a discriminator threshold; autonomous within a DOM; described in Section \ref{sec:Discriminators}.

	\item[DOM Launch:] The initiation of the process of capture and digitization of waveforms and data preparation for transmission	organized by the FPGA according to the operation mode. Normally, a DOM is launched by the discriminator trigger, but the trigger can also be generated internally by the FPGA, for example for a baseline determination (Section \ref{subsubsec:minbias}).  To be launched the DOM has to be in the `ready' state (see the discussion of dead times below). 	
	\item[Hit:] The basic information recorded upon a DOM launch is referred to as a `hit'. 
	\item[Local Coincidence (LC):] Each DOM in IceTop  is connected by a network of cable connections to neighboring DOMs in the same station.  The presence of a logical level set by the discriminator trigger in neighboring DOMs, the so-called Local Coincidence condition, can be used to set the content of hits sent to the IceCube Lab.  Described in Section \ref{subsubsec:LC}.
	\item[Software trigger:] A series of software algorithms formed in the IceCube Lab using DOM Launches satisfying the LC condition determine when data from the entire IceCube detector should be recorded into events.
	\item[Online filter:] A subset of events are selected for immediate transfer north by satellite link using a set of filtering algorithms that select the physically most interesting events for later detailed analysis.
All events are archived on magnetic tape storage.

\end{description}

\paragraph{Operation modes} How much information a hit contains when being  transferred to the IceCube Lab, is determined by the operation modes for the readout of DOMs. Two different modes were foreseen for IceCube \cite{DOMPaper}:  
\begin{description}
	\item[SLC mode:] In the `soft local coincidence' mode data from all launched DOMs are transferred to the IceCube Lab. The data format of a hit contains a
	header with coarse information about the hit and in addition the full waveform data for `HLC hits', that are hits with a fulfilled LC condition ('LC tag').
For hits which have no LC tag (`SLC hits') only the header with the coarse information
	 is transmitted. For IceTop the coarse information is a timestamp and the integrated ATWD charge. See Section \ref{subsubsec:SLC}.
	\item[HLC mode:]  In the `hard local coincidence' mode only those DOMs transfer data which have an LC tag.
	\end{description}
In both modes additional conditions can lead to full waveform transfer, such as in case of the calibration triggers described below in Section \ref{subsubsec:minbias}.	
The baseline-operation of IceCube is the SLC mode \cite{DOMPaper} which, however, was not applied before 2009. For IceTop the standard SLC mode has been modified so that the ATWD digitization is not aborted if there is no LC tag because the integrated ATWD charge is always transmitted in the SLC mode (for in-ice DOMs only fADC information is used for the SLC transmission so that the ATWD digitization can be aborted if there is no LC tag).

\subsection{Decisions at detector level}	\label{subsec:detector_decisions}
\subsubsection{Local Coincidence} \label{subsubsec:LC}
For standard data taking the full-waveform readout of a DOM is initiated if a local coincidence (LC) of a station is fulfilled. This results in a single station trigger rate of about 30 Hz compared to about 1600 Hz of a single high-gain DOM at a threshold of about 0.1 VEM. The single DOM rate is dominated by small showers with a good fraction caused by single muons passing a tank (see Section \ref{subsubsec:calibration_daq}).

\begin{figure}
  \centering
\includegraphics[width=0.45\textwidth]{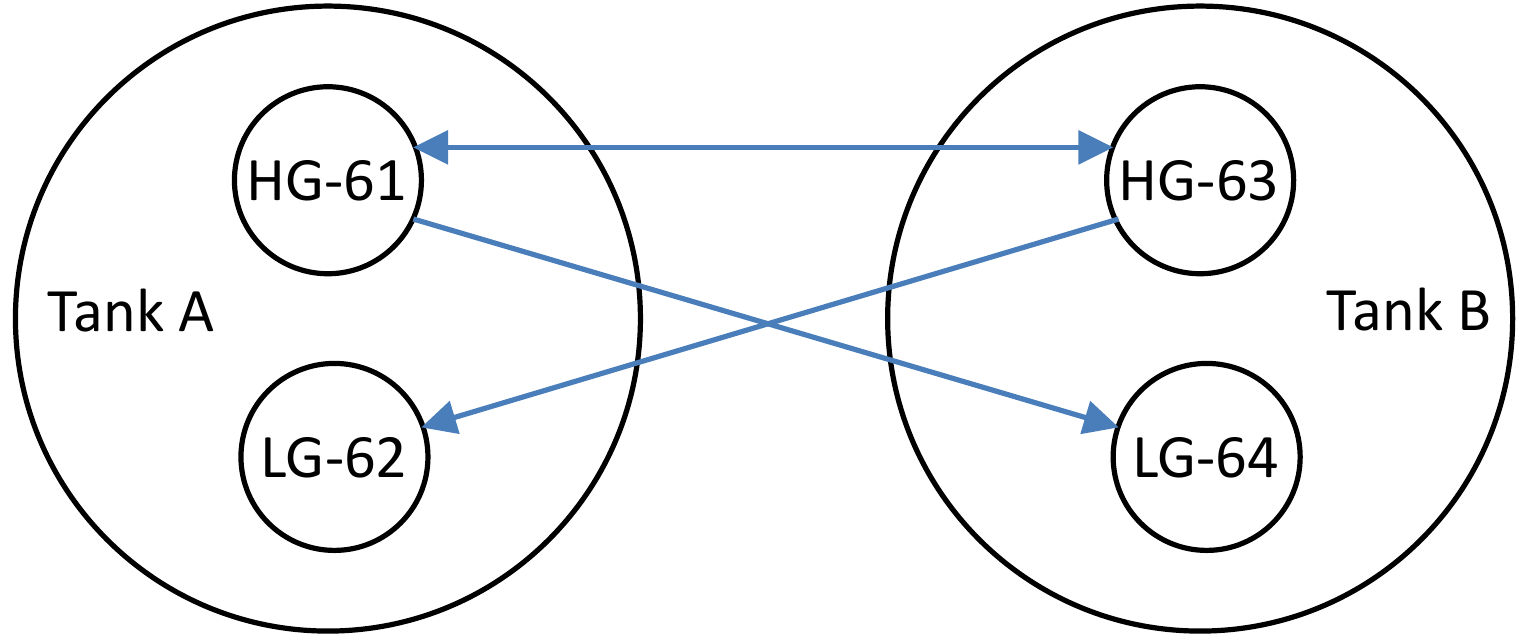}

  \caption[Schematic of the Local Coincidence configuration in IceTop]
    {Schematic of the Local Coincidence configuration of high-gain (HG) and low-gain (LG) DOMs in an IceTop station. The coincidence requirements are described in the text. The DOM numbering is explained in Fig.\ \ref{fig:Tank-DOM-positions}.}
  \label{fig:icetop_lc}
\end{figure}

To establish a local coincidence the high-gain and low-gain DOMs of a station are interconnected via cables as shown in Fig.\ \ref{fig:icetop_lc}. 
The following simplified description of the local coincidence decision focuses on the IceTop applications (details can be found in \cite{DOMPaper}). If a high-gain DOM passes the discriminator threshold a logical signal is sent to both DOMs in the other tank, which is used to define a \unit{1}{$\mu$s} coincidence time window.  In the other tank, a DOM of either gain is read out if it had a discriminator trigger within this time window.
  Thus, in the normal configuration, station data always consist of signals  from both high-gain DOMs and optionally additional signals from low-gain DOMs. However, due to the fact that the LC condition is only based on the discriminator trigger, dead time (see below) can cause data from a high-gain DOM to be missing in the rare cases where both ATWDs are busy.

The connections between high-gain and low-gain DOMs and an additional, normally unused connection  between the two low-gain DOMs (not shown in Fig.\ \ref{fig:icetop_lc}) add also redundancy for the case that one or both of the high-gain DOMs stops functioning. In such an event  low-gain DOMs can be switched to high-gain and an LC connection can be reestablished. 

A fulfilled LC condition will result in  storing the digitized waveforms into the local DOM memory from where it will be transfered to the IceCube Lab as described in Section \ref{subsubsec:ICL_transfer}.

\subsubsection{Soft Local Coincidence} \label{subsubsec:SLC} 
As explained above,  `SLC mode' refers to an operation mode which reads out every DOM for which a signal passed the discriminator threshold, at minimum recording the sparse `SLC hit' information of integrated charge and timestamp. Normally, only HLC hits include the full waveform information. Including SLC hits in the analysis is useful for detecting single muons in showers where the electromagnetic component has been absorbed (low energies, outer region of showers, inclined showers) as well as for vetoing air shower background for in-ice studies.

\subsubsection{Calibration launches} \label{subsubsec:minbias}
Under certain conditions full waveform readout can be initiated without the LC condition being fulfilled. Currently there are two such special DOM launches, both serve calibration or monitoring purposes. 
{VEMCal launch:} This launch of single high-gain DOMs collects single muon hits for calibration of the VEM unit. It is described in Section \ref{subsubsec:calibration_daq}.
{Beacon launch:} In order to determine the ATWD and fADC baselines
from `empty events' all DOMs are launched at a rate of \unit{1}{Hz} without requiring that a discriminator trigger has fired. 

\subsubsection{Dead time and unusual hit patterns} \label{sec:deadtime}
In order to achieve a DOM launch
the DOM must be in the "ready" state. Several conditions may cause a DOM to be not ready. If the DOM got a valid LC the DOM is in a "busy" state for \unit{6.4}{$\mu$s} (the time the fADC records the waveform). If the DOM did not get a valid LC a new DOM launch is disabled until the DOM knows it did not receive a valid LC. This time is currently set to \unit{2.5}{$\mu$s} for IceTop.
For a launch rate of 1600 Hz, this creates a dead time probability of 0.4\%. After that time the DOM is reenabled if there is a free ATWD, but if both ATWDs are busy the DOM remains disabled until an ATWD becomes available. For IceTop the SLC mode requires always the ATWD digitization to be completed (see the definiton of the SLC mode above), which takes about \unit{30}{$\mu$s} (Section \ref{subsec:signal_capture}). Assuming the 1600 Hz of launches is uncorrelated, there is about 5\% probability of overlapping the processing from two DOM launches, and a dead time probability of about 2.5\% that a discriminator crossing will not result in a DOM launch. When operating in HLC mode, if there is no LC condition, the ATWD launch can be aborted, in which case the ATWD in question can be re-enabled after about \unit{6.5}{$\mu$s} (the LC decision time plus about \unit{4}{$\mu$s}), and the LC-coincidence dead time is reduced to 0.5\%. In HLC mode, other cases occur under unusual circumstances; for example, if the high-gain DOM in one tank (say, A) and the low-gain DOM in the other tank (B) are the only DOMs with a discriminator trigger, then a full report for the low-gain DOM will be generated, but no report will be made for the DOM launch in tank A. This might occur if a single low energy electron occurs in proximity to the low-gain DOM in tank B, so that the light signal in the high-gain DOM is below threshold. 

\subsection{Data acquisition and processing} \label{subsec:daq}

\begin{figure}
	\centering
	
\includegraphics[width=0.70\textwidth]{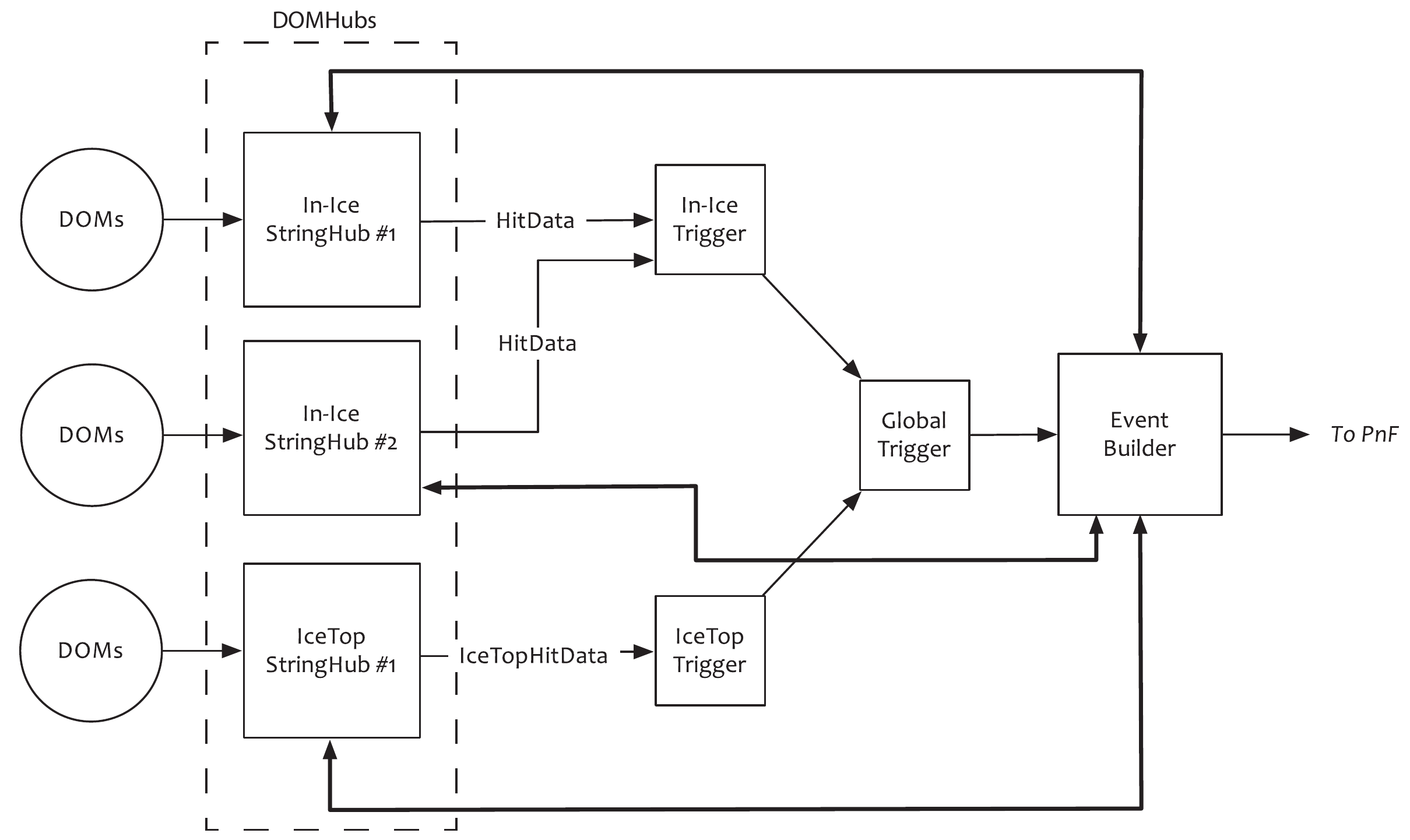}
	\caption{Schematic of the data flow and the  software components of IceCube data acquisition system. The data from the DOMs are collected by the StringHub modules (residing in the DOMHub computers) which send data to the trigger modules. Upon a global trigger the Event Builder gathers the corresponding data of the trigger window and sends the data to the online processing and filtering system PnF.}
	\label{fig:IceCube_DAQ}
\end{figure}

An overview of the data acquisition system of all IceCube components is given in Fig.\ \ref{fig:IceCube_DAQ}. In the following we explain the features relevant to IceTop operation.

\subsubsection{Data transfer to the IceCube Lab} \label{subsubsec:ICL_transfer} 

Each IceTop DOM is connected to a computer, called DOMHub (Fig.\ \ref{fig:IceCube_DAQ}), by a single twisted pair of wires. This differs from the setup for in-ice DOMs, where the lower counting rate allows each twisted pair to carry the signals from two adjacent DOMs. 
The DOMHubs are standard personal computers equipped with PCI digital optical module readout (DOR) cards. Each DOMHub accommodates eight IceTop stations (32 DOMs). 

A fulfilled DOM launch condition, usually the LC condition, will result in  storing
the digitized waveforms of the highest-gain ATWD channel which is not saturated (the lowest-gain channel if all are saturated) and of the fADC into the local DOM memory. In regular intervals the DOMHubs request these data to be transferred.
In SLC mode, time and integrated charge information is always transmitted and for HLC hits (and some special DOM launches, see Section \ref{subsubsec:minbias}) in addition the full waveform is transmitted. 

\subsubsection{Data processing in the IceCube Lab}  \label{subsec:ICL_processing}
 In the DOMHubs the data are prepared for further processing by the StringHub modules. First the timestamps of each hit, which are calibrated to a common time reference by the RAPcal system (Section \ref{subsec:calibration_dom}), are converted to UTC time. The 
   hits are then time sorted and grouped together within time windows of a few microseconds. Subsequently the HLC hits are sent to the trigger units of the in-ice detectors and of IceTop where trigger algorithms are applied (see below). 
 The triggers define readout windows, which differ for the different sub-detectors and triggers. In IceTop the window for the normal air shower events is \unit{\pm 10}{$\mu$s} before the first and after the last hits which produced the trigger. The IceTop calibration trigger has a window of \unit{\pm 1}{$\mu$s}.  The global trigger merges all overlapping individual trigger windows for all sub-detectors. Based on this global readout window, the event builder collects all launches from all sub-detectors that happen during the global trigger window.

  In the common IceCube event builder the data selected by the global trigger are transformed into the DAQ data format, sent to the online processing and filtering (PnF) and are stored on tapes. The tapes can only be shipped to the north during the Antarctic summer. Therefore part of the data, as defined by the filters, are sent via satellite transmission.  In addition,  the data for monitoring transient events (Section \ref{subsubsec:transient_rate_monitoring}) are also sent via satellite.

\subsubsection{IceTop trigger} \label{subsubsec:trigger_settings}
 The current basic trigger for IceTop air shower physics is the IceTop Simple Multiplicity Trigger (IceTopSMT) which requires 6 HLC hits (usually from 3 stations) in IceTop DOMs within \unit{6}{$\mu$s}:
\begin{equation} \label{eq:IceTopSMT}
{\rm	IceTopSMT: \qquad \geq 6\ HLC\ hits}.
\end{equation}
  The readout window starts \unit{10}{$\mu$s} before the trigger window and lasts until
 \unit{10}{$\mu$s} after the last of the 6 hits. Upon an IceTop trigger the whole detector is read out 
   and similarly, when an in-ice HLC trigger is formed, the
    previous interval is recorded to capture any activity in IceTop.

In addition a `minimum bias' trigger, IceTopMinBias, triggers on any HLC launch in IceTop with a `prescale' factor of $10^4$, that means only every $10^4$th HLC launch triggers.

All IceTop VEMCal launches (Section \ref{subsubsec:minbias}) are selected by the trigger IceTopCalibration. 
The treatment of calibration triggers is described in Section \ref{sec:calibration}.

A summary of various trigger rates is given in Table \ref{tab:ITrates}. 

\begin{table}
	\centering
	\caption{Summary of IceTop rates. }
\begin{tabular}{llc}
name & condition& rate [Hz] \\
\hline
\multicolumn{3}{c}{hit rates}\\[1mm]
DOM (single) & DOM launch & 1600 \\
HLC (all DOMs) &LC fulfilled & 3000 \\
\hline\multicolumn{3}{c}{triggers}\\[1mm]
IceTopSMT &${\rm \geq 6\ HLC\ hits}$ &30 \\ 
IceTopMinBias&HLC prescale $10^4$ &0.3 \\   
IceTopCalibration&VEMCal launch&30\\

\hline
\end{tabular}%
	\label{tab:ITrates}
\end{table}

\subsubsection{IceTop online filter} \label{subsubsec:ITfilters}
For satellite transmission the data have to be filtered for data reduction. The filters, summarized for IceTop in Table \ref{tab:filter_scheme}, select events according to various physics cases or for monitoring purposes. For some filter classes only a certain fraction of events is transmitted (prescaling).

\paragraph{The basic IceTop filter} The most important IceTop filter, called IceTopSTA3, is based on events triggered by the simple multiplicity trigger, IceTopSMT, and requires that at least 3 stations have an HLC hit.
During the construction phase events with less than 8 stations were prescaled for satellite transmission. From 2012 on all events with 3 or more stations are transmitted without prescale, though with partly only condensed information employing the so-called SuperDST format (see below). The three-station condition  has an energy threshold of about 300\,TeV.

To allow adjusting to different transfer and/or analysis modes the events passing the IceTopSTA3 filter are further broken down by the IceTopSTA5 and IceTop\_InFill\_STA3 filters. The IceTopSTA5 filter requires at least 5 stations to be hit. The IceTop\_InFill\_STA3 filter, which aims to select events with energies as low as 100 TeV,  requires at least 3 stations from the infill tanks (Section \ref{subsec:detector_layout}) to be hit.

There is an additional IceTop filter, called InIceSMT\_IceTopCoin (\ref{tab:filter_scheme}), which is independent of the IceTopSMT trigger. It requires the in-ice simple multiplicity 
trigger to be fulfilled and at least one HLC hit in IceTop. 
Events which pass the InIceSMT\_IceTopCoin filter consist of coincident events triggering in-ice IceCube with mostly just one IceTop station. These events will be used for vetoing high energy inclined cosmic ray showers in addition to being used to test and calibrate the entire IceCube detector (see Section \ref{subsec:perf_calib}).

\subsubsection{Data formats and satellite transfer} \label{subsubsec:SatTransfer}
For all events passing the filter conditions IceTopSTA3 or InIceSMT\_IceTopCoin
the total charge and the leading edge time of all DOM launches is recorded in a condensed format called SuperDST and transmitted via satellite. 
Using a logarithmic representation the IceTop charges, which span 5 orders of magnitude,  have to be encoded by 14 bit to ensure a precision of about
 $3\times10^{-3}$ throughout the whole dynamic range. A total of 5 byte per DOM is necessary for encoding charge, time, station number, DOM number and LC bit. For the observed mean number of 22 launches per event the average event size yields 110 bytes. The estimated IceTop contribution to the data rate for all events in SuperDST format is less than 500 MB per day. 

In addition to the transmission in SuperDST format, the full waveform will be transmitted for selected or prescaled events
 according to the scheme summarized in Table \ref{tab:filter_scheme}. For the IceTopSTA5 and IceTop\_InFill\_STA3 filters all events will have the full waveform information, while for the others only some fraction of events will have full waveform information for control studies.

\begin{table}
	\centering
	\caption{IceTop filters with their inclusive rates. The rates for full waveform transmission via satellite after associated prescaling (4th column) and the corresponding data loads are reported in the last 2 columns. All events including the prescaled ones are transmitted in the condensed SuperDST format as described in the text.}
\begin{tabular}{llcccc}
filter & condition& rate & \multicolumn{2}{c}{full waveform subset}& satellite data\\
& & [Hz] & prescale & rate [Hz] & [GB/day] \\
\hline
IceTopSTA3 & IceTopSMT \& \# Stations $\geq$ 3     & 22.3  & 10    & {2.2} &  0.9 \\
IceTopSTA5 & IceTopSMT \& \# Stations $\geq$ 5      & 6.1   & 1     & {6.1} & 3.1 \\
IceTop\_InFill\_STA3 & IceTopSMT \& \# In-fill  Stations $\geq$ 3      & 3.7   & 1     & {3.7} & 1.3 \\
InIceSMT\_IceTopCoin & InIceSMT \& \# IceTop HLCs $\geq$ 1      & 100   & 100   & 1.0 &0.5\\
\hline
\end{tabular}%
	\label{tab:filter_scheme}
\end{table}

\paragraph{Transient rate monitoring} \label{subsubsec:transient_rate_monitoring}

In \unit{1}{s} intervals, the rates of the SPE and MPE discriminators (see Section \ref{subsubsec:DiscrThresholds}) as well as of the Local Coincidences are written out for monitoring and detection of transient events. These so-called scaler rates are also transmitted for each run via satellite. There is a delay of typically a day until the data are accessible in the North.
\section[Calibration]{Calibration}
\label{sec:calibration}
%
%
%

The tank signals are calibrated in two steps: First, the PMT and the DOM electronics are calibrated in order to obtain signal charges in units of photoelectrons and a time reference which is the same for all DOMs in all parts of IceCube. This part is common to all DOMs in IceCube. The second step is the calibration of the tank signal charges in units ``vertical equivalent muon'' (VEM) which is the signal a vertical energetic muon generates in a tank.

\subsection{Calibration of DOM electronics and PMT} \label{subsec:calibration_dom}
Calibration of PMT gain and electronic components of a DOM are performed during special calibration runs by software called `DOMCal' running on the DOM CPU \cite{achterberg06}. The basic calibration quantity is the charge recorded for single photoelectrons which are produced by cathode noise and radioactive decays in the glass pressure sphere (in the following referred to as `cathode noise').

\paragraph{Charge calibration}
The charge calibration of the ATWDs starts by converting the ADC counts of each bin to voltages. The ADC-count to voltage relation is determined by varying the ATWD reference voltage and applying a linear fit to the resulting amplitudes yielding slopes and intercepts for each ATWD bin. The `pedestal pattern', given by the intercepts, is equalized for all bins to a common baseline by the
 DOM firmware. This common baseline is continuously determined during the data taking runs as described in Section \ref{subsubsec:minbias}. Since the pedestals of all ADC cells are equalized, only one number for a channel's baseline has to be obtained by averaging for a certain time period, usually for a run of about eight hours duration. The baselines are continuously monitored and the calibration constants are updated when needed.

The ATWD front-end amplifiers are then calibrated in a two-step process. 
First, the channel with the $16\times$ pre-amplifier is calibrated by injecting single photoelectron-like pulses with the pulser on the mainboard, and comparing the measured pulse charge to the known true pulse charge from the pulser.
The other channels are then calibrated consecutively relative to the highest-gain channel using LED light injected into the PMT.

The PMT gain is calibrated using single photoelectron pulses from cathode noise.
The charge distribution obtained from these pulses can be described by a Gaussian single-photoelectron peak and an exponential contribution at low charges \cite{PMTPaper}.
The relation between high voltage setting and single-photoelectron peak is determined by varying the bias voltage between $1200$ and \unit{1900}{V}. These voltages are higher than those used for the low gain PMTs in IceTop (\unit{750}{V} and \unit{1250}{V} for low gain and high-gain PMTs, respectively, Table \ref{tab:FrontEndComponents}) so that an extrapolation is necessary. The observed single-photoelectron pulses allow the absolute calibration of charges in units of PE with an estimated precision of 10\%. Note, however, that for IceTop the final precision is given by the VEM calibration as described below.

While for in-ice DOMs the discriminator thresholds are calibrated in terms of single photoelectron charges the IceTop discriminator thresholds are set at fixed voltages as described in Section \ref{subsubsec:DiscrThresholds}. The voltages can be related to a charge in numbers of PE or VEM for average IceTop pulses as demonstrated in Fig.\ \ref{fig:2662-120014-qmoni-thr_upper}. The charge values are obtained from the calibrated ATWDs.

\paragraph{Timing calibration}
The pulse transit times from the PMT to the recording are determined using pulses from the LED on the DOM mainboard (Fig.\ \ref{fig:icecube:dom}\,b). 

The master clock of the IceCube experiment is a GPS clock in the IceCube Lab. The local $20\un{MHz}$ oscillator of each DOM is compared to this master clock in a procedure called ``reciprocal active pulsing calibration'' (RAPcal) \cite{DOMPaper} at a frequency of \unit{1}{Hz}. Each signal from any IceCube detector component gets a timestamp with the same time reference expressed in UTC times.  From experimental data the precision of the time calibration was found to be better than \unit{3}{ns} \cite{achterberg06}. 

The RAPCAL process involves analog measurements on pulses exchanged
between the IceCube Lab and individual DOMs, and was optimized for
the dispersion and attenuation characteristics of in-ice cables.
Since the IceTop cables are shorter, electronic filters are used to
make the corresponding pulse shapes similar to in-ice pulses.

 \begin{figure}
   \includegraphics[width=0.45\textwidth]{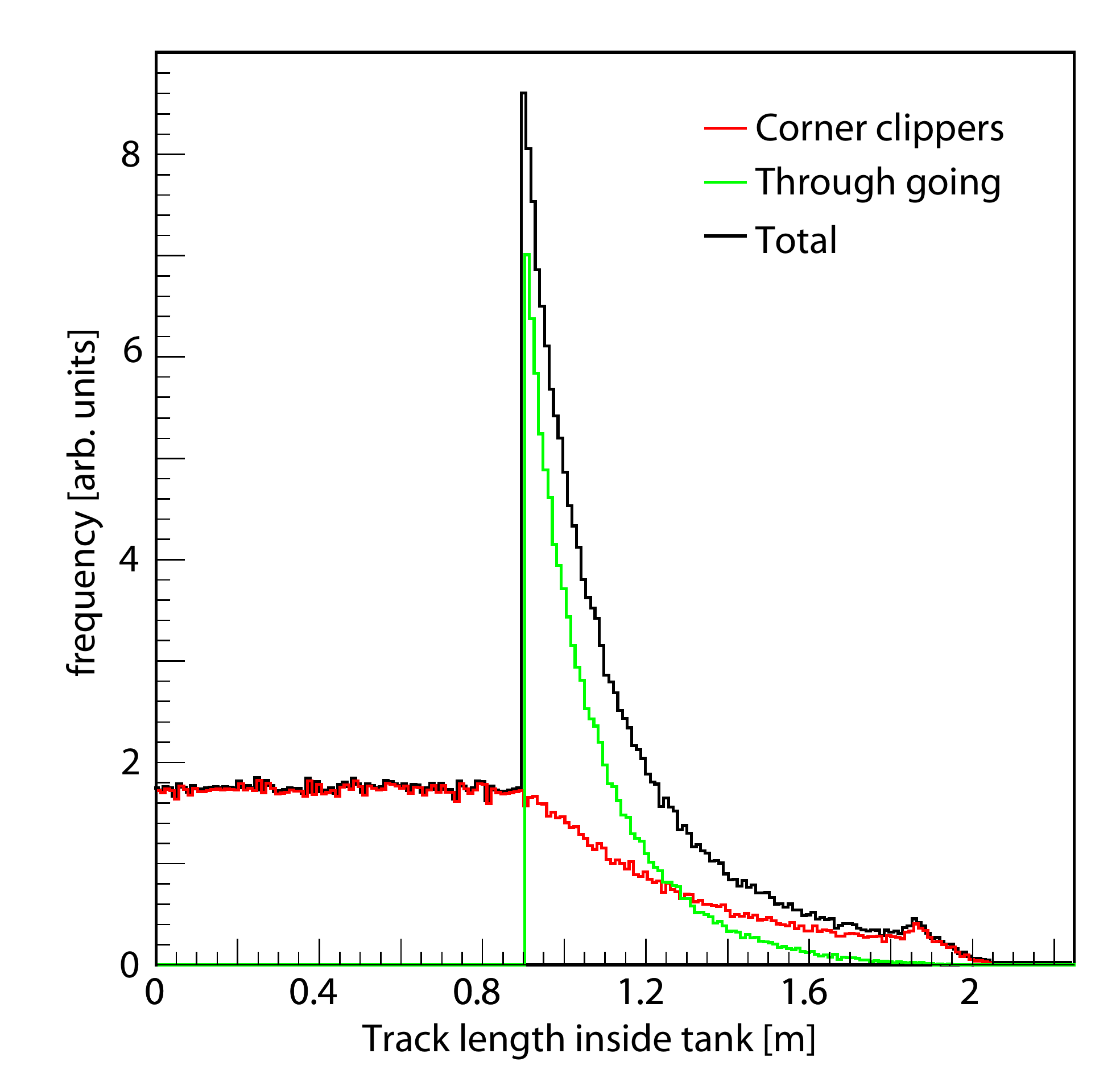}
   \caption{Distribution of muon track lengths in a tank for a $\cos^{2.3}\theta$ zenith angle distribution.}
   \label{fig:MuonTrackLength}
 \end{figure}

\subsection{Tank calibration using atmospheric muons}\label{sec:tank-calibration}

The DOM response to the same energy deposition in a tank varies from tank to tank because of different optical properties of the tanks and DOM efficiencies.
Therefore, the tank signals are calibrated by normalizing to the signal charge of a vertical muon defining the unit `Vertical Equivalent Muon' (VEM) as described in Section \ref{subsubsec:pulsesVEMunit}. 

The VEM calibration procedure uses the natural flux of atmospheric muons. Since 2009 a single-muon calibration trigger is run continuously together with the normal data taking, which has several advantages as compared to employing special calibration runs as done before. In particular, there is no detector downtime, and the calibrations can be done more regularly.

\subsubsection{Single muon spectra}   The average muon energy at the detector level is in the order of \unit{2\,-\,2.5}{GeV}.
Since muons at these energies are minimum ionizing, they always lose about the same amount of energy for a given track length in the tank. For instance, the mean energy deposit of a minimum ionizing vertical muon in a tank of \unit{90}{cm} ice (density \unit{0.92}{g/cm$^2$}) is about \unit{165}{MeV}. 
Cherenkov light is emitted either by the muon itself or by the secondary electrons along the muon track. The total number of Cherenkov photons scales with the energy deposit, and thus with the muon track length, in the tank. Therefore, the spectrum of muon signals in a tank is mainly determined by the angular distribution of the muons and the tank geometry.   Figure \ref{fig:MuonTrackLength} shows a purely geometrical calculation of the track length distribution in a tank, using a realistic muon zenith angular distribution ${dN}/{d\Omega} \sim \cos^{2.3}\theta$ \cite{Gaisser2002285}. 
The sharp peak at \unit{90}{cm} is mainly due to vertical tracks passing through top and bottom of a tank whereas shorter track lengths are entirely due to edge-clippers. Because of the relation between the track length and the number of Cherenkov photons, the experimental muon spectrum (Fig.\,\ref{fig:61A_muonSpec_61_61}\,a) exhibits the same features but they are smeared out due to statistical fluctuations of the Cherenkov photons, the optical properties of the tanks, and the DOM electronics. Furthermore, there is a background under the muon peak mainly from electromagnetic cascade particles from low energy air showers.

\subsubsection{Taking calibration data} \label{subsubsec:calibration_daq}  
The calibration of the IceTop tanks is based on signal spectra from events with preferentially only one muon hitting the tank.  These spectra are recorded for high-gain DOMs in parallel to the normal data taking by special calibration DOM launches, called VEMCal launches, which require the LC condition between the DOMs of the same station not to be fulfilled (see section \ref{subsubsec:minbias}). 
With this anti-coincidence condition, contributions from extensive air showers are largely suppressed and those of single muons are enhanced. The DOM resident firmware pre-scales the discriminator trigger rate by a factor of 8192~($=2^{13}$) and launches the DOM readout if there is no LC fulfilled, otherwise the following discriminator trigger is taken. This results in a DOM launch rate of about \unit{0.2}{Hz} per DOM. For each of these VEMCal launches the ATWD waveforms are sent to the IceTop DOMHub (Section \ref{subsubsec:ICL_transfer}). When a hub receives such data it generates a calibration trigger which initiates event building around the hit (Section \ref{subsec:ICL_processing}) and passes the data to the online processing.


VEMCal launches are only enabled for the high-gain DOMs since the gain and discriminator thresholds of the low-gain DOMs do not allow triggering on single muons. For low-gain DOMs the VEM calibration has to be determined from the overlap of charge measurements by each DOM in a tank, as will be explained below.

The VEMCal hits are calibrated and extracted during the online processing at South Pole exactly the same way as the physics data which are (currently) processed in the North. 
The raw waveform, given in ATWD channel counts, 
is 
calibrated using
the DOMCal  calibration constants (see Section \ref{subsec:calibration_dom}). Corrections for residual baseline offset and a droop correction are also applied.  
Finally, the charge, given in PE units, is calculated by summing up all the waveform bins. 
The extracted VEMCal hits are stored in a special data container
 and transferred to the North within the normal physics data stream, while the original VEMCal events are only stored on tape (unless they satisfy other filter conditions). 

\subsubsection{Processing of calibration data} 
In the North the VEMCal hit data for each individual DOM are 
collected for one week and analyzed. The analysis involves fitting of two plots : The fit to the muon spectra to extract the muon peak of the high-gain DOMs and a fit of the correlation between high-gain and low-gain charges to cross-calibrate the low-gain DOMs. 

\begin{figure}
  \centering

\includegraphics[width=0.5\textwidth]{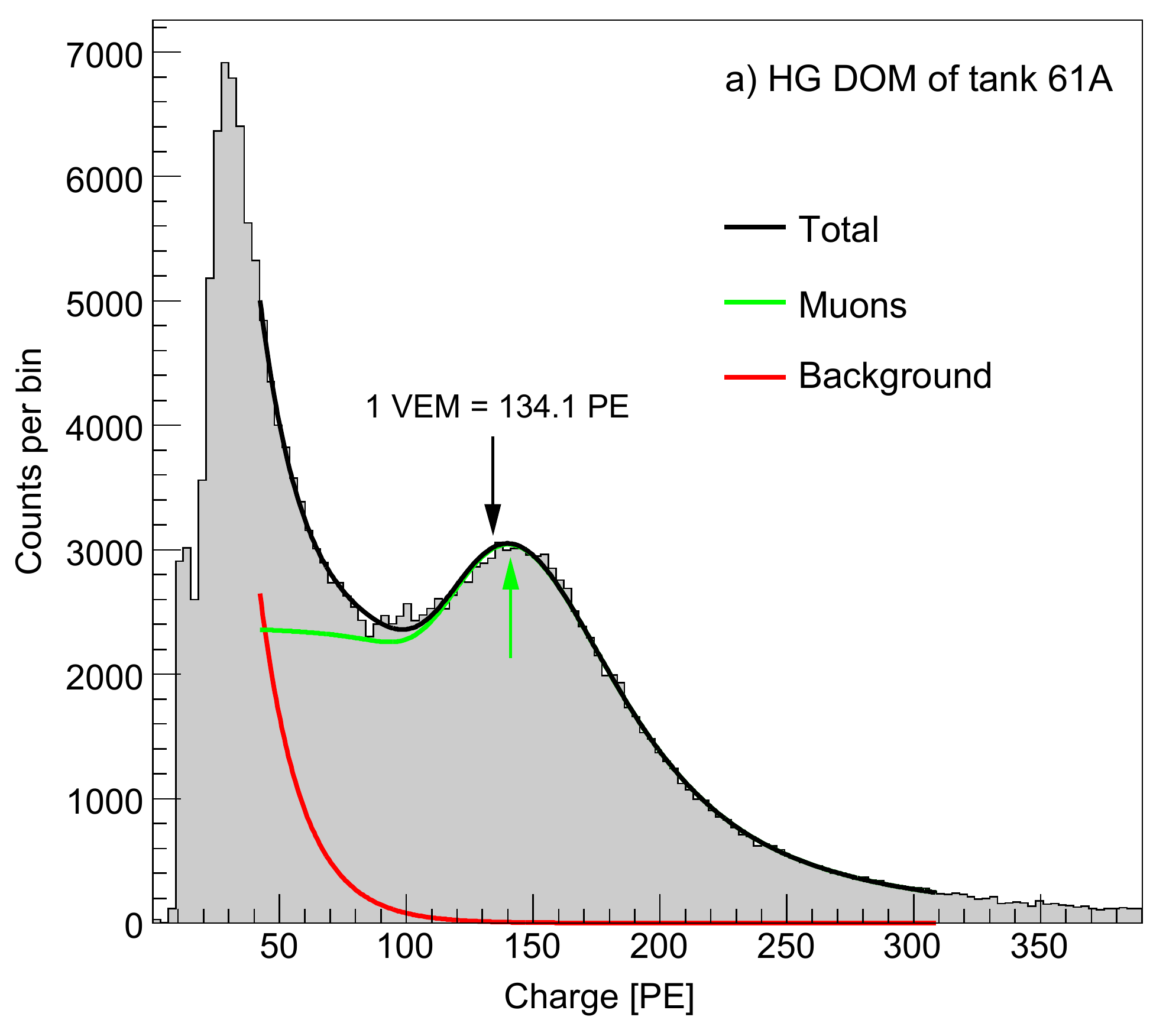}\hfill \includegraphics[width=0.5\textwidth]{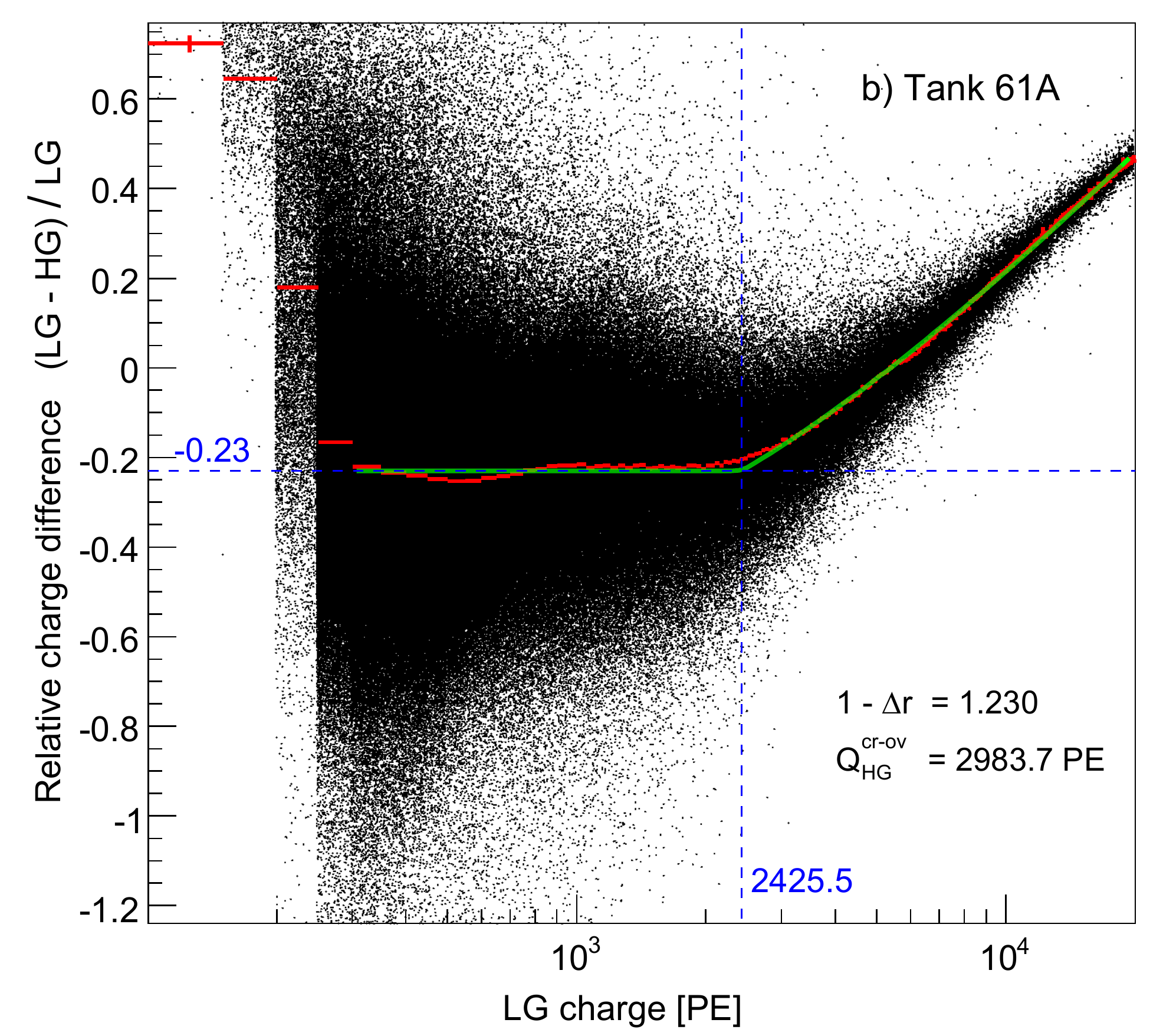}

  \caption{Calibration monitoring plots of the DOMs in tank 61A  from April 15, 2012. a) Charge spectrum of the high-gain DOM in Tank 61A. The black curve corresponds to a fit employing the function \eqref{eq:VEMCal_fit_function}  with the muon (green) and electromagnetic (red) contributions. The maximum of the muon contribution is indicated by the green arrow while the black arrow points to the value of 1 VEM at 95\% below the maximum. b)~Relative charge difference of the high-gain (HG) and low-gain (LG) DOMs in tank 61A. The red curve is the average charge difference for a given LG charge slice which is fitted by the green curve according to the function \eqref{eq:LG-HG_crossover}. }
\label{fig:61A_muonSpec_61_61}
 \end{figure}

The muon spectra are fitted with an empirical formula which includes vertical muons, edge clippers and the electromagnetic background \cite{BeimfordeDiplom}:
\begin{equation} \label{eq:VEMCal_fit_function}
  f(x) = \underbrace{ p_{0}\, \left[ {\rm L}\left(x; p_{1}, p_{2}\right) +  \frac{1.85}{p_{1}}\cdot\frac{1}{\exp{\left(\frac{x-p_{1}}{p_{2}}\right)} + 1} \right]}_{\displaystyle f_{\mu}} + \underbrace{p_{3}\cdot \exp{(p_4\cdot x)}}_{\displaystyle f_{em}}
\end{equation}
The first term in the muon part, ${\rm L}\left(x; p_{1}, p_{2}\right)$, describes through-going muons by a normalized Landau distribution (from ROOT library \cite{ROOTpage}), the second the edge-clipping muons by a Fermi-like function and the $f_{em}$ term describes the electromagnetic background by a single exponential function with two parameters ($p_3,\ p_4$). 
%
The parameter $p_0$ is the number of muons which are not edge-clipping, $p_1$ and $p_2$ are the location and width parameters of the Landau distribution, respectively. The number of edge-clippers is $1.85\,p_0$ determined from the purely geometrical considerations in Fig.\ \ref{fig:MuonTrackLength}; the Fermi function is normalized for positive $x$ values by the factor $1/p_1$.
Figure \ref{fig:61A_muonSpec_61_61}\,a shows a fit of the function \eqref{eq:VEMCal_fit_function} to a charge spectrum. 

\subsubsection{VEM definition} \label{subsubsec:VEMdefinition}

\begin{figure}
  \centering
  
\includegraphics[width=0.5\textwidth]{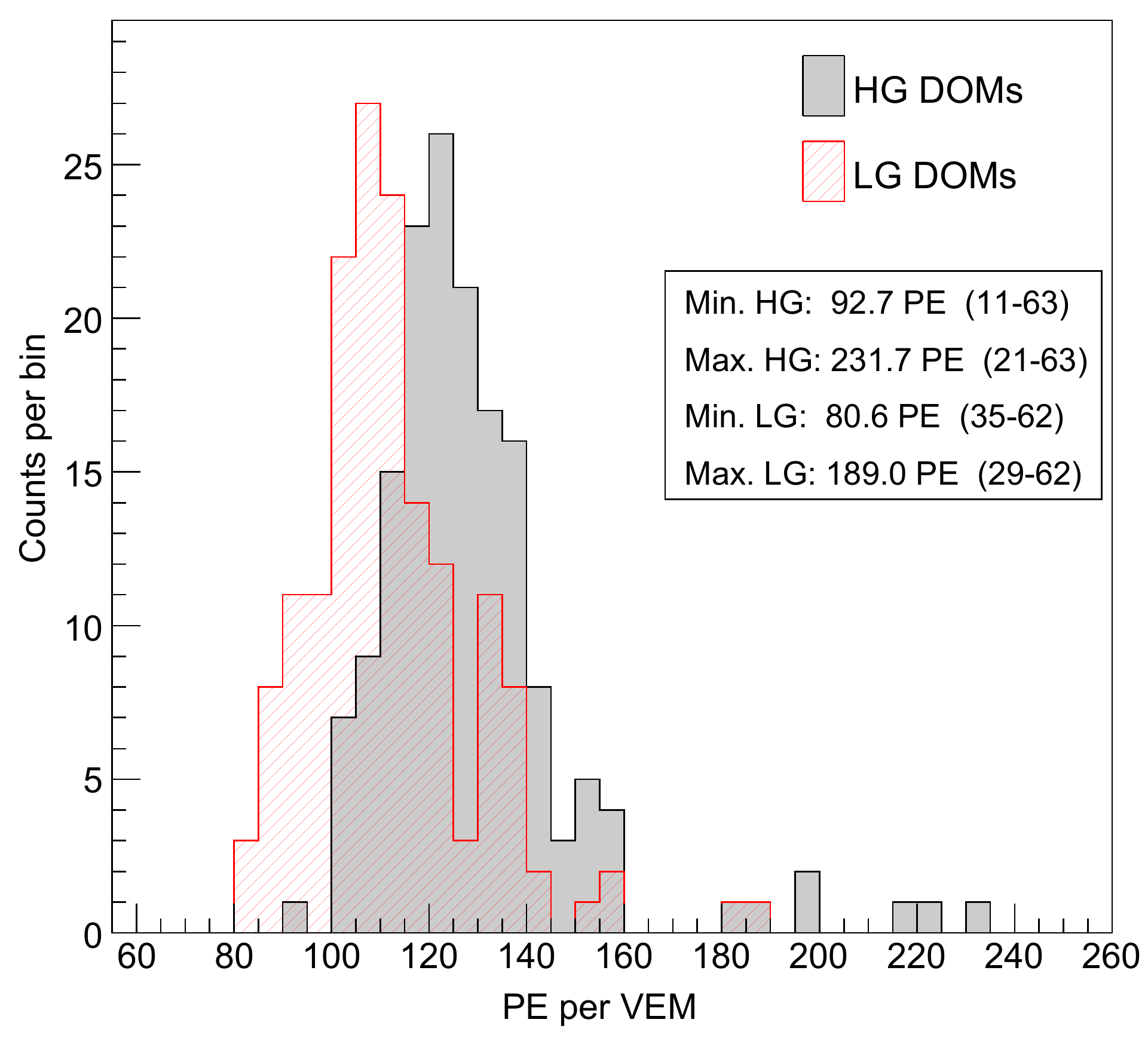}

  \caption{Histogram of the VEM values of all the high-gain and low-gain DOMs from VEM calibration April 15, 2012. }
  \label{fig:summary_vem_2011-11-13}
 \end{figure}

The unit \unit{1}{VEM} is defined as the charge value at $95\%$ of the muon peak which is the maximum of the muon contribution in the fit (term $f_{\mu}$ in \eqref{eq:VEMCal_fit_function}).  The measured muon peak position, obtained with a well-defined fitting procedure, is the essential calibration reference. Shifting the VEM definition to  $95\%$ of the peak value has been chosen because the measurement of tagged vertical muons resulted in a $5\%$ lower muon peak (see Section \ref{subsec:muon_telescope}). However, the choice of a 5\% shift does not affect physics results because it also appears in simulations. The charge \unit{1}{VEM} corresponds to about \unit{125}{PE} for most high-gain DOMs except for those commissioned in 2005 (the tanks are listed in Table \ref{tab:ICIT_configs})  which have about \unit{200}{PE}  (Fig.\,\ref{fig:summary_vem_2011-11-13}). The low-gain DOMs have systematically lower VEM charges of about \unit{105}{PE} with correspondingly higher values of the 2005 tanks. This systematic shift between high-gain and low-gain DOMs is not fully understood (but accounted for by the VEM calibration), see discussion in the next Section \ref{subsubsec:LG_cross-calibration}.
 
The energy of air showers can only be determined by comparison with simulations. It is therefore essential that the simulation uses the same definition of VEM to quantify the tank response to air showers. Here the merit of the VEM calibration becomes particularly obvious: the details of each tank do not have to be simulated, but one typical tank behavior can be used in the simulation, and the simulation can be approximately applied to all tanks when signals are expressed in units VEM. The  `calibration' of the simulated signals in units of VEM is described in Section \ref{sec:CalibrationSimulation}.

%



\subsubsection{Low gain cross-calibration} \label{subsubsec:LG_cross-calibration}
Since the muon spectra are only recorded and fitted for the high-gain DOMs the low-gain DOMs are cross-calibrated assuming that the two DOMs of a tank record on average proportional light yields per particle. Therefore the signal ratio of the two DOMs should be constant in an overlap region where saturation does not play a role.   Figure \ref{fig:61A_muonSpec_61_61}\,b shows the relative charge differences between high gain and low gain versus the low-gain charge in units of PE. The mean values of the scatter plot at a given low-gain charge are constant up to about 2200 PE where  saturation of the high-gain DOMs starts.

For each tank the average relative charge differences as a function of the low-gain charge are fitted by the expression
\begin{equation} \label{eq:LG-HG_crossover} 
f(x) = \left\{ \begin{array}{lll} p_0 & {\rm for} & x < p_3 \\ p_0 + p_1\,\log_{10}(1 + p_2\,(x - p_3)) & {\rm for} & x \geq p_3 \end{array}   \right.
\end{equation}
with the free parameters $p_0$, $p_1$, $p_2$ and $p_3$. The cross-over between the two function parts in \eqref{eq:LG-HG_crossover} is at $x=p_3$.

The horizontal line to the left of the cross-over has an offset $p_0=\Delta r$ from zero corresponding to the relative  difference in the efficiencies of both DOMs which are systematically lower for low-gain DOMs as can be seen in Fig.~\ref{fig:summary_vem_2011-11-13}. The efficiency  includes the quantum efficiency of the PMT as well as the optical efficiencies of the tank which may vary within a tank due to inhomogeneities in absorption and scattering.  Furthermore a potential mis-calibration of the gain of the low-gain DOMs, likely due to extrapolation of the PMT gain below the single-photon calibration range (Section \ref{subsec:calibration_dom}), could be the cause for the systematic shift between low-gain and high-gain DOMs. The offset from zero can be converted to a correction factor for the low-gain DOM for the charge of a VEM (or any other fixed signal) according to: 
\begin{equation} \label{eq:HG-LG-offset} 
Q_{LG}({\rm VEM}) = \frac{1}{1-\Delta r}\,Q_{HG}({\rm VEM})\, .
\end{equation}
The left side is the VEM value of the low-gain DOM in units of PE if $Q_{HG}({\rm VEM})$ is expressed in units of PE. 
The cross-over at $x=p_3$ (in terms of low-gain charge) indicates the beginning of the saturation of the high-gain PMT and determines the switching from high-gain charges to low-gain charges for the waveform measurements. In terms of high-gain charge (in units PE) the cross-over is at
\begin{equation} \label{eq:HG-crossover} 
Q_{HG}^{cr-ov} = (1-\Delta r)\, p_3\, .
\end{equation}

\subsubsection{Monitoring VEM calibration}

\begin{figure}
	\centering
		\includegraphics[width=0.495\textwidth]{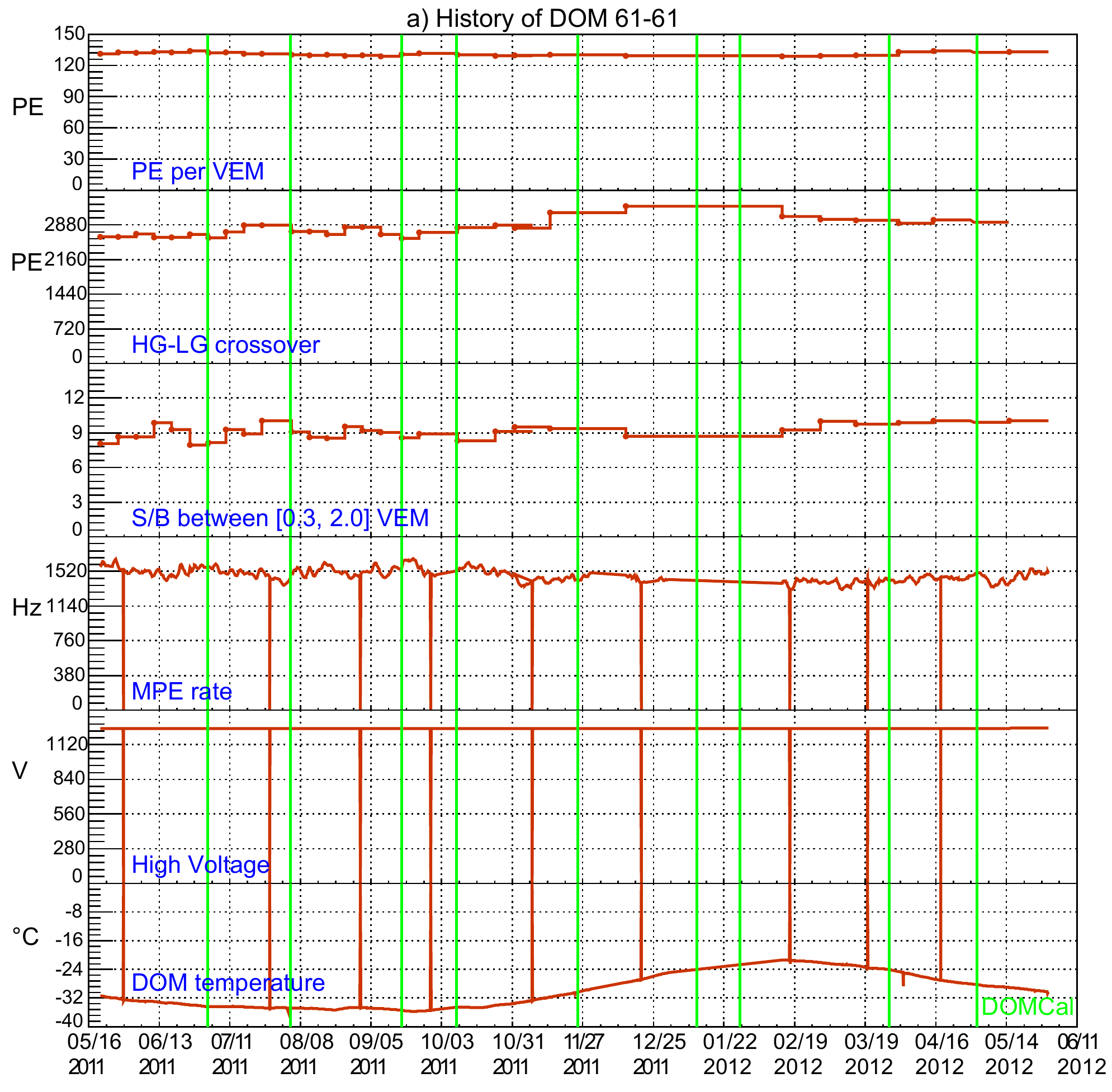}	
		\includegraphics[width=0.495\textwidth]{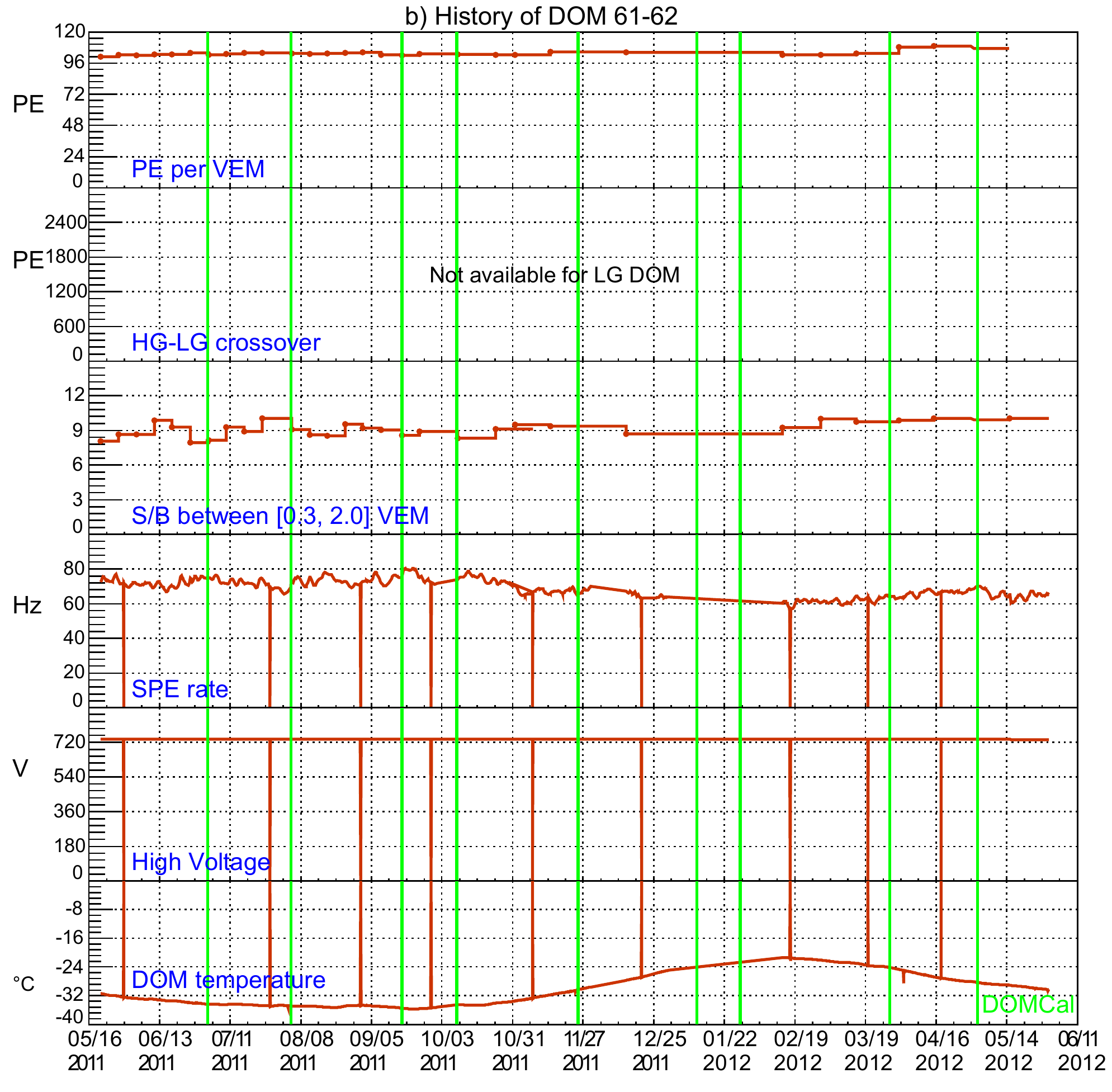}
	\caption{Calibration history of the high-gain DOM (a) and low-gain DOM (b) in tank A of station 61 (same as in Fig.\ \ref{fig:61A_muonSpec_61_61}). From top to bottom are displayed: The number of photoelectrons per VEM, the high-gain - low-gain cross-over used in the calibration of low-gain DOMs, the ratio of the muon signal to the electromagnetic background from the fit of the muon spectrum, the discriminator trigger rate (MPE for high-gain DOMs and SPE for low-gain DOMs), the high voltage of the PMT and the temperature measured on the DOM board. The green vertical lines indicate the time of DOMCal runs where in general DOMCal constants can change.}
	\label{fig:61A_history_61_61}
\end{figure}

The VEM calibration as shown in Fig.\ \ref{fig:61A_muonSpec_61_61} is done for all DOMs in an automated way. Thus stability checks and long time DOM response changes can be monitored, as depicted in Fig.~\ref{fig:61A_history_61_61} for the DOMs in tank  61A (for DOM numbering see Fig.\ \ref{fig:Tank-DOM-positions}).
 The variations of the number of photoelectrons per VEM for all DOMs between two successive calibrations are
within 10\% with an rms spread of 3\%.  The increase of the $S/B$ ratio of the muon signal to the electromagnetic background (see definition in Eq.\ \eqref{eq:SmuBEM}) indicates increasing absorption of the electromagnetic shower component and is related to the increase of the snow height on the tank (Section \ref{sec:Snow_on_tanks}). 
The discriminator trigger rates plotted below the $S/B$ ratio are mostly determined by the electromagnetic shower component and thus decrease with increasing snow.

\begin{figure}
   \includegraphics[width=0.3\textwidth]{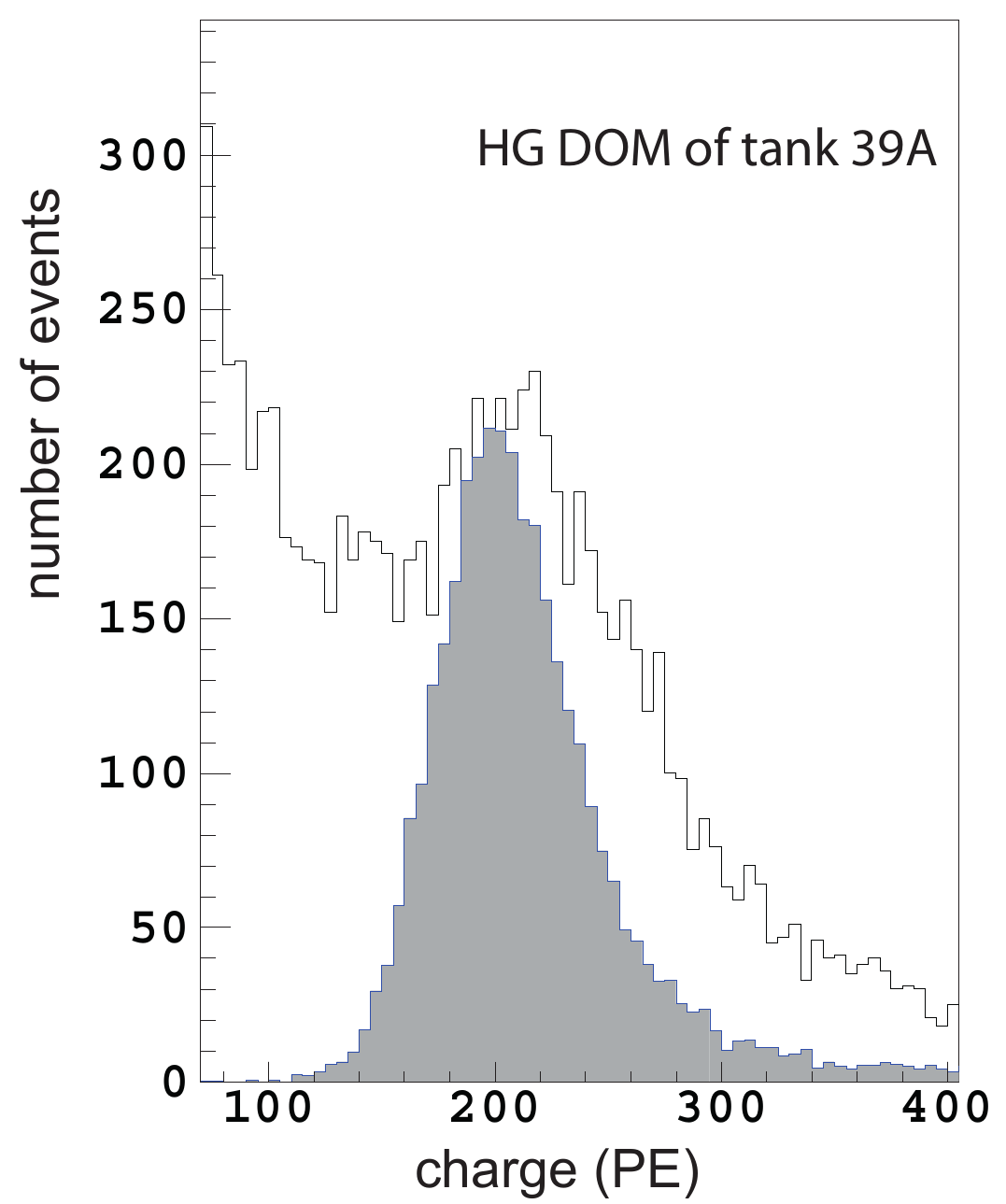}
   \caption{Charge spectra of a high-gain DOM (a) and low-gain DOM (b) in tank 39B  collected by the calibration procedure (VEMCal, black lines) compared to the spectrum tagged by the Muon Telescope (filled histogram) \cite{Demirors:2007zz}.}
   \label{fig:chargehisto_MT}
 \end{figure}

\subsection{Comparison to runs with muon taggers}
\label{subsec:muon_telescope}
\paragraph{Muon telescope at the South Pole}
A portable, solar-powered muon telescope based on scintillation detectors was employed
in order to tag nearly vertical muons for calibration studies.  Comparing the tagged muon events to simulations allows verifying the shower generation and the tank simulation. The variation of the tank response as a function of muon position and the tank-to-tank variations have also been investigated \cite{Demirors:2007zz}. 

Measurements were done in the polar season 2005/2006 on tanks deployed
one year earlier. The muon tagger was placed on top of an IceTop tank, and all hits in both DOMs were recorded for six hours.
Figure
\ref{fig:chargehisto_MT} shows the charge spectra for the two DOMs
 in tank 39B and superimposed the tagged muon charge spectra with the muon telescope placed above the center of the tank. The muon telescope spectrum peaks about 5\% lower than the calibration spectrum because in the latter case the accepted zenith angle range is wider leading to on average longer particle trajectories in the tank. These measurements are the justification behind choosing the definition of one VEM to be 95\% of the muon peak, as described in Section \ref{subsubsec:VEMdefinition}.

\paragraph{Zenith dependence of muon flux} The muon flux at the South Pole was measured for five zenith angles,  0\degree , 15\degree , 35\degree , 82.13\degree\ and 85.15\degree , with a scintillator muon telescope incorporating an IceTop prototype tank as the absorber \cite{Bai:2006mf_muon_flux}. Besides the flux measurements also the zenith angle dependence of the recorded waveforms were analyzed and compared to simulations. 

The comparison of waveforms from vertical and nearly horizontal muons showed two features in the data: (i)~The average amplitude of horizontal muons is about two times as big as that of the vertical muons. (ii)~The average rise time of the these pulses in the tank is 3 ns longer than that of vertical muon pulses while the decay time of the average waveform is nearly the same as that of the average waveform of vertical muons. Both features were well reproduced by a Geant4 simulation of the tank response.

\paragraph{Laboratory calibration measurements} To study the response
of an IceTop tank in more detail a laboratory setup has been
developed. It comprises an IceTop tank filled with 76~cm of 
water (not frozen), equipped with two optical modules with analog readout. Tiled
plastic scintillator plates, mounted around the tank, are
operated in a coincidence mode. They allow the selection of muon tracks
passing through the tank at defined angles and positions. For each
position and angle, typically $10^{4}$ muons were measured and the mean
value and width of the distribution of the number of photoelectrons in
each optical module was determined. Since the dimensions (\unit{76}{cm} height
vs.\ \unit{90}{cm}) and optical properties of the active medium (water vs.\ ice)
were not the same as at South Pole, the aim of the test was (i)~to confirm
that the results are correctly reproduced by the simulation program
`Tanktop' (described in Section~\ref{subsubsec:tanktop}) which is the
reference for the IceTop simulation (Section~\ref{sub:detsim}), and (ii)~to gain a better understanding of possible systematic uncertainties
of the calibration measurement.

\begin{figure}
  \centering
  \includegraphics[width=0.4\textwidth]{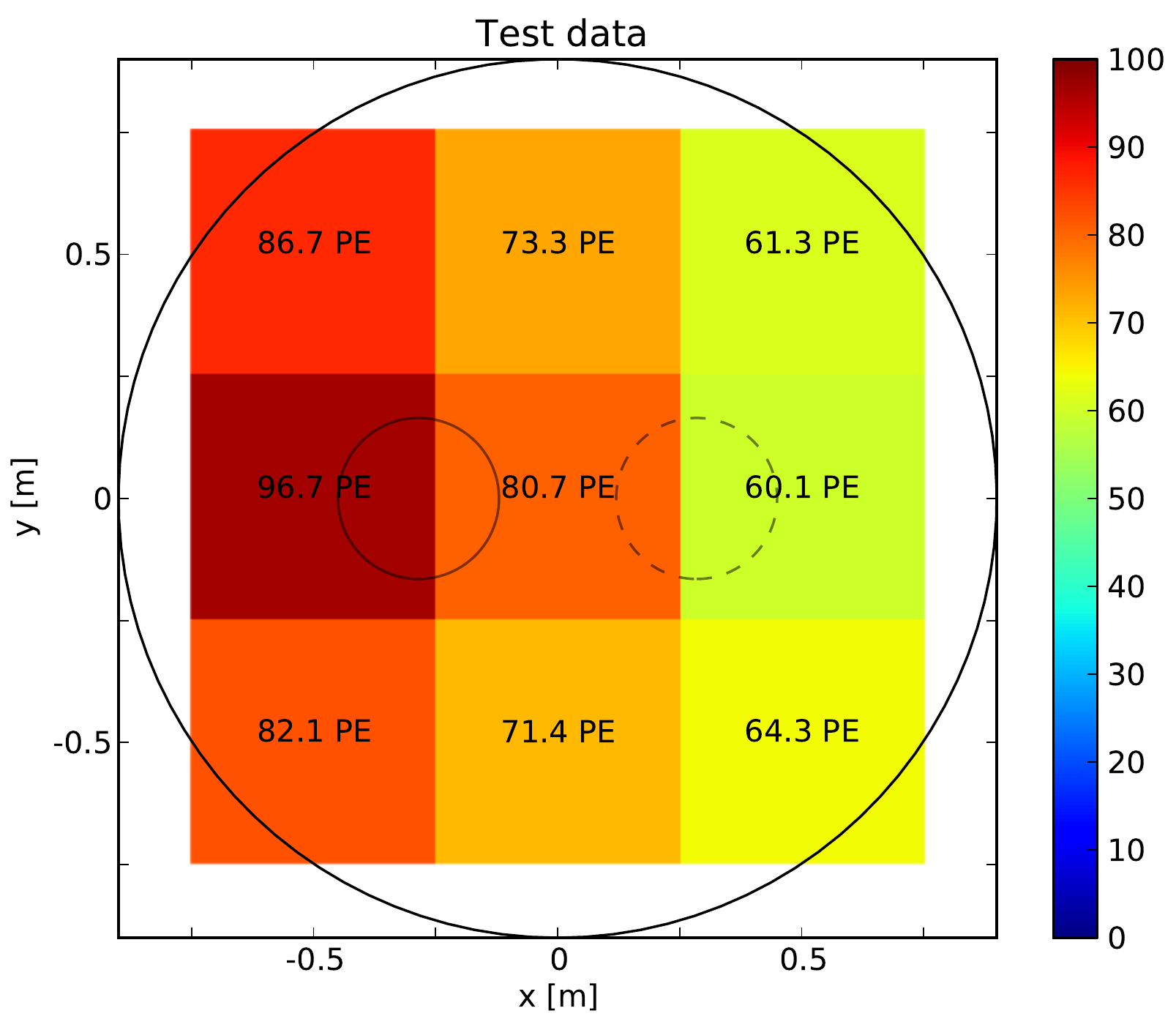}
  \includegraphics[width=0.4\textwidth]{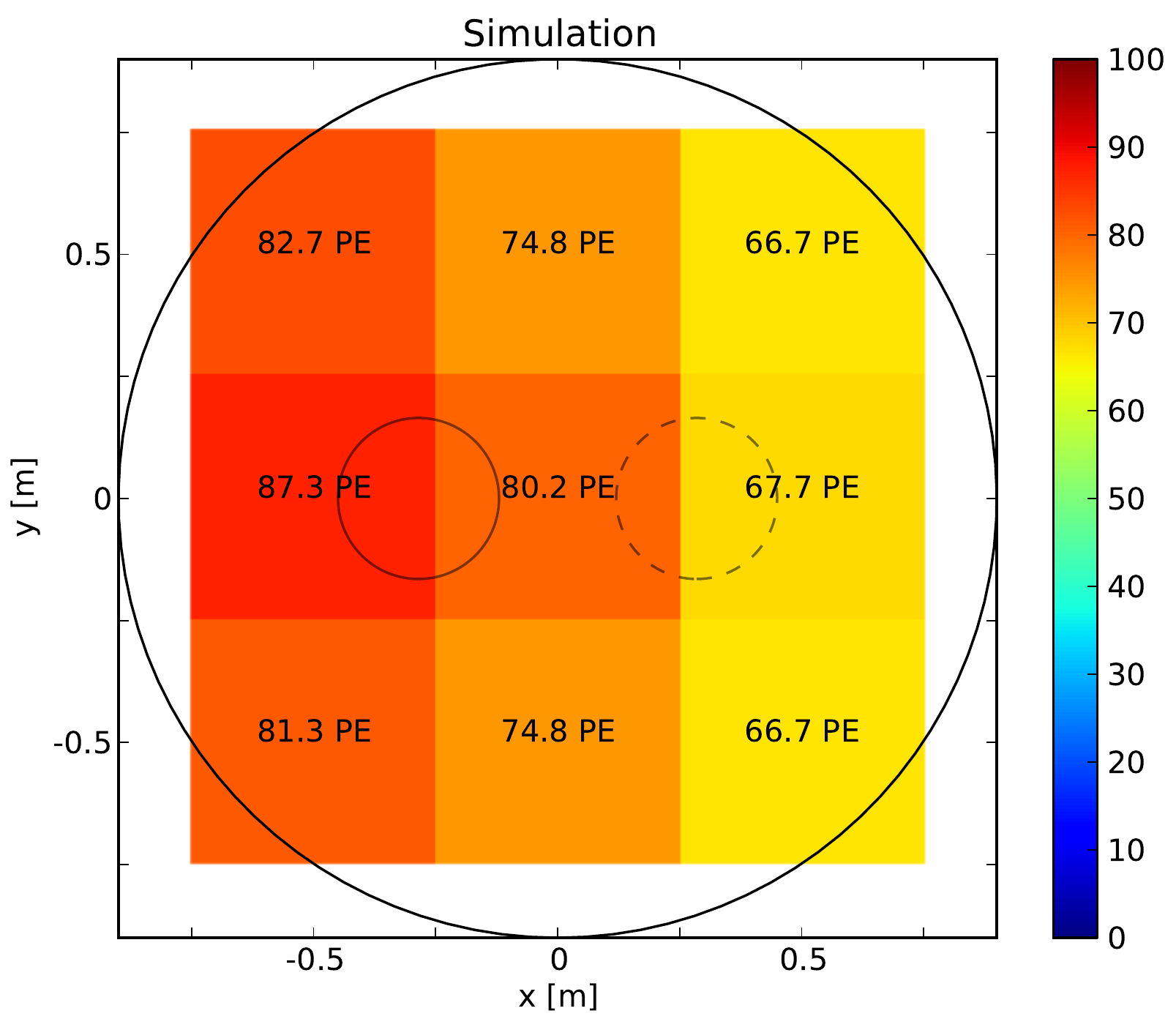}
  \caption{Top view of the tiled scintillator centered on the IceTop
    tank. The mean number of photoelectrons generated by nearly
    vertical muons entering through each segment and measured in an
    optical module located at $(x = -0.285~\textrm{m}, y = 0)$ is
    shown. The measurements (left) are compared to simulations with
    the simulation code `Tanktop' (right).}
  \label{fig:muon_position}
\end{figure}

A scan of nearly vertical muons passing through the tank at different
$x$-$y$-positions has been performed. In Fig.~\ref{fig:muon_position}
a top view of the tiled scintillator centered on the IceTop tank is
presented. The left plot shows for each segment the mean number of photoelectrons
generated by muons which enter through the segment and exit through
the corresponding segment below the tank. 
The signal variations due to different distances of the muon track to the optical module, indicated by the small circles, amount to about 40\% maximum and are described to within 10\% by the simulation shown in the right plot.

\begin{figure}
  \centering
  \includegraphics[width=0.6\textwidth]{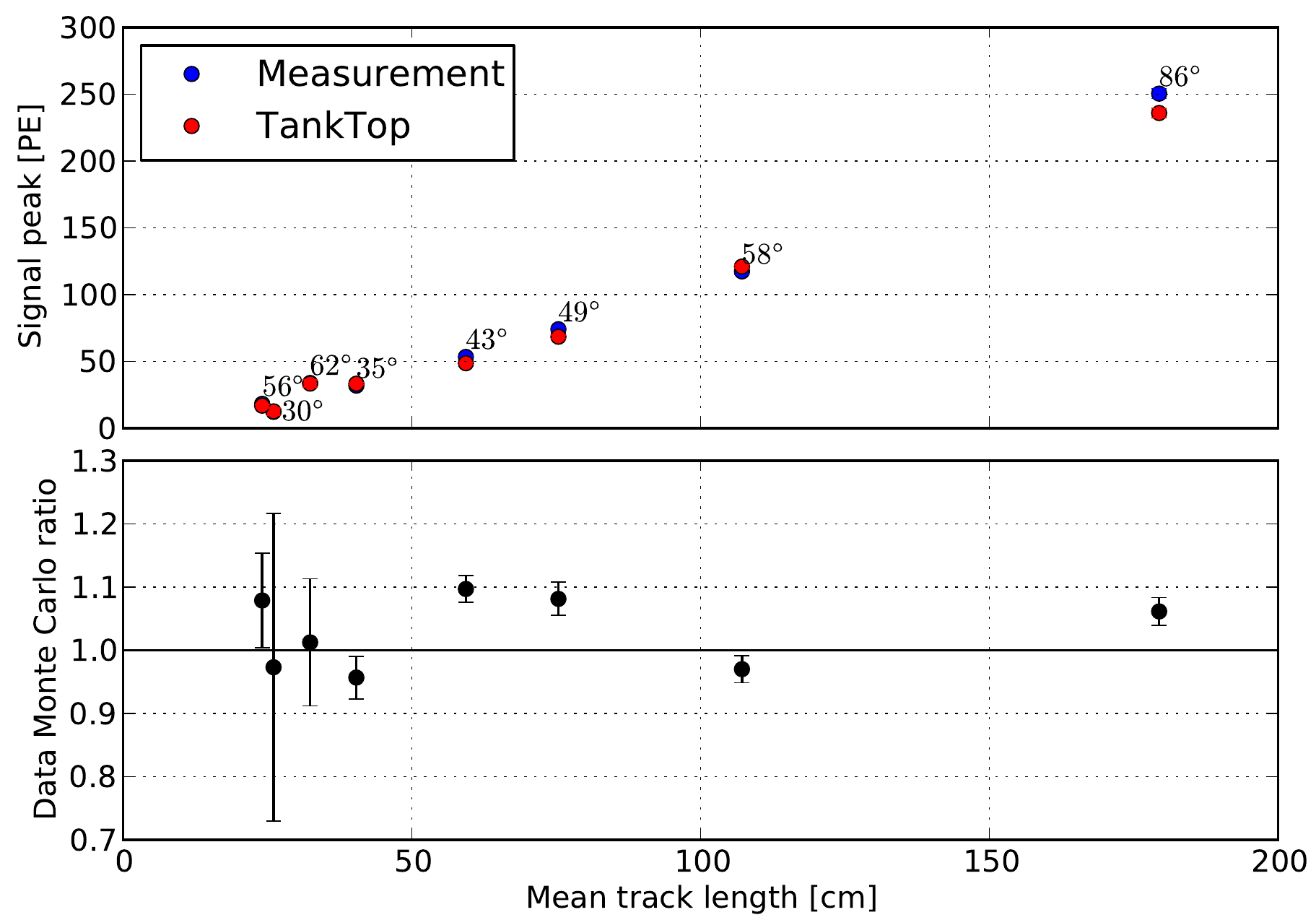}
  \caption{Mean measured number of photoelectrons for muons passing with different track length through a water-filled tank. The track lengths are selected under different zenith angles by the scintillator coincidences as described in the text. The same track length can be obtained under different zenith angles. The measurements (blue) are compared to simulations.}
  \label{fig:track_length}
\end{figure}

Figure~\ref{fig:track_length} shows the mean measured signal strength
in units PE for different muon track lengths in the tank. The signal strength scales nearly linearly with the effective track length. Since the track lengths have been selected by the scintillator coincidences each measurement corresponds to a zenith angular range. The average zenith angle for each point is also displayed in the figure.  The measurements are well reproduced by the simulation. 

Further investigations of the zenith angle dependence of signal sizes are anticipated, since the test tank has not the same geometrical dimensions as the deployed tanks. In addition, a data-based study, e.\,g.\ exploiting isolated muons on the periphery of well-reconstructed showers, would allow better control of angle-dependent systematic uncertainties in physics analyses.

\subsection{SLC charge calibration} \label{subsec:SLCchargeCalibration}
For SLC hits (Section \ref{subsubsec:SLC}) integrated charges and time stamps are delivered by the DOM firmware. Since the baseline calibration as described in Section \ref{subsubsec:minbias} is not available at this point, the baseline correction is applied off-line using the same baseline determination as for HLC hits.
Remaining differences to the HLC charges are corrected by comparing the correlation between HLC and SLC pulses when both are available.


\section[Environmental conditions at South Pole]{Environmental conditions at South Pole} \label{sec:icecube:environment}

\subsection{Overview}\label{sec:Environment_overview}
Since cosmic ray measurements with IceCube use the Earth's atmosphere as a converter medium, variations of the atmosphere will affect air shower measurements. The shower development is mainly determined by the overburden,  $X(h)=p(h)/g$, at a height $h$ and a pressure $p(h)$ ($g$ is the Earth acceleration, here assumed to be constant). Pressure measurements are available at the South Pole from balloon flights. In addition to the total overburden, as given by the ground pressure (at South Pole the average is about \unit{680}{hPa}), also the density profile of the atmosphere influences the air showers. A change in density changes the balance between strong interactions and weak decays of mesons and thus the rate of lepton production. The high-energy muon rate observed in the deep ice is mainly influenced by the density (or temperature for a given pressure) profile in the stratosphere.

\begin{figure}
	\centering
		\includegraphics[width=1.00\textwidth]{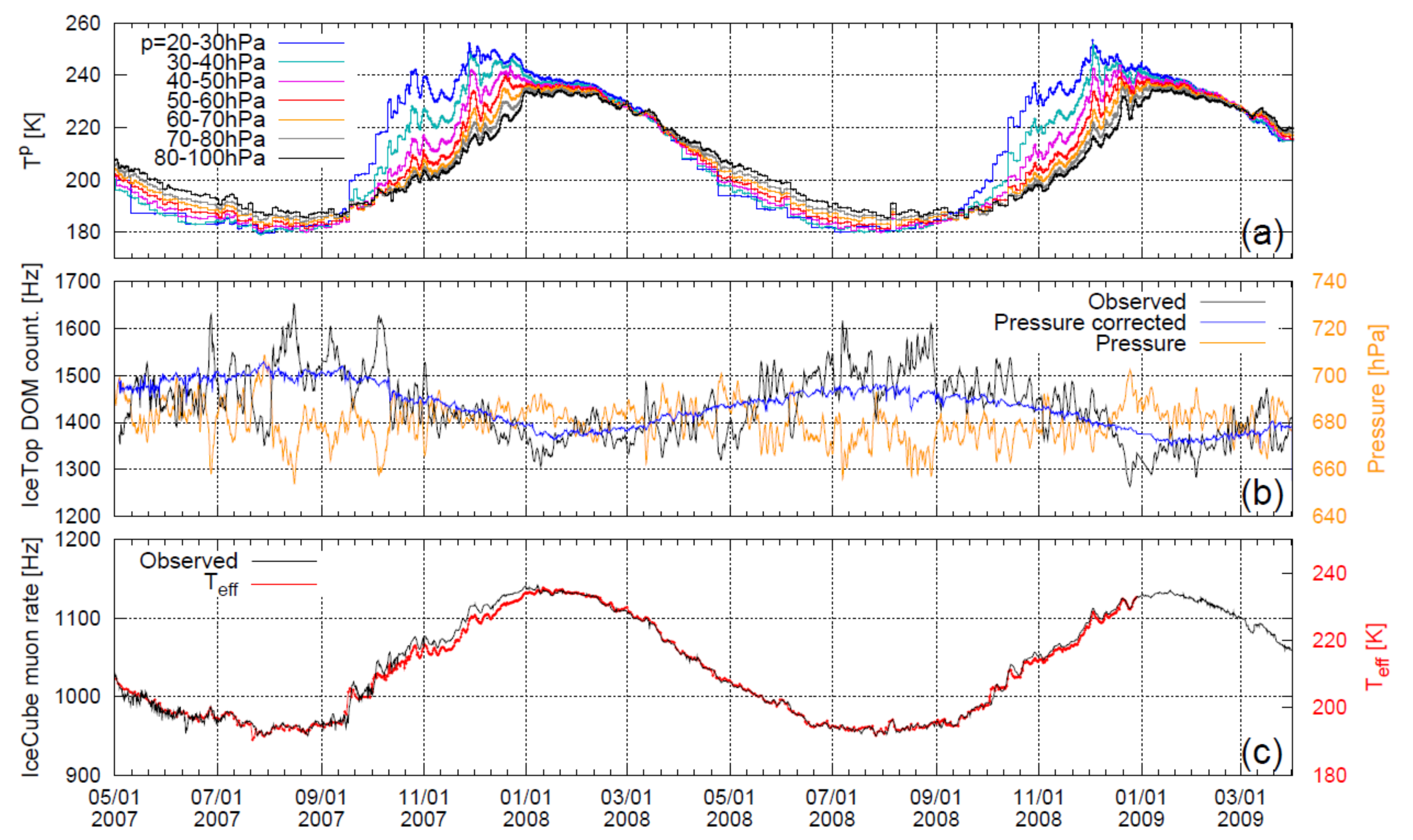}
	\caption{The temporal behavior of the South Pole stratosphere from May 2007 to April 2009 is compared to IceTop DOM counting rate and the high energy muon rate in the deep ice. (a) The temperature profiles of the stratosphere at pressure layers from 20 hPa to 100 hPa where
the first cosmic ray interactions happen. (b) The IceTop DOM counting rate (black -observed, blue -after barometric correction) and the surface pressure (orange). (c) The IceCube muon trigger rate and the calculated effective temperature (red) \cite{Tilav:icrc2009}.}
	\label{fig:atmosphere_effects}
\end{figure}

The South Pole atmosphere undergoes a pronounced annual cycle.
In the winter months, April to September, when the sun is below the horizon, the air is much colder than in summer, with surface temperatures ranging from \unit{-20}{\degC} down to \unit{-70}{\degC}.
Therefore, the winter atmosphere is much denser than in summer. 
Figure \ref{fig:atmosphere_effects} shows the seasonal variation of the density in the upper atmosphere and the correlation with IceTop DOM rates and in-ice muon rates \cite{Tilav:icrc2009}.  The high-energy muon rate in the deep ice follows closely these density changes with an approximate $\pm$(8-9)\% variation (Fig.\ \ref{fig:atmosphere_effects}c), while the variation of the IceTop DOM rates is about $\pm 5\%$ (Fig.\ \ref{fig:atmosphere_effects}b). 

In contrast to the strong annual variation of the density profile, there is no cyclic variation of ground pressure. Instead changes happen on a much shorter time scale, as seen in Fig.\ \ref{fig:atmosphere_effects}b. While the high-energy muons rates are not influenced by the ground pressure changes, there is a strong correlation between pressure and IceTop DOM rates. An increase in air pressure causes a stronger attenuation of air showers leading, for a given shower, to a decrease of the observed shower size in IceTop. The related shift of primary energy thresholds makes the IceTop rates anti-correlated to air pressure. The pressure effect can empirically be corrected for (blue curve in Fig.~\ref{fig:atmosphere_effects}\,b),  which is useful for rate monitoring. For air shower reconstruction it is necessary to find a correction for the shower size $S$, the observed shower signal at a reference radius,   which is the most important observable of an air shower (the determination of $S$ is described in Section \ref{sec:reconstruction}).

Another effect which influences the shower development is the snow accumulating on the tanks. While the atmosphere affects the overall profile of an entire shower, snow accumulation affects individual tanks depending on how deeply they are buried.

\subsection{Atmosphere}\label{sec:atmosphere}

\subsubsection{Barometric effect on rates}

Empirically, the change in the IceTop rates with the surface pressure $P$ as shown in Fig.\ \ref{fig:atmosphere_effects} depends almost linearly on the absolute pressure change:
\begin{equation} \label{eq:rate_change_with_pressure} 
\frac{dN}{N} = \beta \, dP\ \ \ \Rightarrow \ \ \ N=N_0\, \exp(\beta \, (P-P_0)).
\end{equation}
The barometric coefficients $\beta$, introduced by this relation, are regularly determined for the various trigger and filter rates by fitting data at different pressures. The fitted $\beta$ values range from about $0.6\ {\rm to}\  0.7\,{\rm \%/hPa}$ for different rates. For monitoring purposes a rate $N^{meas}_i$ measured at a particular pressure $P_i$ is normalized to a rate at the reference pressure $P_0$  according to 
\begin{equation}\label{eq:rate_corrected}
	N^{corr}_i	= N^{meas}_i \, \exp(-\beta \, (P_i-P_0))\,.
\end{equation}

After correcting the DOM rates for surface pressure variations (blue curve in Fig.\ \ref{fig:atmosphere_effects}\,b) seasonal effects are clearly visible in the DOM rates, although not as strong as for the in-ice muon rates in Fig.\ \ref{fig:atmosphere_effects}\,c.

\subsubsection{Effect of the atmosphere on energy measurements} \label{subsubsec:atm_on_energy}

At different pressures air showers are differently absorbed in the atmosphere so that an effect on the measured shower size is expected. Corrections are derived by comparing measured shower size spectra 
from periods with different surface pressure and finding the correction which makes the spectra coincide.  Experimentally it is found that the shower size parameter $S$ changes with the surface pressure $P$ with respect to a reference pressure $P_0$ as 
\begin{equation} \label{eq:S_pressure_corr} 
S(P)= S(P_0)\, \exp\left( -\frac{P-P_0}{\lambda_P(S, \theta)}\right)\qquad{\rm or}\qquad   \frac{d\ln S(P)}{dP} = -\frac{1}{\lambda_P(S, \theta)} .
\end{equation}
The absorption parameter $\lambda_P(S, \theta)$ is studied as a function of the shower size $S$ and the zenith angle $\theta$. For vertical showers above 1 PeV one finds  $1/\lambda_p\approx 0.25 \times 10^{-5}\, {\rm Pa^{-1}}$ or, in terms of overburden,  $1/\lambda_p\approx 0.25 \times 10^{-4}\, {\rm cm^{2}/g}$ \cite{FabianPhdThesis}. 

The seasonal effect on the shower size measurements was estimated to be $\Delta \log_{10} S \approx\pm 0.01 $ for the two extreme atmosphere densities in winter and summer \cite{FabianPhdThesis}.
Corrections of atmospheric effects on energy determination from reconstructed showers are discussed in Section~\ref{sub:likelihood_fit}.

\subsection{Snow on tanks}\label{sec:Snow_on_tanks}

\begin{figure}
  \centering
\includegraphics[width=0.5\textwidth]{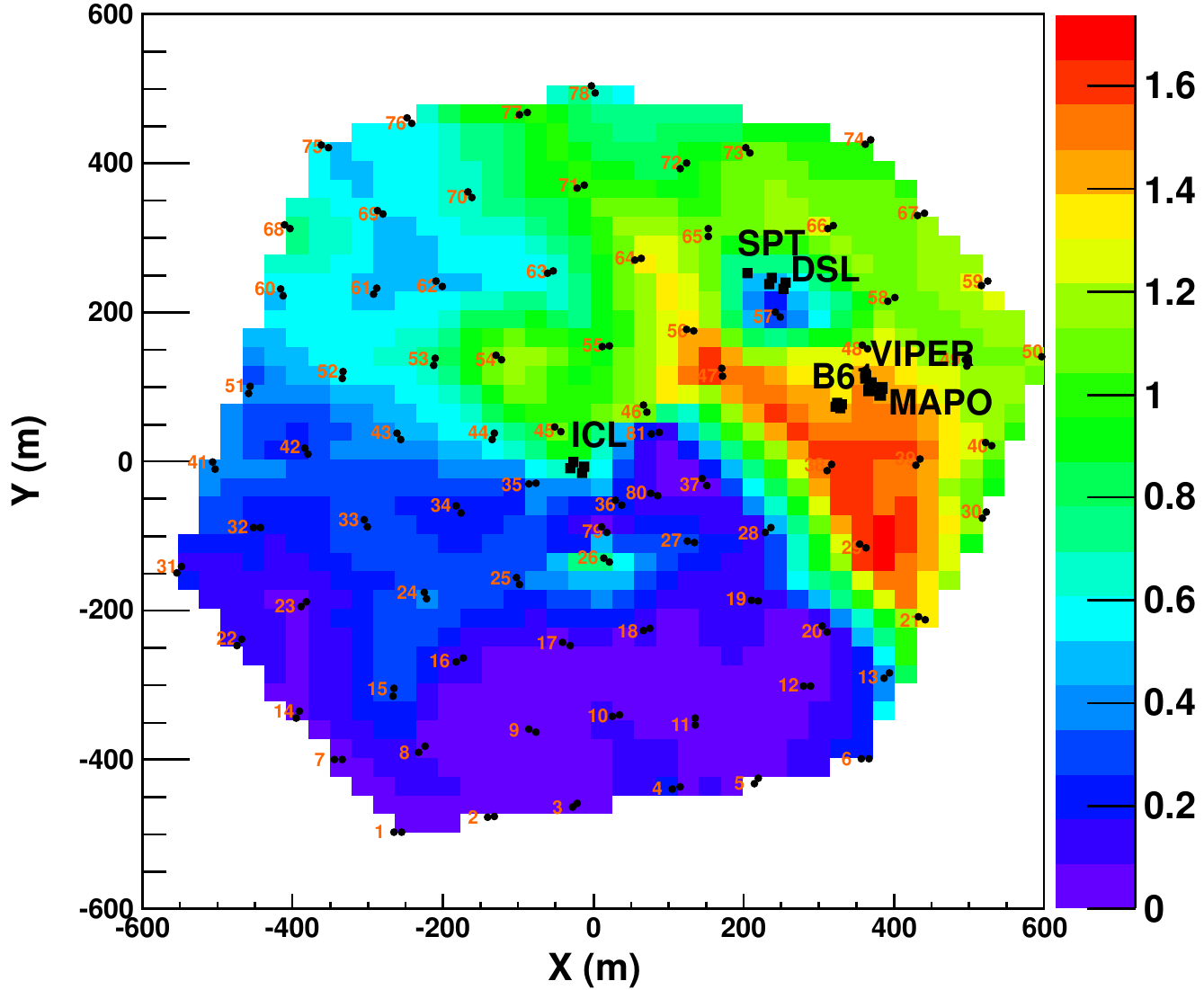}  
\caption[Distribution of snow heights on top of IceTop tanks in January,~2011]
    {Distribution of snow heights on top of IceTop tanks measured in February, 2012, and interpolated between tanks. The labels point at buildings which affect snow drifting and may give rise to special snow management.}
  \label{fig:icecube:snowdistribution}
\end{figure}

While precipitation at South Pole only amounts to about~$2\un{cm}$ per year, there is a significant snow accumulation due to drifting.
The snow height on top of the tanks increases on average by~$20\un{cm}$ every year with a density of $0.35$ to $0.4\un{g/cm^3}$ depending on depth. However, accumulation strongly depends on surrounding terrain and buildings.
At South Pole the wind has a predominant direction and a relatively constant average speed of about \unit{5}{m/s}.
This causes an increased accumulation of snow leeward of buildings and slopes. Figure~\ref{fig:icecube:snowdistribution} shows the distribution of snow on top of IceTop stations in February 2012. 

Snow depths on the tanks can directly be measured only during the Antarctic summer. However, the development of snow heights during the whole year is important for the analyses of the data. For the determination of the snow heights between direct measurements the calibration spectra as shown in Fig.~\ref{fig:61A_muonSpec_61_61} are employed.
The effect of snow is mainly an absorption of the electromagnetic shower component while the muon spectrum remains nearly unaffected by growing snow. Therefore, the ratio of the number of muons around the muon peak in Fig.~\ref{fig:61A_muonSpec_61_61} to the underlying mainly electromagnetic background depends on the snow height. This ratio is defined by integrating over the muon part $f_{\mu}$ and the electromagnetic part $f_{em}$ of the calibration fitting function \eqref{eq:VEMCal_fit_function}:
\begin{equation} \label{eq:SmuBEM} 
{\displaystyle S_{\mu}/B_{EM}={ \int_{S_{min}}^{S_{max}} f_{\mu}\,dS}\ \left/\ {\int_{S_{min}}^{S_{max}} f_{em}\,dS}\right. }
\end{equation}
Suitable integration boundaries have been found to be $S_{min}=0.3$\,VEM and $S_{max}=2.0$\,VEM.
\begin{figure}
	\centering
\includegraphics[width=0.495\textwidth]{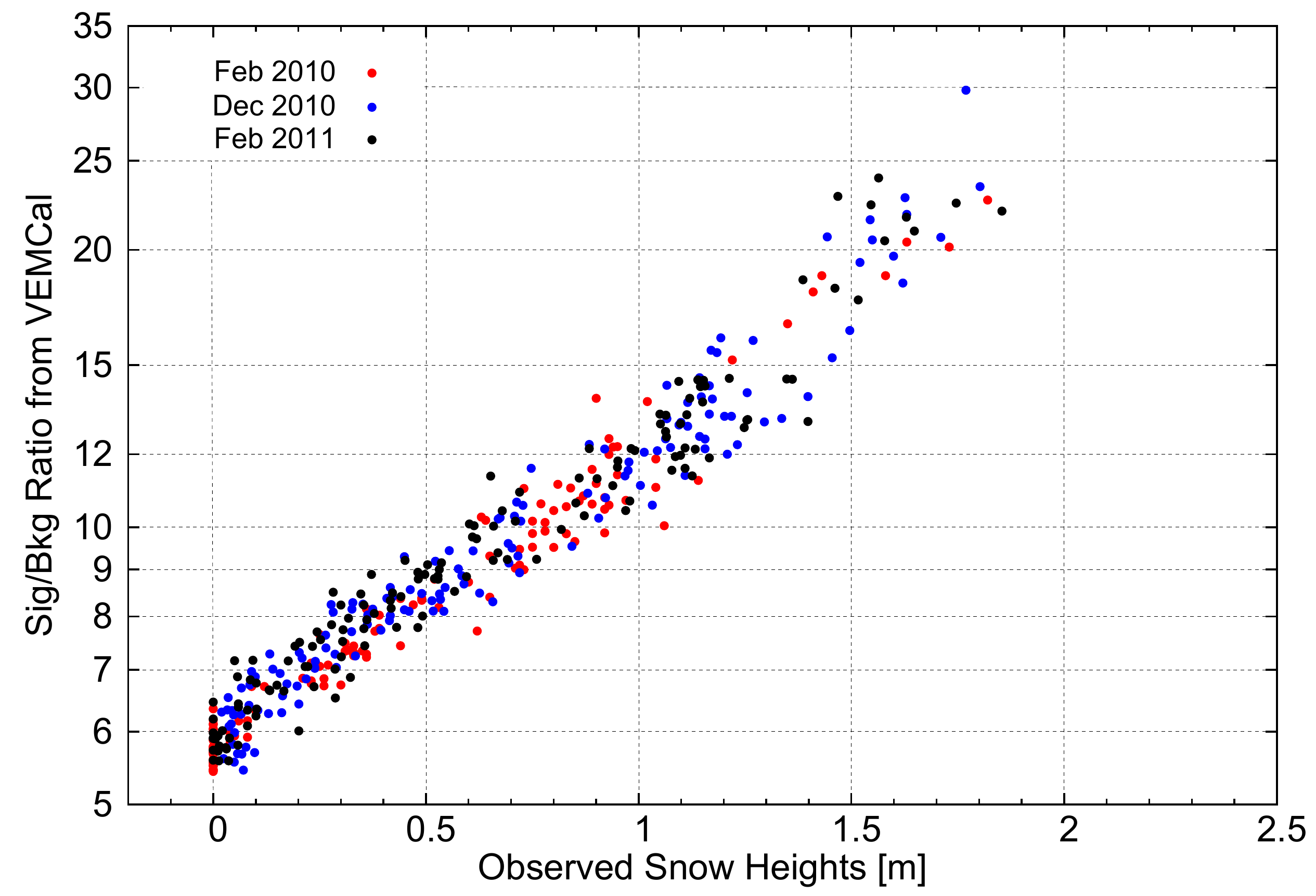}			\includegraphics[width=0.495\textwidth]{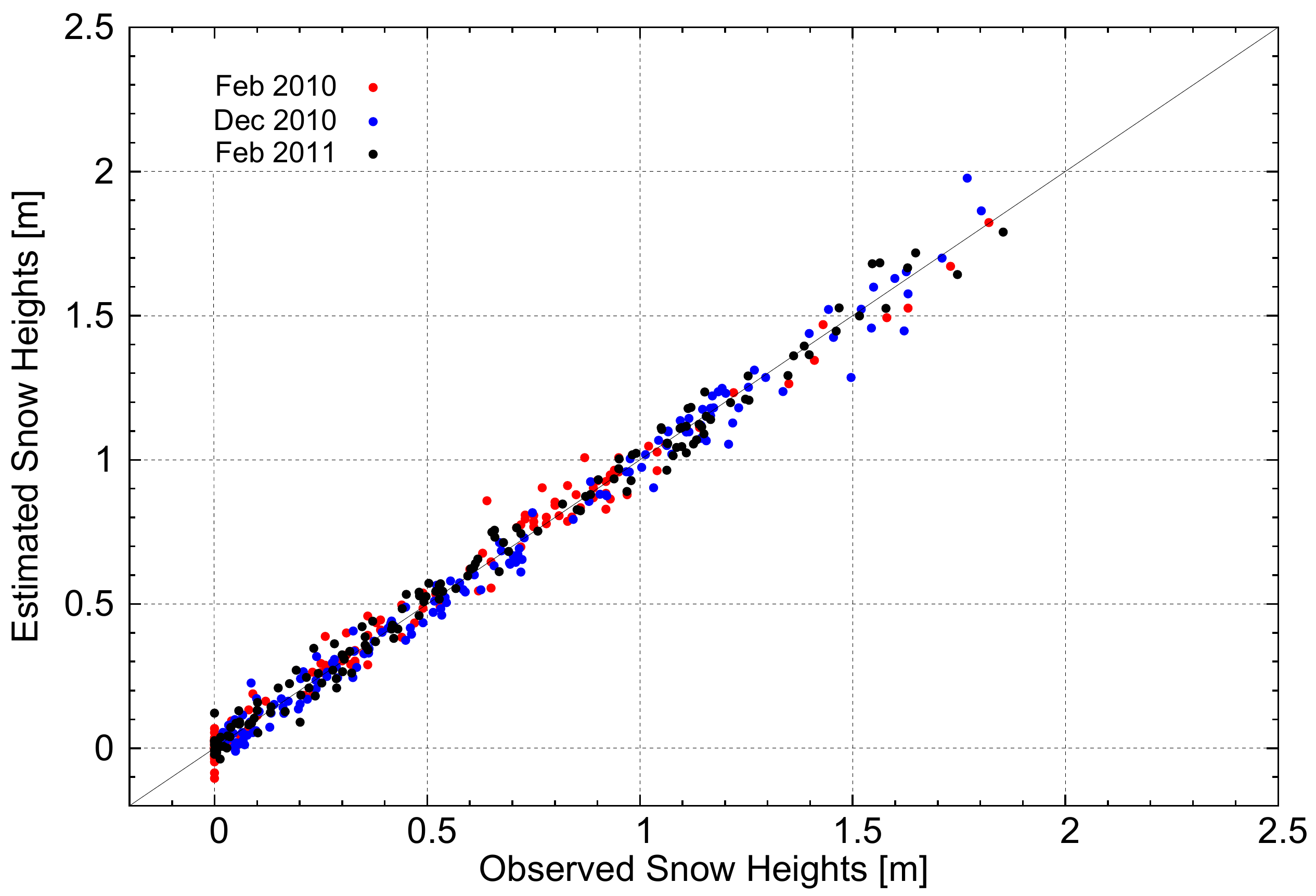}	
	\caption{Snow heights measured in Feb.\ 2010, Dec.\ 2010 and Feb.\ 2011: a) The ratio $S_{\mu}/B_{EM}$ versus the measured snow height. b)~Comparison of measured and extrapolated snow heights. The extrapolation is done by using the signal-over-background ratio of the VEM calibration with formula  \eqref{eq:VEMCal_fit_function}.}.
\label{fig:snow_height}
\end{figure}

 Figure \ref{fig:snow_height}\,a shows the correlation between measured snow heights and this ratio. The points can be well fitted by the function:
\begin{equation} \label{eq:h_snow} 
h_{s}=a_s \times (\ln(S_\mu/B_{em})+b_s)
\end{equation}
typical values of the parameters are $a_s \approx 1.37\,{m}$ and $b_s \approx 1.75$. 

Formula \eqref{eq:h_snow} is used to estimate snow heights in between direct measurements. 
For a correct extrapolation the slope $a_s$ is fixed to the average of all snow heights but the intercept $b_s$ is determined for each tank separately. A comparison of snow heights determined in this way with direct measurements is shown in Fig.\ \ref{fig:snow_height}\,b.
 The fitted snow heights are used to account for the snow effects in the reconstruction of shower parameters (Section \ref{subsec:show_correction}).

\section[Signal processing and data preparation]{Signal processing and data preparation} \label{sec:signal_proc}

Figure \ref{fig:waveforms} shows typical waveforms measured in IceTop.
For air shower reconstruction, these waveforms are processed by a series of software modules to perform three steps: calibration, extraction of total pulse charge and time, and pulse cleaning. Only the unsaturated ATWD channel with the highest gain is used in the data analysis.

First, the ATWD baseline (common for all bins, see Section \ref{subsec:calibration_dom}) is subtracted from the digitized ATWD waveforms. The waveforms are then converted from ADC counts into voltage taking into account the gain of the ATWD amplifier and the calibration of each individual ATWD bin.
In case of SLC hits, where no waveform is available, the average ADC calibration of bins 5 to 60 is used.
The start time of the waveform is corrected for the transit time of the pulse through the PMT and the electronics (Section \ref{subsec:calibration_dom}).
Finally, all waveforms are corrected for the effects of droop caused by the transformer that couples the PMT output to the recording electronics (Section \ref{subsec:droop}). The correction uses the inversion of formula  \eqref{eq:icecube:droop} in Section \ref{subsec:droop} in a discretized way. 
This is done by correcting the voltage in each time bin for the droop in the same bin and for the summed voltage responses from all preceding bins.
When a readout window contains consecutive hits from the same DOM, the 
waveform of the last hit is used to correct additionally for the residual droop in the follow-on
 hit. Droop correction cannot be applied to SLC hits but is on average accounted for by the SLC calibration (Section \ref{subsec:SLCchargeCalibration}).

The total HLC pulse charge is then extracted by integrating the calibrated
 waveform 
and converting it into units of PE and VEM to be used in reconstruction.
The pulse time is defined by the crossing of the leading edge slope of the first pulse with the baseline. The slope is taken between 10 to 90\% of the leading edge (see Fig.\ \ref{fig:waveforms}\,a). All times are expressed in UTC time.

For each tank, charge and time of the pulse in the high-gain DOM are used, unless the charge surpasses a saturation threshold determined during the VEM calibration (expression \eqref{eq:HG-crossover} in Section \ref{subsubsec:LG_cross-calibration}).
In that case, the charge is used from the low-gain DOM. The pulse time is always based on the high-gain DOM.
If there is no pulse in a low-gain DOM within~$\pm 40\un{ns}$ of a saturated high-gain pulse, the pulse is marked as saturated to be treated accordingly by reconstruction algorithms.

In addition, for air shower reconstruction, events are cleaned by requiring the following conditions:
\begin{itemize}
 \item A station is discarded if the following condition on the signal times in the two tanks A and B is not met:
  \begin{equation}
    |t_A - t_B| < \frac{|\boldsymbol{x_A}-\boldsymbol{x_B}|}{c} + 200\un{ns},
  \end{equation}
  where $t_A$ and $t_B$ are the signal times in the two tanks located at $\boldsymbol{x_A}$ and $\boldsymbol{x_B}$.
 \item Stations are grouped in clusters which in principle could belong to the same shower, such that any pair of stations $i$
and $j$ in the cluster fulfills  the condition
  \begin{equation}
    |t_i - t_j| < \frac{|\boldsymbol{x_i}-\boldsymbol{x_j}|}{c} + 200\un{ns}.
  \end{equation}
  The station position $\boldsymbol{x_i}$ is the center of the line connecting its two tanks, and $t_i$ is the average time of both tank signals. 
  In each event, only the largest cluster of stations is kept. 
\end{itemize}
This selection is done in order to remove obviously unrelated pulses from events. In the IT26 analysis \cite{IT26-spectrum_Abbasi:2012wn} it only affected about 4\% of events, where on average 2.3 tanks were removed.

\section[Air Shower Reconstruction]{Air Shower Reconstruction}
\label{sec:reconstruction}
%
%
%

\begin{figure}
  \centering%
  \includegraphics[width=0.5\textwidth]{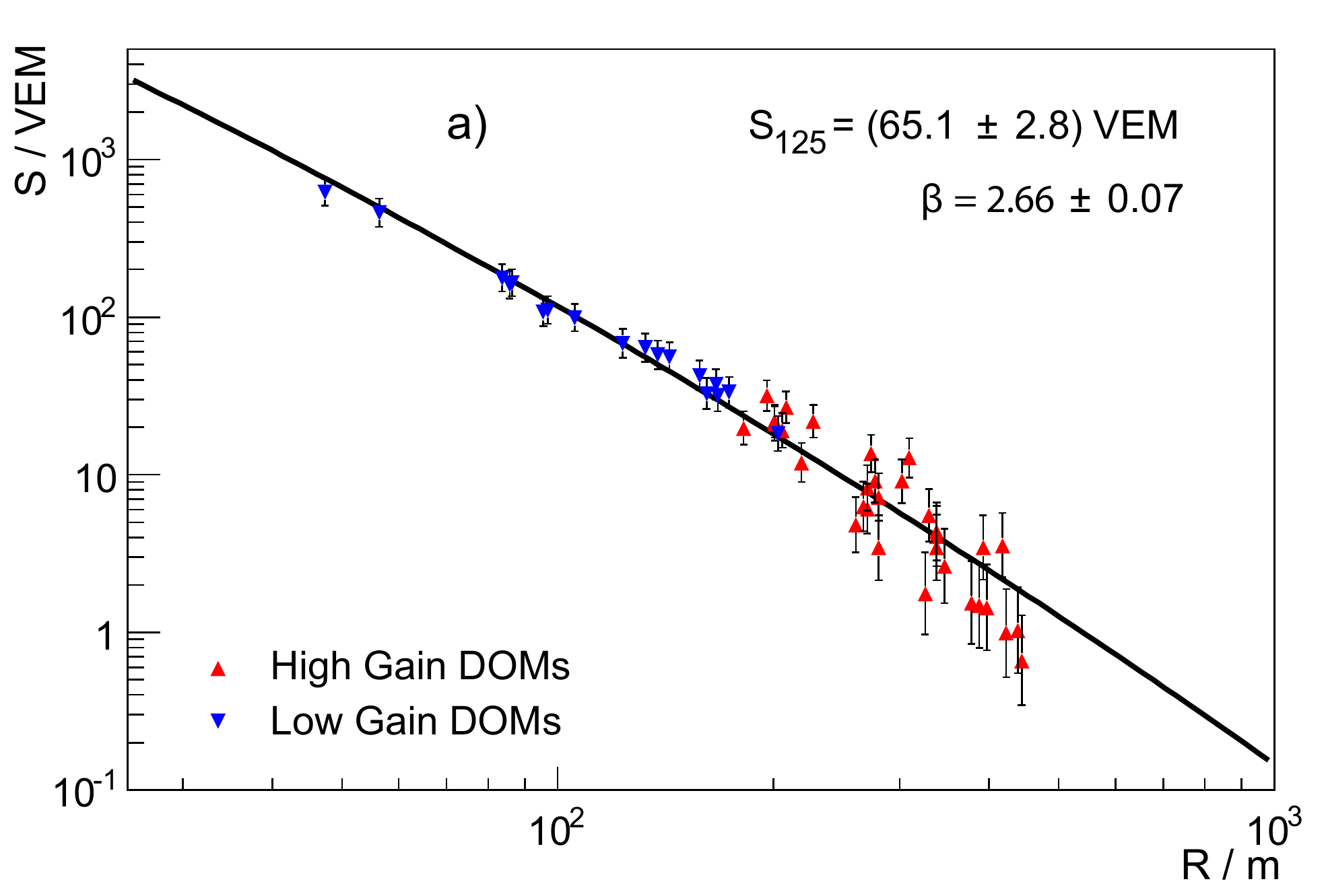}
  \includegraphics[width=0.49\textwidth]{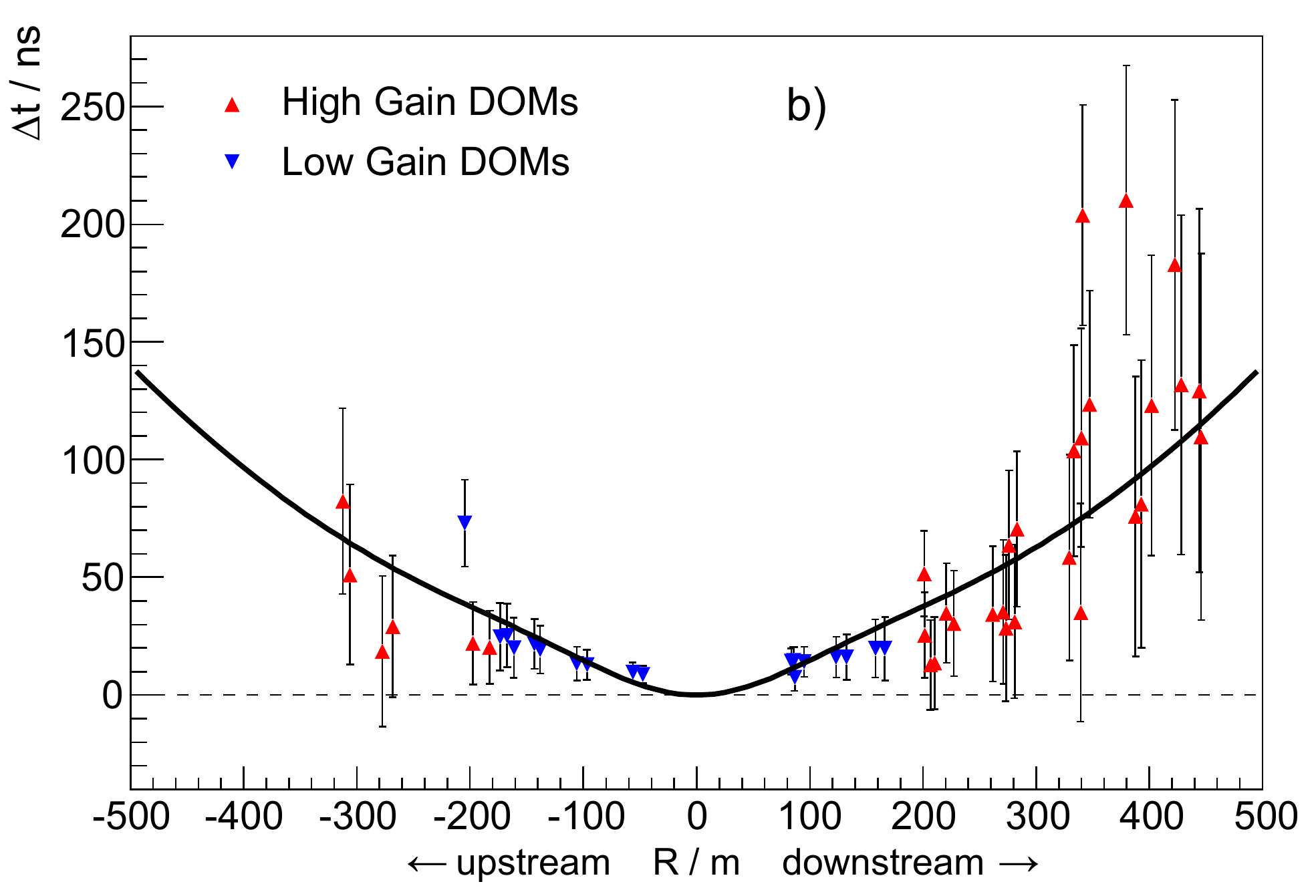}%
  \caption{Example of the IceTop air shower reconstruction. a) Lateral fit to signals from  $25$ triggered stations with a reconstructed shower size $S_{\!125} = (65.1 \pm 2.8)\un{VEM}$. b) Time residuals with respect to a plane perpendicular to the shower direction given by Eq.~\eqref{eq:curvature}. ``Upstream'' and ``downstream'' refer to tanks being hit before and after the shower core reaches the ground.}
  \label{fig:ldf}
\end{figure}

Properties of a primary particle are inferred from the  air shower parameters reconstructed from the IceTop signals. The reconstructed parameters include the shower core position and direction, and the shower size. 
The latter is a  measure of primary energy and is  defined as the signal $S_{\!\rm ref}$ measured at a certain distance $R_{\rm ref}$ from the shower axis.  
These properties are reconstructed by fitting the measured charges with a lateral distribution function and the signal times with a function describing the geometric shape of the shower front (Fig.\,\ref{fig:ldf}). Currently only HLC hits are used in the reconstruction.

\subsection{Time and charge distribution of air shower signals} \label{sub:expect_shower}

\paragraph{Lateral charge distribution}
IceTop tanks are not only sensitive to the number of charged particles, but also detect photons via electron-positron pair production. That is different from a scintillator array, which is relatively insensitive to the photonic part of the signal resulting in a lateral charge distribution which can be described by the NKG function \cite{NKG1,NKG2}.
Therefore, for IceTop an empirical lateral distribution function was derived from simulations \cite{Klepser:2007zz,StefanThesis}. 

The charge expectation value in an IceTop tank at distance $R$ from the shower axis is described by:
\begin{equation}\label{eq:ldf}
  S(R) = S_{\!\mathrm{ref}} \cdot \left(\frac{R}{R_{\mathrm{ref}}}\right)^{-\beta -\kappa\, \log_{10}(R/R_{\mathrm{ref}})}.
\end{equation}
which is a second order polynomial in the logarithm of $R$:
\begin{equation}\label{eq:logldf}
  \log_{10}\, S(R) =  \log_{10}\, S_{\!\mathrm{ref}} - \beta\, \log_{10}\left(\frac{R}{R_{\mathrm{ref}}}\right) - \kappa\, \log_{10}^2\left(\frac{R}{R_{\mathrm{ref}}}\right).
\end{equation}
This function behaves unphysically at small distances to the shower axis ($R \lesssim 1\un{m}$). 
However, as described in the next subsection, all signals that are closer than a minimal distance to the core, are excluded from the fit. 
The free parameters of the function, in addition to the shower size, $S_{\!\mathrm{ref}}$, are $\beta$ and $\kappa$, corresponding to the slope and curvature in the logarithmic representation at $R=R_{\mathrm{ref}}$. 
Simulation studies suggest that $\kappa$ can be fixed without a significant impact on the fit result. The current default is $\kappa = 0.303$.

At medium energies, the distance from the shower axis corresponding to the average logarithm $\langle\log_{10} R/{\rm m}\rangle$ of signals participating in the fit is about \unit{125}{m}. The current standard fit uses a constant $R_{\mathrm{ref}} = 125\un{m}$, which was found to minimize the correlation between the parameters $S_\mathrm{ref}$ and $\beta$ on average for the IT26 configuration. The shower size parameter is thus referred to as $S_{\!125}$. An optimization of this parameter for the full detector is under study.

\paragraph{Time distribution}
The arrival times of the signals map out the shower front. 
The expected signal time of a tank at the position $\boldsymbol{x}$ is thus parametrized as
\begin{equation} \label{eq:time_expect}
  t(\boldsymbol{x}) = t_0 + \tfrac{1}{c}(\boldsymbol{x} - \boldsymbol{x}_c) \cdot \boldsymbol{n} + \Delta t(R).
\end{equation}
Here, $t_0$ is the time the shower core reaches the ground, $\boldsymbol{x}_c$ is the position of the shower core on the ground and $\boldsymbol{n}$ is the unit vector in the direction of movement of the shower. 
For fixing the core position, `ground' is defined as the~\mbox{$\sqrt{S}$-weighted} average of participating tank altitudes, which varies by about \unit{6}{m} over the whole array. 
The term $\Delta t(R)$ describes the shape of the curved shower front as a function of distance $R$ to the shower axis and is the time residual with respect to a plane perpendicular to the shower axis which contains $\boldsymbol{x}_c$. A plane through $\boldsymbol{x}_c$ is described by $\Delta t(R)=0$, used as start value for the fit, see Section \ref{sub:likelihood_fit}. 
Studies on experimental data showed that the shower front can be described by the sum of a parabola and a Gaussian function, both symmetric around the shower axis \cite{StefanThesis}:
\begin{equation}\label{eq:curvature}
  \Delta t(R) = a \, R^2 + b \left( 1 -\exp\left(-\frac{R^2}{2\sigma^2}\right)\right),
\end{equation}
with the constants
 \[ a = 4.823 \, 10^{-4}\un{ns/m^2}, \qquad  b = 19.41\un{ns} , \qquad \sigma = 83.5\un{m}. \]
The energy, zenith angle and mass dependence of these parameters is under study and presents an opportunity for further improvement of the reconstruction quality.
Furthermore, inclined showers will not be symmetric with respect to the shower core, which may require a modification of the signal time and charge distribution.

Function \eqref{eq:time_expect} is fitted to the measured signal times with five free parameters: two for the core position, two for the shower direction and one for the reference time $t_0$. 
Hence, the complete air shower reconstruction has the following parameters: position of the shower core~\mbox{$(x_c, y_c)$}, shower direction $\theta$ and $\phi$, shower size $S_{\!125}$, slope parameter $\beta$, and time at ground $t_0$.

\subsection{Shower fluctuations} \label{subsec:shower_fluc}
Fluctuations of signal size and arrival time for each tank have to be included in the fit procedure.
The signal size fluctuations are described by a normal distribution of $\log_{10} S_{\!i}$ around the fitted expectation value $\log_{10} S^{\rm fit}_i$, with standard deviations $\sigma_{\log_{10} S}$ depending on the signal charge. 
The charge dependence of $\sigma_{\log_{10} S}$ has been determined experimentally from the local shower fluctuations between the two tanks of a station and are reasonably well reproduced by simulation \cite{FabianThesis}. 
For the shower
reconstruction a functional description is used: \begin{equation}\label{eq:reconstruction:sigma_S}
 \log_{10}\left( \sigma_{\log_{10} S}\right) = 
    \begin{cases}
      0.283 - 0.078 \log_{10}\left(S/\mathrm{VEM}\right)                               &  \log_{10} \left(S/\mathrm{VEM}\right)   < 0.340 \\
      {-0.373 - 0.658\log_{10}(S/\mathrm{VEM}) + 0.158\log_{10}^2(S/\mathrm{VEM})}  & 0.340 \leq \log_{10} \left(S/\mathrm{VEM}\right) < 2.077 \\
      0.0881       & 2.077 \leq \log_{10} \left(S/\mathrm{VEM}\right) 
    \end{cases}
\end{equation}
%
The arrival time fluctuations are described by a normal distribution with a standard deviation $\sigma_t(R_i)$, 
\begin{equation}\label{eq:sigma_t}
  \sigma_t(R_i) = 2.92\un{ns} + 3.77 \cdot 10^{-4}\un{ns} \cdot (R_i/\mathrm{m})^2\, ,
\end{equation}
depending on the distance $R_i$ of tank $i$ to the shower axis as found in experimental data \cite{FabianThesis}. No energy dependence of $\sigma_t$ has been included so far.

\subsection{Likelihood fit} \label{sub:likelihood_fit}

\paragraph{Likelihood function}
The functions~\eqref{eq:logldf}, \eqref{eq:time_expect} and \eqref{eq:curvature} describing the expectations for the charge and time of air shower signals are fitted to the measured data using the maximum likelihood method (Fig.\ \ref{fig:ldf}). 
In addition to the signal charges and times, the likelihood function also takes into account stations that did not trigger so that the full log-likelihood function consists of three terms:
\begin{equation}\label{eq:llh}
  \mathcal{L} = \mathcal{L}_q + \mathcal{L}_0 + \mathcal{L}_t.
\end{equation}

The first term,
\begin{equation}\label{eq:llh_q}
  \mathcal{L}_q = -\sum_i\frac{\bigl(\log_{10} S_{\!i} -
      \log_{10} S^{\rm fit}_i\bigr)^2}{2\,\sigma^2_{\log_{10} S}(S^{\rm fit}_i)} 
    - \sum_i\ln\bigl( \sigma_{\log_{10} S}(S^{\rm fit}_i)\bigr),
\end{equation}
describes the probability of measuring the charges $S_{\!i}$ if the fit expectation value at the position of the tank is $S^{\rm fit}_i$ as given by the lateral distribution function \eqref{eq:ldf}. The standard deviation $\sigma_{\log_{10} S}$ is given by \eqref{eq:reconstruction:sigma_S}.
The sum runs over all tanks that have triggered. 
The second sum in $\mathcal{L}_q$ accounts for the proper normalization of the signal likelihood and has to be added because the standard deviations  depend on the fitted signals.

The next term of the log-likelihood function~\eqref{eq:llh},
\begin{equation}\label{eq:llh_0}
  \mathcal{L}_0 = \sum_j\ln\Bigl(1-\bigl({P^{\rm hit}_j}\bigr)^2\Bigr),
\end{equation}
accounts for all stations $j$ that did not trigger. 
The probability that one tank in station $j$ delivers a signal at a given charge expectation value is
\begin{equation}
	\label{eq:p_hit}
	  P^\mathrm{hit}_j = \frac{1}{\sqrt{2\pi} \sigma_{\log_{10} S}(S^\mathrm{fit}_j)} \cdot
	  \int\limits_{\log_{10} S^\mathrm{thr}_j}^{\infty}
	    \exp\left(-\frac{\bigl(\log_{10} S_{\!j}-\log_{10} S^\mathrm{fit}_j \bigr)^2}{2\sigma^2_q(S^\mathrm{fit}_j)}\right)\dd \log_{10} S_{\!j}.
\end{equation}
The lower integration limit is defined through the charge threshold  $S^\mathrm{thr}_j$ for the tank signal.
The charge expectation value, $S^\mathrm{fit}_j$, is evaluated for the center of a line joining the centres of the two tanks.
Since the two tanks of one station operate in coincidence and tanks with unusually missing partners, as described in Section \ref{sec:deadtime}, are removed by the cleaning described in Section \ref{sec:signal_proc}, there are no single untriggered tanks. There is a natural correlation in the signal expectation values of two nearby tanks because they have a similar value of the lateral distribution function. However, in  
Eq.~\eqref{eq:llh_0} the no-hit probability of the are assumed to be independent because
the fluctuations about the average expectation value, $S^\mathrm{fit}_j$,  of both tanks are uncorrelated. 

The third term of~\eqref{eq:llh}, $\mathcal{L}_t$, describes the probability for the measured set of signal times,
\begin{equation}
	\label{eq:llh_time}
	  \mathcal{L}_t= -\sum_i \frac{(t_i - t_i^{\rm fit})^2}{2\,\sigma_t^2(R_i)} - \sum_i\ln(\sigma_t(R_i)),
\end{equation}
where the index $i$ runs over all tanks, $t_i$ is the measured signal time of tank $i$,  $t_i^{\rm fit}=t(\boldsymbol{x}_{i})$ is the fitted expectation value according to function~\eqref{eq:time_expect}, and $\sigma_t$ is defined in \eqref{eq:sigma_t}. 

\paragraph{Fit procedure}
The likelihood fit is seeded with first-guess calculations for the core and the direction of the shower. 
As a first estimate of the core position the centre-of-gravity of the positions $\boldsymbol{x}_i$ of the $m$ tanks with the highest signals weighted with the square root of the charges is calculated:
\begin{equation}\label{eq:cog}
  \boldsymbol{x}_{\rm COG} = \frac{\sum_{i}\sqrt{S_{\!i}}\,\boldsymbol{x}_i}{\sum_{i}\sqrt{S_{\!i}}}.
\end{equation}
The weight $\sqrt{S_{\!i}}$ is chosen based on a study of the achievable fit accuracy. The restriction to the $m$ tanks with the highest signals (current default is $m=7$) yields a better estimate, in particular near the boundaries.
The starting value for the direction is obtained by fitting to the signal times of the  $m$ tanks with the highest signals a plane, that means setting $\Delta t(R)=0$ in the function \eqref{eq:time_expect}. 

The likelihood minimization is then done in several iterations to improve the stability of the fit. 
At first the shower direction is fixed and only the lateral fit of the charges is iterated with the free parameters $S_{\!125}$, $\beta$, and core position. 
After each iteration those tanks are removed from the fit that are closer than a minimal distance $R_{min}$ to the shower axis. 
Iteration is stopped when no more tanks are removed from the fit. The parameter $R_{min}$ is optimized to improved core resolutions by mitigating the effect of saturated pulses and of biases in the determination of core positions\footnote{In the IT26 analysis \cite{IT26-spectrum_Abbasi:2012wn} $R_{min}=11$\,m was used.}. 
Then, a third iteration step is done in which the shower front curvature is included and the shower direction is varied.
Finally the direction is fixed again and $S_{\!125}$, $\beta$, and core position are allowed to vary for a fine tuning.

\begin{figure}
  \centering%
\includegraphics[width=0.48\textwidth]{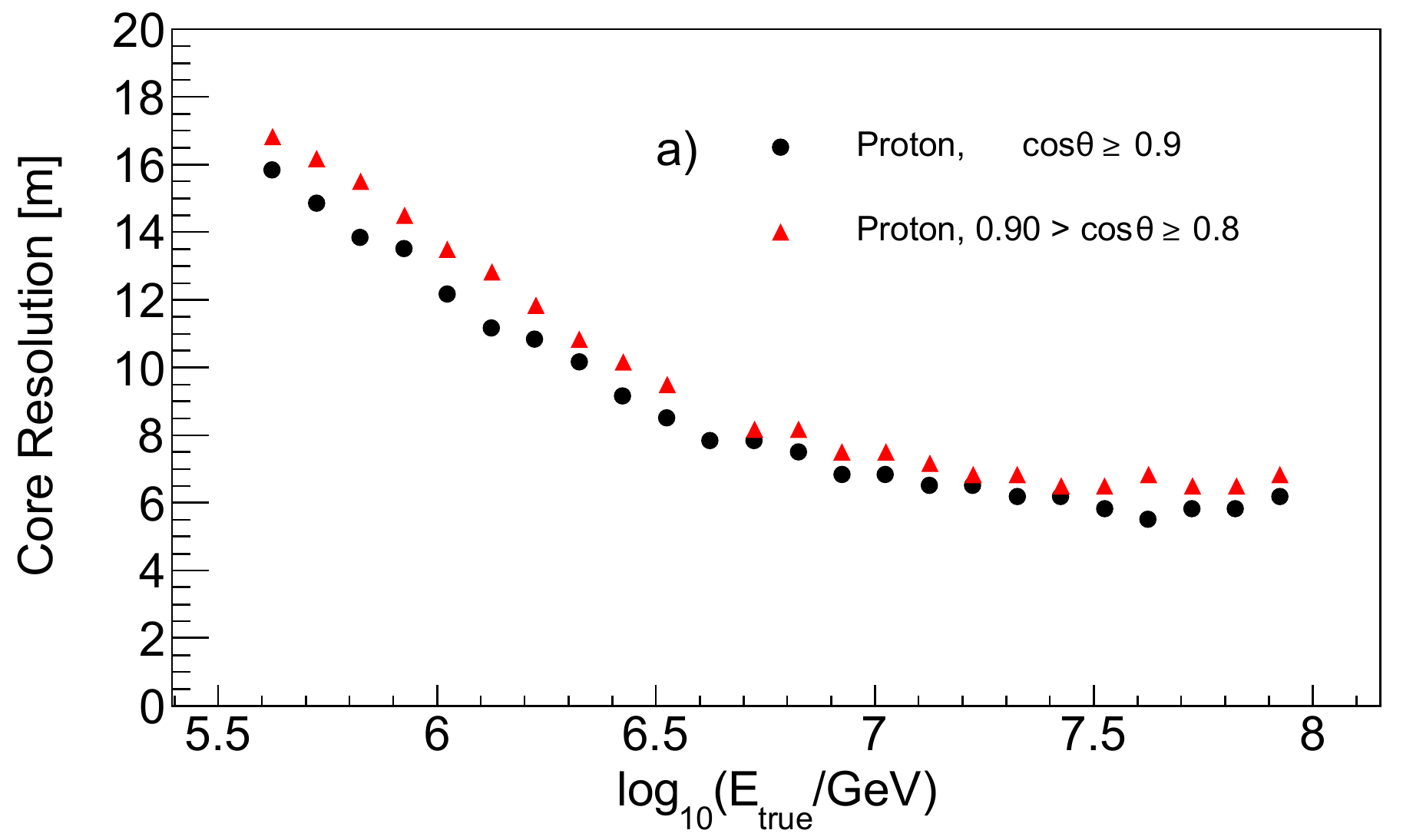}\hfill%
\includegraphics[width=0.48\textwidth]{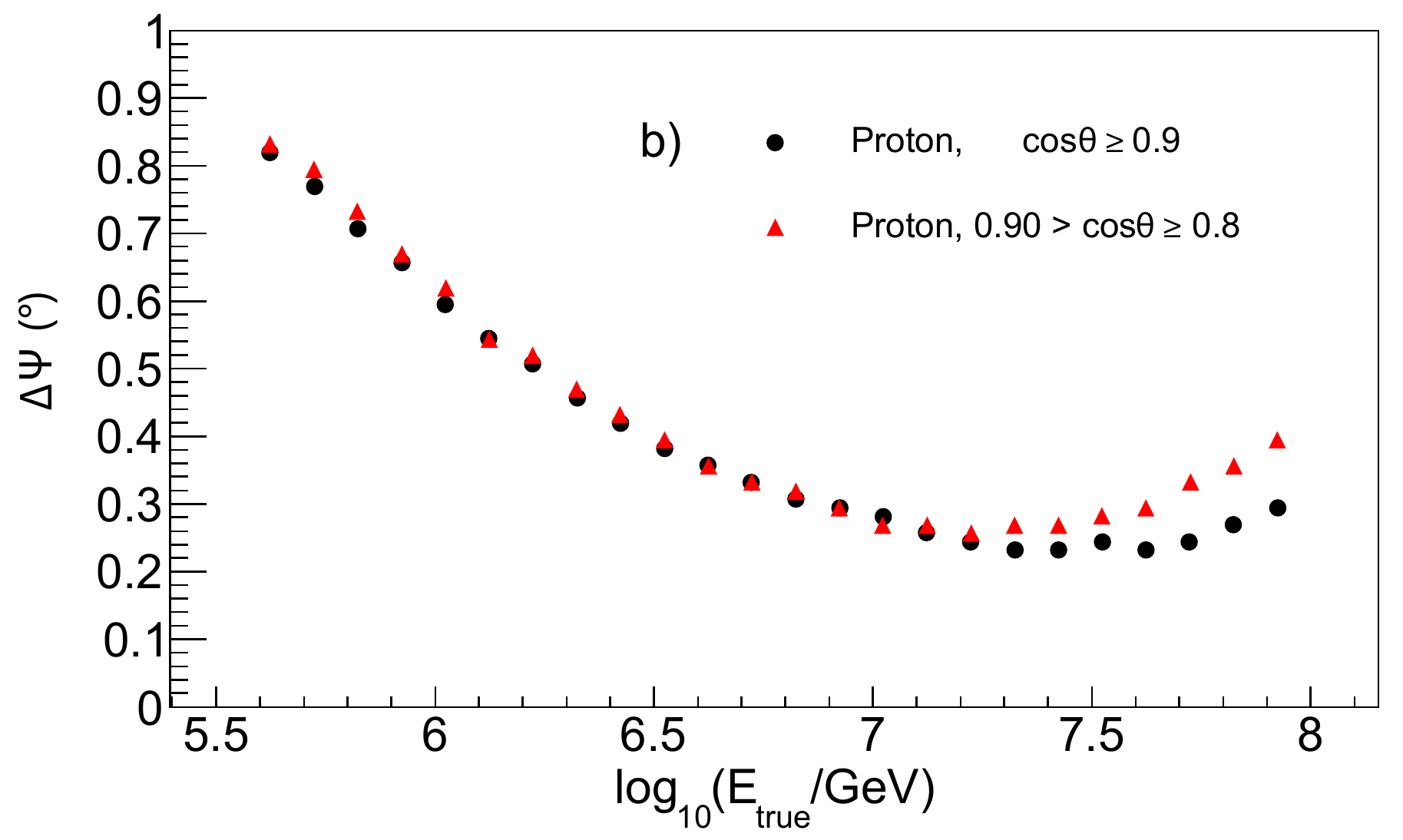}
  \caption[Core position and direction resolution]
    {Reconstruction quality of showers in two zenith angle ranges for proton showers with 5 or more stations triggered in the IT73 configuration. Shown are the 68\% resolutions as a function of primary energy. a) Distance between true and reconstructed shower core. b) Angle between true and reconstructed shower direction. }
  \label{fig:simulation:resolution}
\end{figure}
\smallskip
The fit performance is demonstrated in 
Fig.\ \ref{fig:simulation:resolution} where the energy dependence of the core and angular resolutions for the full likelihood fit are displayed. The performance for the energy reconstruction is discussed in Section \ref{subsec:perf_airshower}.

\paragraph{Small Shower reconstruction}
If only 3 or 4 stations are included in the fit, the last two iteration steps are dropped because the small number of stations with small separation between the tanks of a station does not offer sufficient redundancy. The shower direction and the $t_0$ parameter are taken from the first guess fit assuming a plane shower front.

\paragraph{Snow correction in reconstruction} \label{subsec:show_correction}
During the fit procedure the absorption of shower particles due to snow accumulation on the tanks and the snow around the tanks is corrected for. The correction is applied to the fit values in each iteration rather than to the measured charges. In this way the `no-hits' probability for signals due to the thresholds can be correctly computed using Eq.~\eqref{eq:p_hit}. For the correction an exponential absorption model is assumed:
\begin{equation} \label{eq:snow_absorption} 
 \hat{S}_i^{corr} = \hat{S}_{\!i}\, \exp\left(-\frac{h_s^i}{\lambda_s\,\cos\theta }\right).
\end{equation}
Here $\hat{S}_i$ is the charge expectation for the tank $i$ given by the lateral distribution function and $\hat{S}_i^{corr}$ the corresponding snow corrected value; $h_s^i$ is the snow height above tank $i$, $\theta$ is the zenith angle of the shower axis, and $\lambda_s$ is an effective absorption length in snow. We are studying if an effective absorption describes correctly the interplay between absorption and regeneration in a shower and the different absorption behavior of different particles or if a more complicated description is required.  From studies of data and simulations this absorption length has been found to lie between \unit{2}{m} and \unit{4}{m}. Dependences of snow attenuation on energy, signal size, zenith angle, and position with respect to the core are still under study.  

\paragraph{Atmosphere corrections of shower size} \label{subsec:atmosphere_correction}
The shower size $S_{\!125}$ is corrected for surface pressure dependence by inverting formula \eqref{eq:S_pressure_corr}. For the analysis of the IT26 energy spectrum \cite{FabianPhdThesis} the following corrections were found (replacing in \eqref{eq:S_pressure_corr} $d \ln S = \ln 10 \, \log_{10} S$  and the pressure $P$ by the overburden $X=P/g$):
\begin{align}
  \notag 0\dg \leq \theta < 30\dg:\quad & \frac{\dd\log_{10} S_{\!125}}{\dd X} = (-1.2 \pm 0.9) \cdot 10^{-4} \un{\frac{1}{g/cm^2}},\\
         30\dg \leq \theta < 40\dg:\quad & \frac{\dd\log_{10} S_{\!125}}{\dd X} = (-9.0 \pm 1.5) \cdot 10^{-4} \un{\frac{1}{g/cm^2}}, \label{eq:systematic_atmosphere_error}\\
  \notag 40\dg \leq \theta < 46\dg:\quad & \frac{\dd\log_{10} S_{\!125}}{\dd X} = (-14.3 \pm 2.9) \cdot 10^{-4} \un{\frac{1}{g/cm^2}}.
\end{align}

The dependence on the atmospheric density profile has not yet been explicitly corrected for. For data of a whole year the effect should cancel out. For more restricted time periods different atmospheres are studied in simulations (see Section \ref{sub:corsika}).

\subsection{Shower size spectra and energy determination} \label{subsec:shower_size_spectrum}
\begin{figure}
	\centering
		\includegraphics[width=0.70\textwidth]{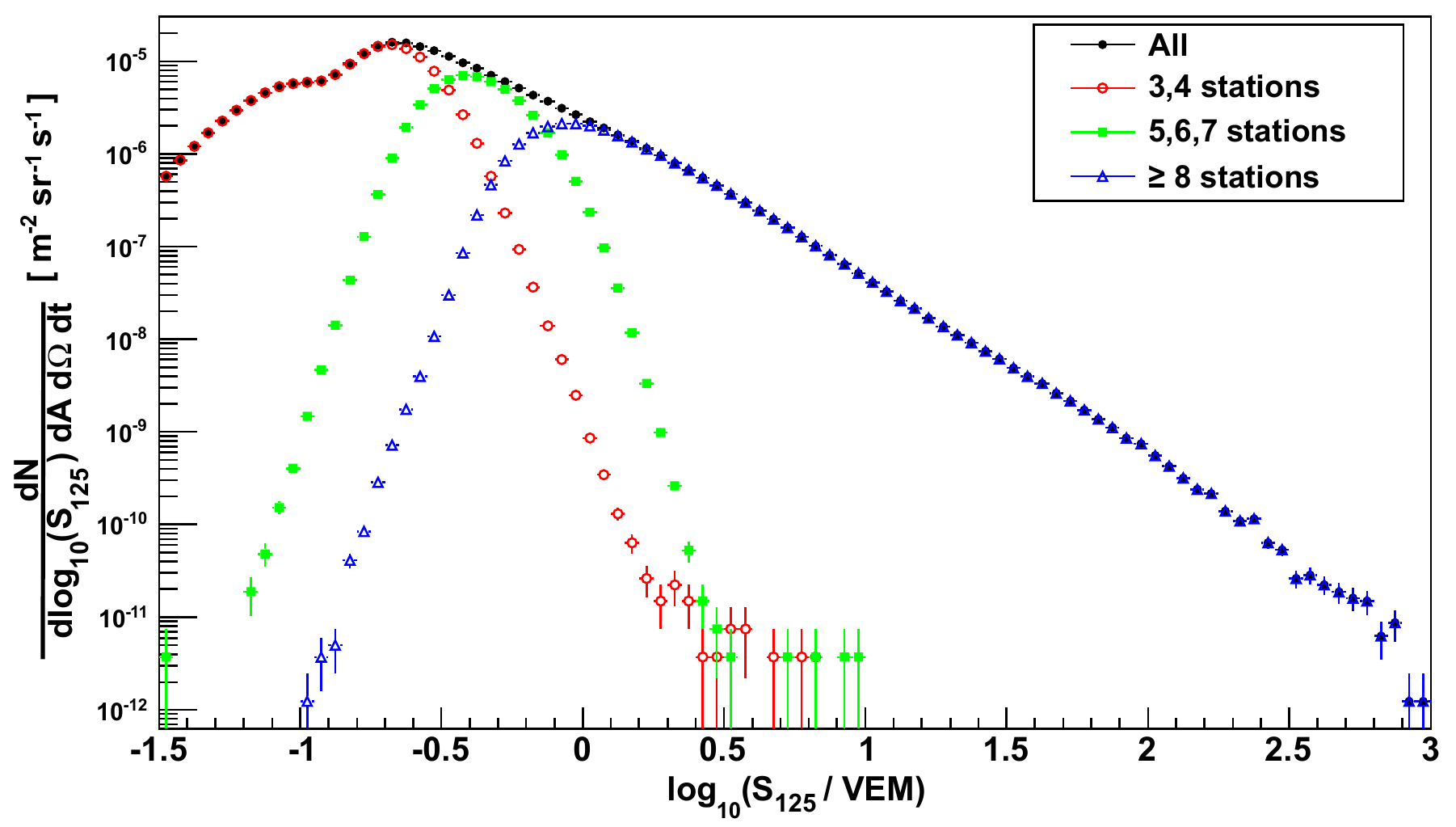}
	\caption{Shower size spectrum measured with the IT73 configuration  in the zenith angular range $\cos\theta > 0.8$ in an area of \unit{5.2\times 10^5}{m$^2$}. The spectrum contains three separate data sets with 3 and 4 stations, 5, 6 and 7 stations and 8 or more stations. }
	\label{fig:s125-filters}
\end{figure}
A shower size spectrum  reconstructed from data taken with the IT73 configuration in 2010/11 is shown in Fig.~\ref{fig:s125-filters}. 
The zenith angular range is $\cos\theta > 0.8$ and the shower core is required to lie within a polygonal fiducial area with a size of \unit{5.2\times 10^5}{m$^2$} excluding all border stations. The spectrum contains three separate data sets with 3 and 4 stations, 5, 6 and 7 stations and 8 or more stations, depicting the different energy ranges. 
 
The relation of the shower size parameter to the cosmic ray energy depends in a non-trivial way on the primary mass composition and the zenith angle and has to be evaluated using simulations (Section \ref{sec:simulation}). A determination of the cosmic ray energy spectrum requires the knowledge of the mass composition. Measurements with IceTop alone can exploit the zenith angular dependence of the shower size on the mass \cite{IT26-spectrum_Abbasi:2012wn}. A better handle on the mass composition is achieved by measuring the air shower on the surface, which is dominated by the electromagnetic component, in coincidence with the high-energy muons in IceCube. The energy spectrum and mass composition have then to be extracted by multi-variate algorithms, for example a Neural Network \cite{KarenThesis,ITIC40-spectrum_Abbasi:2012} or two-dimensional unfolding \cite{ChenThesis}.

To get a rough idea of the corresponding energies of the shower size spectrum in Fig.\ \ref{fig:s125-filters}, one may use above $S_{\!125}\approx 1\,{\rm VEM}$ the approximate relation $E/S_{\!125}\approx 1\,{\rm PeV/VEM}$ for proton primaries at small zenith angles. 

\section[Simulation of air showers and the IceTop detector]{Simulation of air showers and the IceTop detector} \label{sec:simulation}
The relation between the measured signals and the energy of the primary particle, as well as detection efficiency and energy resolution of the IceTop detector are obtained from simulations of air showers generated by the  CORSIKA program \cite{CORSIKA,corsika_UsersGuide} and then traced through the detector by a Geant4-based \cite{Geant4_1,Geant4_2} program. 

\subsection{Air shower simulation}\label{sub:corsika}

Air showers in the atmosphere  are simulated using the code CORSIKA. The hadronic component of the air showers is simulated using the models SIBYLL2.1 \cite{SIBYLL1,SIBYLL2} or QGSJET-II \cite{QGSJET1,QGSJET2} for high energy interactions and FLUKA~2008.3~\cite{Fluka1,Fluka2} for low energy interactions (the current default is  \unit{< 80}{GeV}). 
The electromagnetic component is simulated using EGS4 \cite{EGS4}.   `Thinning' (reduction of the number of traced particles \cite{Hillas_thinning}) can optionally be applied. 

For the South Pole, CORSIKA provides six different pre-defined atmosphere profiles \cite{corsika_UsersGuide}.  Four of these atmospheres (CORSIKA labels 11 to 14) are based on the MSIS-90-E model \cite{MSIS}, one for each austral season. The two others, modeled by P.\,Lipari, are alternatives for austral summer and winter. There is also the option to define custom profiles, for example using actual measurements of the South Pole atmosphere  by satellites and balloons.  


\subsection{Detector simulation}\label{sub:detsim}
The output of the CORSIKA program which includes the shower particle types, positions and momenta at the observation level of $2835\un{m}$, are injected into the IceTop
 detector simulation which uses the Geant4 package. 
The simulation starts with the generation of the amount of light produced by the shower particles in the tanks followed by the simulation of the PMT, the DOM electronics and the trigger chain.

To make the Geant4-based detector simulation more efficient, optical tank properties are studied in a stand-alone, single tank simulation (Tanktop). Some of these results are used in a parametrized form for the full detector simulation.

\subsubsection{Study of optical tank properties with Tanktop}\label{subsubsec:tanktop}
A standalone program, Tanktop, has been developed allowing detailed studies of the optical properties of the tanks and DOMs.
%
The optical properties are described by wavelength-dependent functions for PMT efficiencies, reflectivities of the tank walls and the ice-Perlite or ice-air boundaries and the absorption in the ice. Reflectivities of the tank wall liners (Fig.\ \ref{fig:DifReflectivity}) and of the perlite filling on top of the ice have been measured under laboratory conditions and have to be tuned for the simulation to obtain agreement with observed signal characteristics. The tuning was done by scaling the reflectivities by a power law $R'(\lambda) = R(\lambda)^p$, with $p$ typically about 0.38 (that means that the tuning requires higher reflectivities than measured in the laboratory).  The absorption length in ice was taken to be \unit{100}{m} from measurements in the deep-ice (this value is not critical). For the wavelength dependent PMT efficiencies averages of sample measurements are used \cite{PMTPaper}. 
Another study concerned the implementation of the snow cover of the tanks into the simulation. 

The simulation studies confirm that an effective, average description of detector properties is sufficient if one employs the VEM concept to calibrate data to a standard signal (Section \ref{sec:tank-calibration}). The essential conclusions from the Tanktop studies for the full detector simulation are:
\begin{itemize}
	\item[-] an effective exponential decay time of pulses due to  reflectivities and absorption;
	\item[-] verification that the number of generated Cherenkov photons in the wavelength interval $300\un{nm}$ to $650\un{nm}$ is proportional to the number of detected photoelectrons;
	\item[-] verification that the latter relation is -- within the assumptions of the simulation --  independent of the direction and location of the simulated particle's trajectory in the tank;
	\item[-] verification that the effect of snow can be included by starting the shower particle tracking just outside the snow volume.   
\end{itemize}

\subsubsection{Tank simulation}\label{sec:TankSimulation}
The Cherenkov emission inside the tanks is simulated using Geant4. 
All structures of the tank, the surrounding snow, including individual snow heights on top of each tank, as well as the air above the snow are modeled realistically (see Tables \ref{tab:IceTopTank} and \ref{tab:g4material} \cite{ThomasThesis}). 
The snow heights used in the simulation are implemented for the specific data period as explained in Section \ref{sec:Snow_on_tanks}.
In order to save computing time, Cherenkov photons are not tracked; only the number of photons emitted in the wavelength interval $300\un{nm}$ to $650\un{nm}$ is recorded. This procedure was studied using the stand-alone  
program Tanktop, described above, 
which includes Cherenkov photon tracking until photons reach the PMT.
The propagation of Cherenkov photons is modeled by distributing the arrival times according to an exponential distribution, which was tuned with Tanktop by varying the optical tank parameters such that simulated waveform decay times match those observed in experimental data (Table \ref{tab:IceTopPulseCharacteristics}). In the simulation the number of Cherenkov photons per VEM is fixed and the same for all tanks (Section \ref{sec:CalibrationSimulation});  the corresponding number of photoelectrons per VEM, in turn, is taken from the individual VEM calibration of the real tanks. While the photoelectrons are simulated with the exponential time distribution of the pulse decay (Table \ref{tab:IceTopPulseCharacteristics}), a realistic leading edge slope results from the simulation of transport through the electronics  (Section \ref{sec:PMT_DOM_Simulation}). 

\begin{table}
	\centering	
	\caption{Table of materials and their properties used in the Geant4 simulation. The column `composition' the fractions are given in parentheses.  }
	\label{tab:g4material}
	\begin{tabular}{|l|c|l|l|}
 		\hline
 		Material	& Composition 						& Density [g/cm$^3$] & Conditions/Properties \\
 		\hline
 	  air			  & N($0.755267$),							&	0.001205					& temperature $T=243.15\,$K,\\
 	  					&	0($0.231781$),							&										& pressure $p=67\,$kPa,\\
 	  					&	Ar($0.012827$),							&										& ionisation potential\\ 
 	  					&	C($0.000124$)								&										& $U_I=85.7\,$eV\\
 	  snow		  &	H$_2$O											&	0.38							& \\
 	  ice				&	H$_2$O											& 0.92							& index of refraction $n=1.31$,\\
 	  					&															&										& optical photon energies:\\
 	  					&															&										& $E_{opt}=[1.91\,$eV, $4.13\,$eV]\\
 	  plastic		&	H$_2$CO											&	1.425							& substitute for the tank materials\\
 	  Perlite		&	N($0.7018$),O($0.2491$),		&	0.1598						& mixture of $92.92\,$\% air\\
 	  					&	H$_2$O($0.0021$),	Ar($0.0119$),		&								& and $7.08\,$\% Perlite\\
 	  					& Si($0.0240$),Al($0.0051$),	&										& \\
 	  					&	K($0.0025$),Na($0.0024$),		&										& \\
 	  					& traces of C, Fe, Ca, Mg, Mn	&										& \\
 	  glass			&	SiO$_2$											&	2.254							& DOM pressure sphere\\
 	  DOM				&	SiO$_2$											&	0.2								& effective DOM material\\
 	  					&															&										& with average density \\
 		\hline
	\end{tabular}
\end{table}

To further reduce the required computing time, particles whose trajectories miss a tank's volume by more than \unit{30}{cm} are excluded from the detector simulation. 
Additionally, when the total charge created inside a tank is larger than \unit{2000}{VEM}, further particles are not injected into the Geant4 simulation, but only counted. 
The final number of photoelectrons at the PMT is then scaled accordingly. 
This significantly reduces the time needed to simulate tank signals very close to the core. 

\subsubsection{PMT and DOM simulation} \label{sec:PMT_DOM_Simulation}
\begin{figure}
	\centering
		\includegraphics[width=0.50\textwidth]{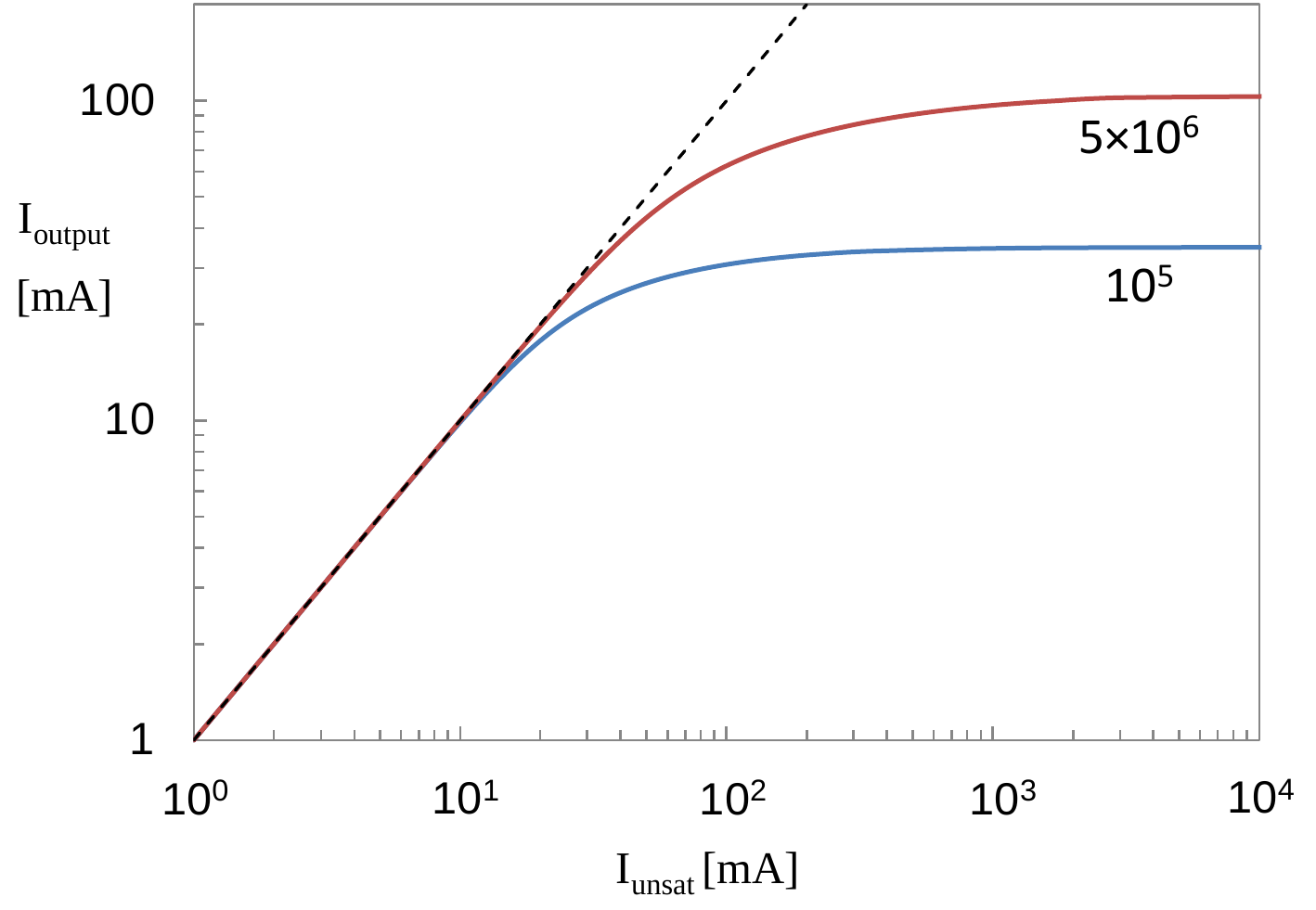}
	\caption{Saturation curves as used for the IceTop simulation for the high-gain (red) and low-gain (blue) PMTs. The the curves are labeled with the corresponding gain value. The saturated current is plotted versus the ideal, unsaturated current. The saturation currents given in Table \ref{tab:IceTopPulseCharacteristics} are the unsaturated currents for which the output current is 50\% lower.}
	\label{fig:PMT_saturation}
\end{figure}

In the next step the generated photoelectrons are injected into a simulation of the PMT followed by the analog and digital electronics of the DOM. 
To simulate the photomultipliers, single photoelectrons with Gaussian shaped waveforms ($\sigma =3.2\, {\rm ns}$) and  a  charge randomly chosen from the measured single photoelectron spectrum \cite{PMTPaper} are superimposed.
Pulse saturation is simulated depending on the instantaneous current according to the gain dependent functions depicted in Fig.\ \ref{fig:PMT_saturation} (PMT saturation is discussed in Section \ref{sec:PMT_gain}). 

In the DOM simulation, the pulse shaping due to the analog front-end electronics is applied to the output of the PMT simulation. 
This includes the simulation of the droop effect using a discretized form of formula \eqref{eq:icecube:droop} (Section \ref{subsec:droop}) and the individual shaping of the signal paths to the ATWD and the discriminators using formula \eqref{eq:atwd_tf} (Section \ref{sec:Discriminators}). 
Next, the discriminators are simulated and the local coincidence conditions are evaluated. Finally, the digitization of the waveforms by the ATWDs and fADCs and the array trigger (IceTopSMT, Section \ref{subsubsec:trigger_settings}) are simulated.

Simulated data have the same format as the experimental data and are reconstructed in the same way, as described in the previous section.

\subsection{Definition of VEM units in simulation} \label{sec:CalibrationSimulation}

The VEM `calibration' of the tank simulation, that is the transfer of the experimental VEM definition to the simulation, is achieved 
by relating the number of Cherenkov photons generated by any particle in the tank to a value in units VEM. The relation of 1 VEM to the number of photoelectrons is taken from the calibration data of a typical tank (for each tank type), see Section \ref{sec:calibration}. However, the specific value of PE per VEM is not important for the absolute calibration because it cancels out, but it has an influence on the fluctuations. 

The value 1 VEM is assigned to a certain number of Cherenkov photons such that the shower simulation yields a charge spectrum with the correct muon peak position.
This assigment is done in an iterative way along the following steps:
\begin{itemize}
	\item[-] Simulate CORSIKA air showers in an energy range in which showers contribute to the VEMCal spectrum with single muons (empirically: \unit{3}{GeV} to \unit{10}{TeV} for an $E^{-2.7}$ spectrum and zenith angles up to 60\degree\ \cite{Arne_icrc2011}) and simulate the number of Cherenkov photons emitted in the tank using Geant4. 
	\item[-] Assign the signal charge of 1 VEM to a certain number of Cherenkov photons (as a start value the Cherenkov photons  which a vertical muon of some energy produces). 
	\item[-] Take the relation of PE per VEM from data of a typical tank.
	\item[-] Simulate the electronic chain for the photoelectrons.
	\item[-] Apply the same calibration trigger condition as in data (VEMCal launch, Section \ref{sec:trigger_daq}).
	\item[-] Generate the single muon spectra as in the VEM calibration and fit with the same function as in data formula \eqref{eq:VEMCal_fit_function} in Section \ref{sec:tank-calibration}).
	\item[-] Compare the position of the muon peak in VEM units to the data of a tank (if VEM units are used it can be any tank). 
	\item[-] If the simulated peak is not at 1/0.95 VEM tune the VEM assignment to the number of Cherenkov photons (for the factor 0.95 see Section \ref{subsubsec:VEMdefinition}).
\end{itemize}
The current implementation of the VEM unit into the simulation is: 1 VEM corresponds to 32\,235 Cherenkov photons in the wavelength range 300 to  650 nm. This is in Geant4 the yield of a  vertical muon with a kinetic energy of about \unit{0.8}{GeV}.

\subsection{Simulation production} \label{sub:simulation_datasets}

\subsubsection{Properties of datasets}

For cosmic ray analysis, output files from CORSIKA simulation are available for showers initiated by different primaries (proton, iron, helium, oxygen, and silicon), energies between 10$^4$ and 10$^9$ GeV and zenith angles of the incoming primaries between 0 and 65\degree . Simulations up to 40\degree\ are used to analyze IceTop events in coincidence with the in-ice detector.  Above 40\degree , simulations are employed for studies of surface events and events with core locations outside the IceTop array area that can still trigger the in-ice detector (for example veto studies, see Section \ref{subsec:IT_veto}). In addition, for energies between \unit{10^7}{GeV} and \unit{10^{9.5}}{GeV}, thinned showers (see below) are simulated.
The overlap with the energy range of non-thinned showers is used to estimate the systematics induced by thinning. 

The simulation datasets  are grouped in energy bins  of size $0.1\times\log_{10}(E/{\rm GeV})$ with an  $E^{-1}$ spectrum, independently normalized in each bin. For comparison with data, the simulations can be weighted to yield a realistic spectrum.  The direction zenith and azimuth angles are generated to yield an isotropic distribution (uniformly distributed in $\cos\theta$, for sampling on a horizontal detector surface the distribution is proportional to $\cos\theta\, d\cos\theta$).

Most of the events are simulated for one specific atmosphere (currently model 12 \cite{corsika_UsersGuide} which is the South Pole atmosphere of  July 1, 1997). Smaller data sets generated with other atmospheres are available for studies of atmospheric effects and their corrections.   The default interaction models used are SYBILL at higher energies and FLUKA in the lower energy range.
Additional datasets generated with the QGSJET model instead of SYBILL are produced for comparison. Compositions according to various models, such as a two-component model \cite{glasstetter} or the polygonato model \cite{hoerandel04,Gaisser:2011cc},  are produced by adding different mass components with an appropriate spectral weighting. 

\subsubsection{Re-Sampling}
Since shower generation is CPU intensive, the same showers are sampled several times before being passed to the detector simulation. Showers are sampled inside a circle with an adaptable, energy dependent radius around the center of the IceTop array. 
The number of samples is chosen such that every shower would remain on average only once in the final sample after applying selection cuts. This  balances effective use of the generated showers with the artificial fluctuations introduced by oversampling.
Typically showers are re-sampled within radii ranging from \unit{600}{m} to \unit{2900}{m} for energies above \unit{10^{5}}{GeV} (usually no re-sampling at lower energies).

\subsubsection{`Thinning'}

Thinning of simulated showers \cite{corsika_UsersGuide} is a method to reduce computation time and storage volume by reducing the number of particles contributing to the shower development. This is achieved by keeping only one particle out of all secondary particles originating from a particle interaction with energies below a threshold energy $E_{thin}=\epsilon\times E_p $, where $\epsilon$ is the `thinning level'.  The kept particle gets a weight assigned which is the sum of energies of the particles below the threshold from which it was selected, divided by the kept particle's energy. This ensures the conservation of energy, although not of the particle number. The kept particles can interact again, leading to a multiplication of weights. Optionally,  to reduce the fluctuations induced by thinning, no more thinning is applied to daughters of that particle if a weight exceeds a maximum weight $w_{max}$.

 For IceTop/IceCube simulations a thinning level of $\epsilon = 10^{-6}$ is used for primary energies up to $E_p = 10^{8.4}$ GeV. Above this energy the thinning level is calculated as $\epsilon = 273\,{\rm GeV}/E_p$ so that muons capable of penetrating deep into the ice (defined as \unit{E>273}{GeV} for vertical muons) are not assigned thinning weights. The weights are limited to $w_{max}= \epsilon\times E_p$.
 
For the detector simulation `un-thinning' has to be performed as follows: a weighted particle which hits a predefined sampling area around a tank generates a number of clones, each with the same energy as the parent, which are uniformly distributed over the sampling area. The number of particles is given by the weight such that the energy remains conserved. Technically only those clones are generated which have a chance to hit the tank. The radius of the sampling area is optimized such that on average only one particle hits a tank. 

\begin{table}
	\centering
		\caption{Some typical CPU times for shower generation with CORSIKA.}
	\label{tab:TypicalCPUTimesCorsika}
		\begin{tabular}{c|cc|cc}
		energy & \multicolumn{2}{|c|}{not thinned} & \multicolumn{2}{c}{thinned} \\
		{\ }[GeV]		& time  [s]& size [MB] & time [s]& size [MB]\\
			\hline
			$10^5$ & 117 & 5 \\
						$10^6$ & 879 & 62\\
			$10^7$ & 8,119 &760&  1\,950 &167\\
			$10^{7.9}$ & 60\,657& 6\,650&3\,314 & 260\\
			\hline
		\end{tabular}
\end{table}

\subsubsection{Computing resources} 
Cosmic ray simulation is performed on different, geographically dispersed computer clusters which are controlled by a GRID-based system. The produced data are transferred to the data center at the University of Wisconsin, Madison and become available for further processing. 

At present, simulated CORSIKA showers for coincident events occupy 145 TB which is a large fraction of the available storage. In Table \ref{tab:TypicalCPUTimesCorsika} typical CPU times and data sizes for single CORSIKA showers are reported. Simulations of the highest energy showers in the range between 10$^7$ and 10$^9$ GeV can take up to 2 days for a single shower with resulting file sizes of up to 10 GB.
 Simulations in the range from 10$^4$ to 10$^7$ GeV result in file sizes of the order of hundreds of MB with computing time of the order of hours. 
 
 The detector simulation needs about 1/3 of the corresponding CORSIKA computing time and adds only 3\% to the disk usage.
\section[Performance]{Performance}
\label{sec:performance}

%
%
%

\subsection{Tank properties and detector stability}  \label{subsec:perf_tank}
\begin{figure}
	\centering
		\includegraphics[width=0.65\textwidth]{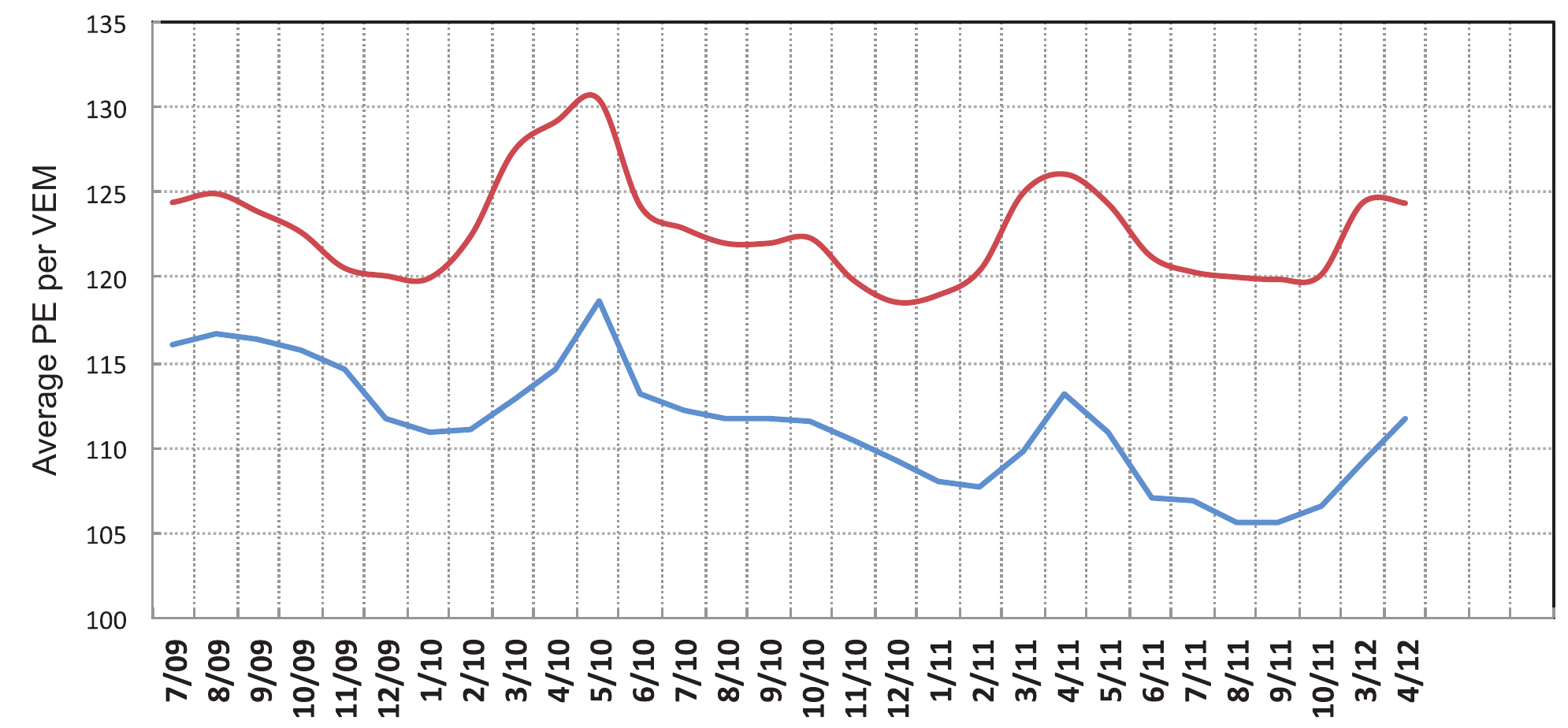}
	\caption{The curves show the average charge in PE per VEM as a function of time between July 2009 and April 2012 for all DOMs which have been operational at the given time. The upper curve is for high-gain DOMs and the lower for low-gain DOMs.}
	\label{fig:Weekly_PE_per_VEM_allDOMs }
\end{figure}

The commissioning history of IceTop tanks is reported in Table \ref{tab:ICIT_configs}. About seven years after the first tanks were commissioned the long-term stability of the tank and ice properties as well as of the electronics can be evaluated. In November 2008, two test tanks that had been deployed in November 2003 were inspected to confirm that the quality of the ice and the bonding between the DOMs and the ice were still good. More quantitatively the stability is demonstrated with Fig.\ \ref{fig:Weekly_PE_per_VEM_allDOMs } showing the average signal charge per VEM as a function of time. The high-gain DOMs do not exhibit an obvious trend.  For the low-gain DOMs a drop of about 10\% over the displayed 30 months or 4\% per year cannot be excluded. The changes per VEM calibration have typically a standard deviation of \unit{\pm (4-6)}{PE}. The PMT gains are usually adjusted by not more than 15\% (the gain precision is estimated to be about 10\%). Only one DOM out of the total of 324 IceTop DOMs shows permanent failure.  The uptime of the detector is typically larger than 99\%.

%

\subsection{Air shower detection in IceTop}  \label{subsec:perf_airshower}

\begin{figure}
	\centering
	\includegraphics[width=0.6\textwidth]{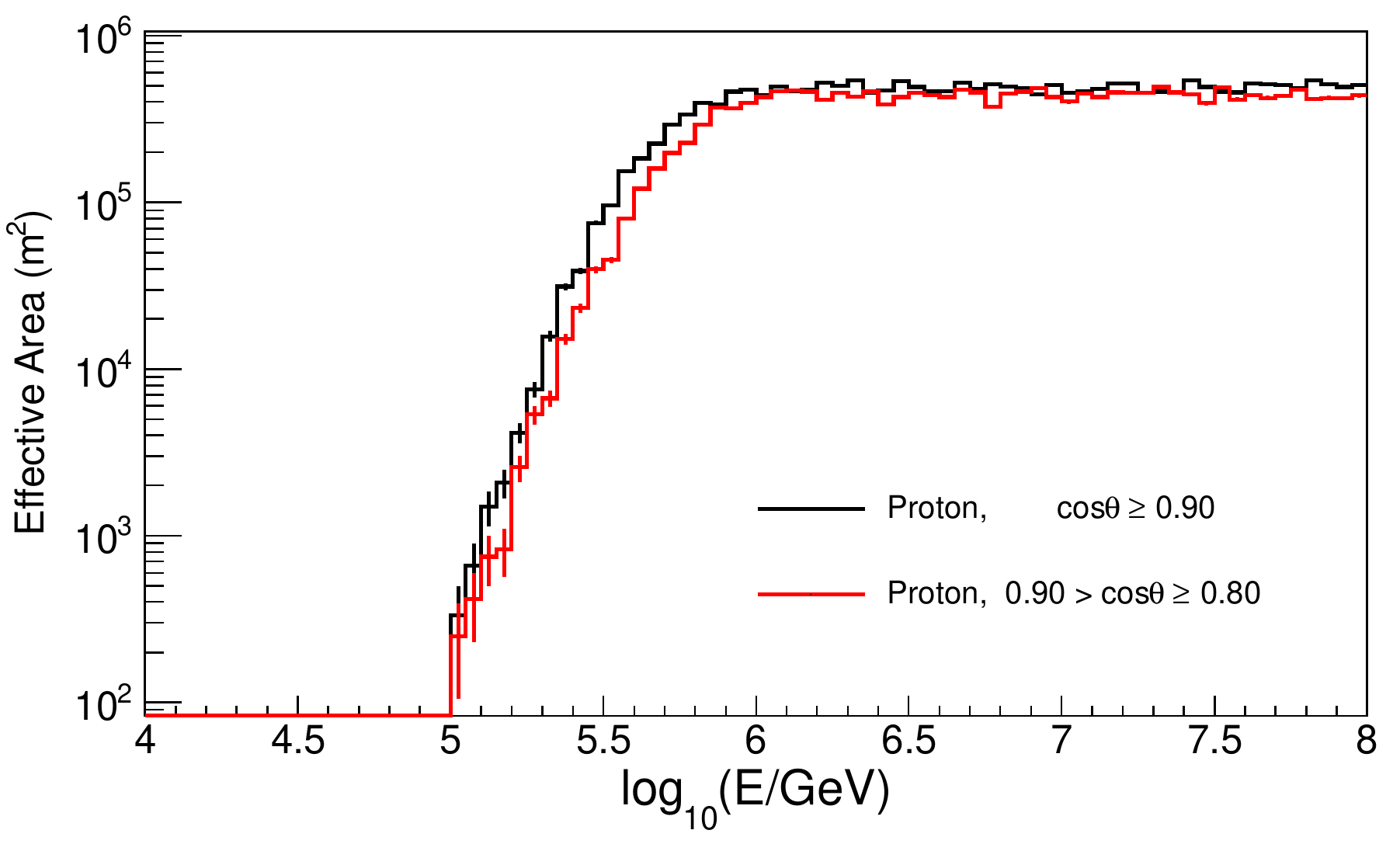}
	\caption{Effective area for proton showers which triggered 5 or more stations as a function of energy for the IceTop array as commissioned in 2010 (IT73).}
	\label{fig:AeffIT}
\end{figure}

\paragraph{Effective area} Air showers are reconstructed if 3 or more stations have triggered as shown in Fig.\ \ref{fig:s125-filters}. For the sub-sample where at least 5 stations have triggered, Fig.\ \ref{fig:AeffIT} shows the effective area as a function of energy  
 from simulations of the
 IceTop array as commissioned in 2010 (IT73). The reconstructed shower cores were required to lie within a fiducial area of about \unit{5.2\times 10^5}{m$^2$} with a polygonal shape excluding the outer stations. The effective exploitable energy range stretches from about \unit{300}{TeV} to \unit{1}{EeV} with 
nearly $3\times10^7$ events per year observed between \unit{300\ {\rm and}\ 400}{TeV}, 
 200 events per year above \unit{300}{PeV} and about 20 events per year above \unit{1}{EeV} in the zenith range from $\cos\theta >0.8$. There is no indication of array saturation up to 1 EeV \cite{TomSerap_icrc2011}. Including the 3 and 4 station events the energy range can be extended to lowest energies of about \unit{100}{TeV} due to the in-fill stations. 

\begin{figure}
	\centering
		\includegraphics[width=0.60\textwidth]{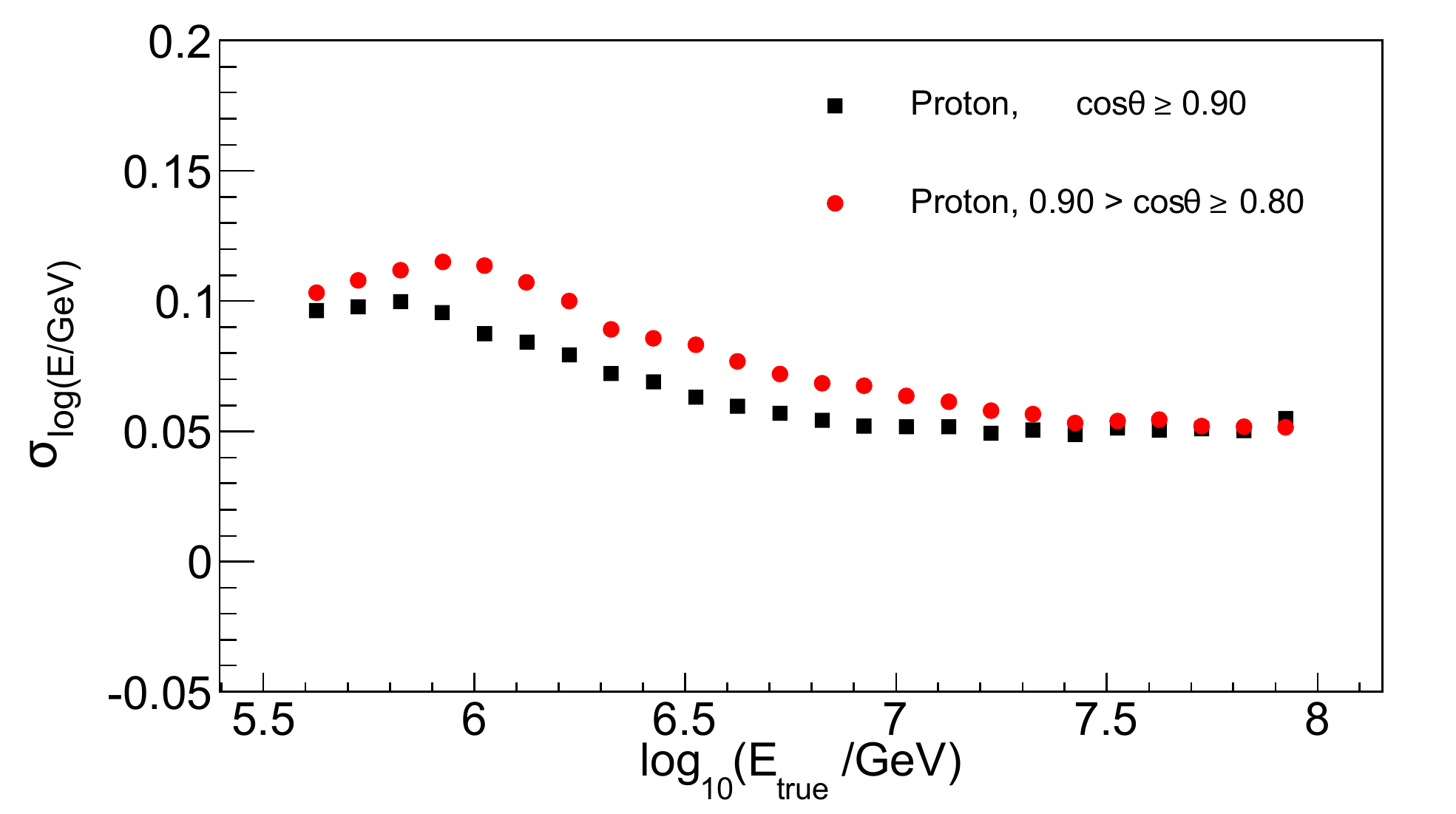}
	\caption{Energy resolution for proton showers which triggered 5 or more stations as a function of energy for the IceTop array as commissioned in 2010 (IT73).}
	\label{fig:energy_resolution}
\end{figure}
\paragraph{Energy, direction and position resolutions} The obtained energy, direction and core position resolutions for some selected energies are summarized in Table \ref{tab:SummaryOfResolutions}.
In Fig. \ref{fig:energy_resolution} the energy resolution as a function of energy is shown for proton showers in the IT73 configuration (same data set as for Fig.~\ref{fig:AeffIT}). The corresponding plots for the core and direction resolutions are displayed in Fig.~\ref{fig:simulation:resolution}.
 
These resolutions were determined in simulations by comparing the reconstructed with the generated variables. In IceTop, resolutions can also be studied experimentally since the same events can be reconstructed independently by the sub-array of A tanks and that of B tanks.  From the comparison one can obtain an experimental measure of the accuracy of  reconstruction.  
The sub-array resolution analysis is in good agreement with the simulation results. However, it should be noted that this method is not sensitive to systematic errors common to the whole detector.
 
\begin{table}
	\centering
		\caption{Summary of air shower resolutions obtained with IceTop IT73 for protons in a zenith angle range $\cos\theta > 0.82$. A resolution is defined such that 68\% of the events deviate less than the listed value.}
	\label{tab:SummaryOfResolutions}
		\begin{tabular}{c|ccc}	
			energy & log-energy & zenith [\degree ] & core [m]\\
			\hline
			\unit{400}{TeV} & 0.1 &0.8 & 9.0\\
			\unit{1}{PeV} & 0.1 & 0.6&7.5 \\
			\unit{10}{PeV} & 0.06 &0.3 & 5.5\\
			\unit{100}{PeV} & 0.05 & 0.4 & 5.5 \\
			\hline
		\end{tabular}
\end{table}

\paragraph{Signal fluctuations} The sub-array method can also be used to determine the signal fluctuations which have to be included in the likelihood fit as described in section \ref{sub:likelihood_fit}. Signal fluctuations are caused by detection fluctuations and fluctuations in the shower development. Both together are measured by comparing signals in the two tanks of a station (with proper corrections for different core distances). The detection fluctuations can be separated from the shower fluctuations by running both DOMs of a tank at the same gain and comparing their signals. In \cite{achterberg06} it was shown that the detection fluctuations are much smaller than the shower fluctuations confirming that the tank design provides adequate (or better) resolution.

\subsection{Cosmic ray anisotropy} 

Cosmic ray anisotropies have been studied by IceCube with high energy muons in the deep ice \cite{Abbasi_anisotropy:2010mf,Abbasi_anisotropy:2011zka}. An analysis using IceTop alone is currently in progress \cite{Santander_IT_aniso}. While the declination angular range is more restricted than for IceCube (to less than about 60\degree ) the much higher energy resolution due to the shower reconstruction allows finer energy binning than for the muons detected with the in-ice detector. Anisotropy measurements can be done in the full IceTop energy range given above; sensitivities to intensity changes of less than $10^{-3}$ can be reached at least up to \unit{10}{PeV}. The angular resolution of better than 3\degree\ is sufficient for studies of anisotropy patterns on a scale above about 10\degree .


\subsection{Primary composition} \label{subsec:composition}

\subsubsection{Primary composition from coincident events} \label{subsubsec:composition_coin}

\begin{figure}
	\centering
		\includegraphics[width=0.40\textwidth]{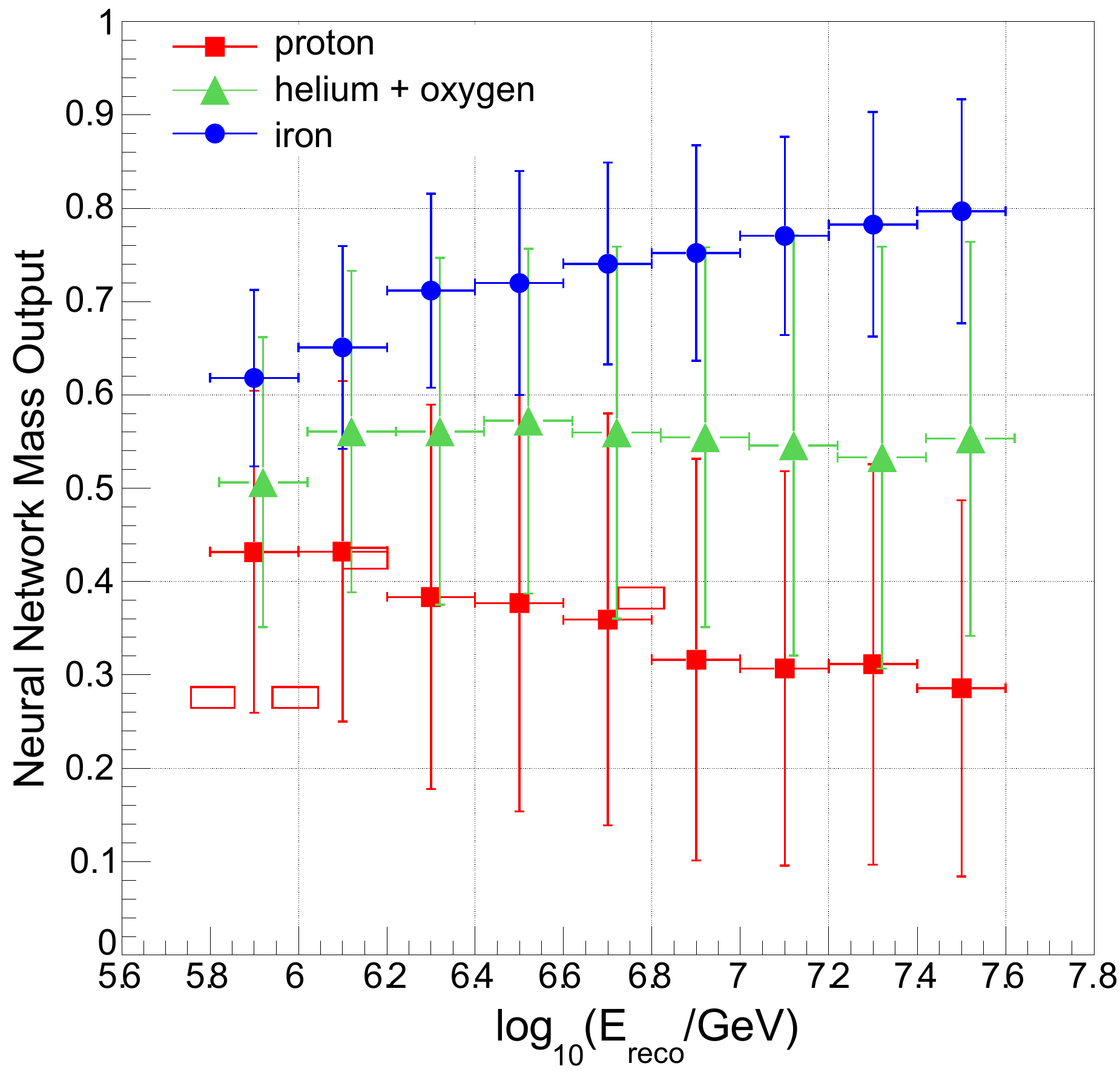}
	\caption{Separation of elements in the primary cosmic ray composition as obtained with a neural net algorithm. The plot shows for proton, iron and a helium-oxygen mixture the separation in terms of a neural network output together with the rms width of the corresponding distributions. From \cite{ITIC40-spectrum_Abbasi:2012}.}
	\label{fig:composition_separation_ITIC40}
\end{figure}

A strength of IceCube is the possibility to measure high energy muons in the deep ice in coincidence with the shower reconstructed in IceTop. The full IceCube detector can cover the energy range from below the knee to EeV. Showers generated by primary cosmic rays in this
energy range produce multiple muons with energy sufficient to reach the depth of
IceCube.  For a primary energy of \unit{10^{15}}{eV}, for example, proton-induced showers near the vertical produce on average about 10 muons with $E_\mu\,>\,500$~GeV and iron nuclei produce about $20$ such muons.  For higher primary energies, the number of muons increases, and the multiplicity in showers generated by nuclei approaches asymptotically a factor of $A^{0.24}$ times the muon multiplicity of a proton shower \cite{gaisser90}, yielding about 2.6 for $A=56$. 
As a consequence of the high altitude of IceTop, showers are observed near maximum so the detector has good energy resolution, which is important when the goal is to measure
changes in composition as a function of energy.

The first analysis of such coincident data is described in \cite{ITIC40-spectrum_Abbasi:2012}. The data set is constrained to a subarray of the detector and a relatively short time period (about 1 month). An example of the separation of elements in this analysis is shown in Fig.\ \ref{fig:composition_separation_ITIC40}. The separation increases with energy and is in the order of the width of the distributions.  The measurement of the average logarithmic mass, $\langle \ln A\rangle$,   from such distributions as presented in \cite{ITIC40-spectrum_Abbasi:2012} is still dominated by systematic uncertainties, mainly from energy reconstruction in IceCube from changing environmental condition in IceTop and the hadronic interaction models for the shower development. In future analyses we expect to reduce the systematic uncertainties and to improve the separability by refining the algorithms and including additional mass sensitive variables. Mass sensitive observables are, for example, muon counts at the surface (see below), shower shape variables or in-ice observables like stochastic energy loss. A combined evaluation of such variables will allow constraining models.

\subsubsection{Primary composition: Muon counting} \label{subsubsec:composition_muon_count}

\begin{figure}
	\centering
		\includegraphics[width=0.5\textwidth]{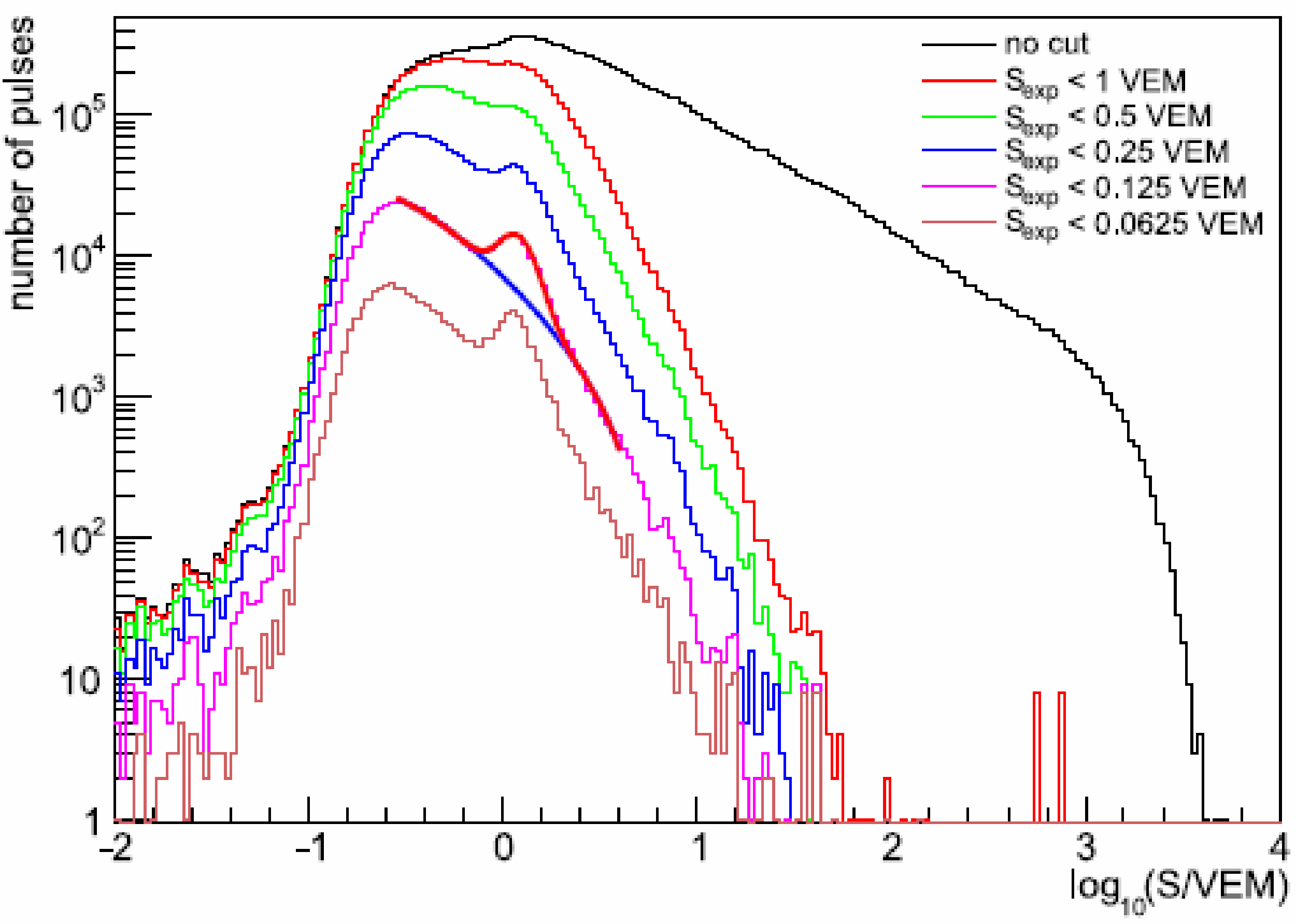}
	\caption{Muon counting in IceTop:  Distribution of tank signals for various cuts on the signal expectation $S_{\!exp}$ in the energy range between 1 and 30 PeV. The average number of muons per event can be statistically extracted by fitting the muon signal together with the background (the example shows a fit for $S_{\!exp}< 0.125\,$VEM).}
	\label{fig:muon_counting}
\end{figure}

Muonic and electromagnetic signals in IceTop can in general not be distinguished. However, muons show up as a relatively constant signal of about 1 VEM at larger distances from the shower axis where the expectation value of a tank signal, $S_{\!exp}$, becomes small compared to a muon signal ($S_{\!exp}$ is obtained from the fit to the lateral shower distribution). As shown in Fig.\,\ref{fig:muon_counting}, the muon signal becomes more and more prominent when requiring smaller $S_{\!exp}$ \cite{Kolanoski_IT_icrc2011}. For $S_{\!exp}< 0.125\,$VEM the figure illustrates that the number of muons can be well fitted. Comparing these muon numbers as a function of energy to simulations of different primary masses, an independent information on the mass composition is obtained.

\subsubsection{Primary photons} \label{subsubsec:photons}
IceCube can efficiently distinguish PeV gamma rays from the background of cosmic rays by exploiting in-ice signals  coincident with an IceTop event as veto against hadronic showers. Gamma-ray air showers have a much lower muon content than cosmic ray air showers of the same energy. Candidate events are selected from those showers that lack a signal from a muon bundle in the deep ice.
 Results of one year of data, taken in the 2008/2009 season when the detector consisted of 40 strings and 40 surface stations (IC40/IT40) are presented in \cite{Stijn_icrc2011}. 
\begin{figure}
	\centering
		\includegraphics[width=0.550\textwidth]{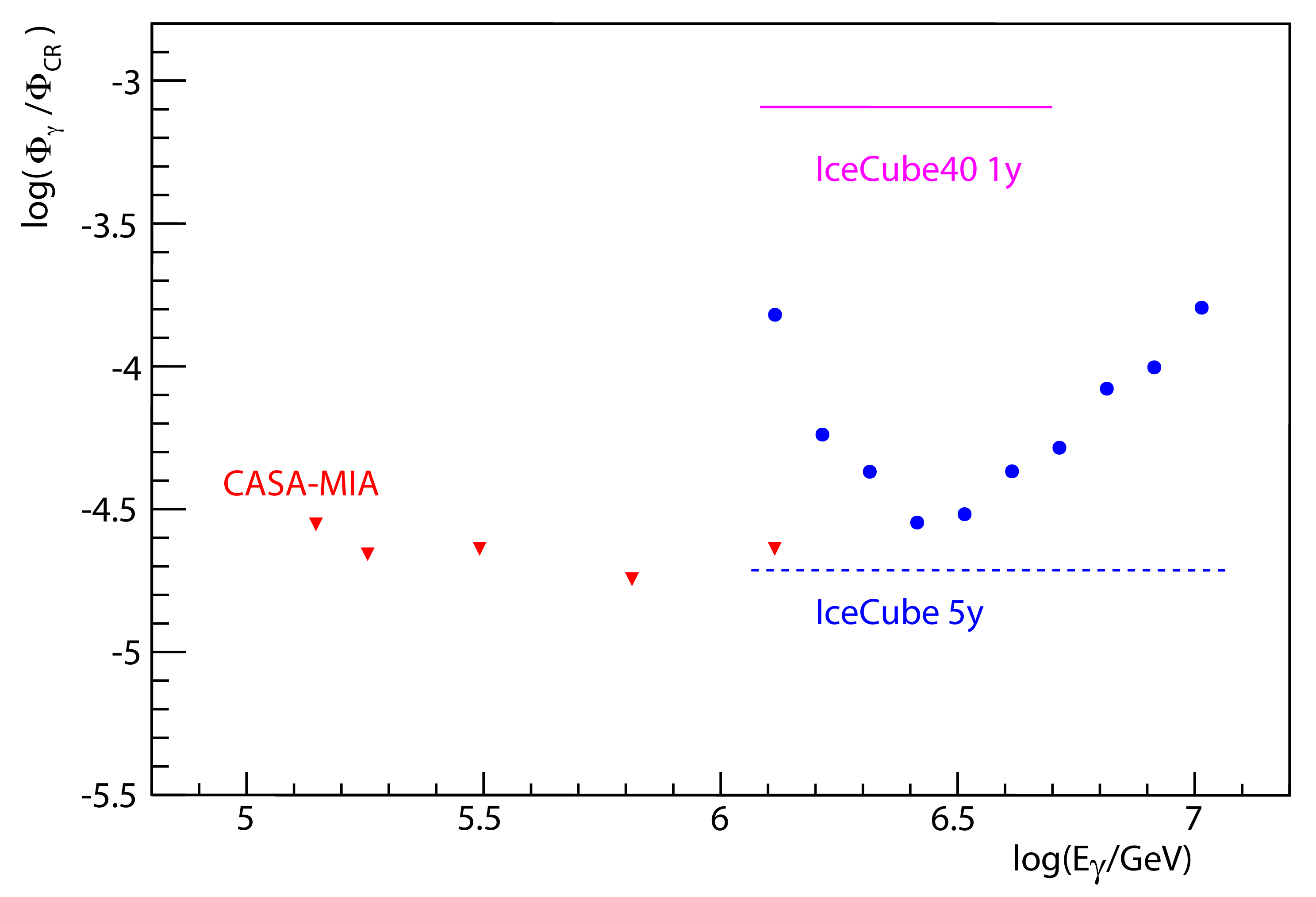}
	\caption{Limits on the ratio of diffuse gamma ray and cosmic ray fluxes from within 10\degree\ of the Galactic Plane. The magenta line is the 90\% confidence level limit obtained from one year of data taken with the IC40/IT40 configuration. The limits are compared to an analysis by CASA-MIA experiment at lower energies \cite{CasaMia_Chantell:1997} and to the projected sensitivity of 5 years data taking with the full IceCube detector. The projected sensitivity is given for the full covered energy range by the dashed blue line and for smaller energy bins by the blue points.}
	\label{fig:diffuse_galactic_gamma}
\end{figure}
Figure~\ref{fig:diffuse_galactic_gamma} shows the IC40/IT40 limits on diffuse gamma ray fluxes together with a projected gamma-ray sensitivity of the final detector (IC86/IT81). The limits are compared to a CASA-MIA analysis at lower energies \cite{CasaMia_Chantell:1997}. Point source fluxes have also been searched for and excluded at a level of about  $10^{-18} - 10^{-17}\, {\rm cm^{-2}s^{-1} TeV^{-1}}$ 
 depending on the location in the sky. The complete
IceCube detector will be sensitive to a flux of more than an order of
magnitude lower, allowing the search for PeV extensions of known TeV
gamma-ray emitters. IceCube/IceTop has currently the highest gamma ray sensitivity in the energy range between about 1 and 10 PeV.

\subsection{IceTop activity as veto for IceCube events} \label{subsec:IT_veto}

The production of extremely high-energetic (EHE) neutrinos is expected in collisions of cosmic rays with the cosmic microwave background. The detection of these cosmogenic neutrinos with IceCube \cite{EHE_Abbasi:2010} requires the ability to discriminate very rare and energetic signal events from an abundant background of cosmic ray induced muons. High energy cosmic ray air showers produce high numbers of muons densely packed around the shower core trajectory. These bundles of muons emit large amounts of Cherenkov light in the IceCube detection volume.

\begin{figure}
	\centering
	
\includegraphics[width=0.49\textwidth]{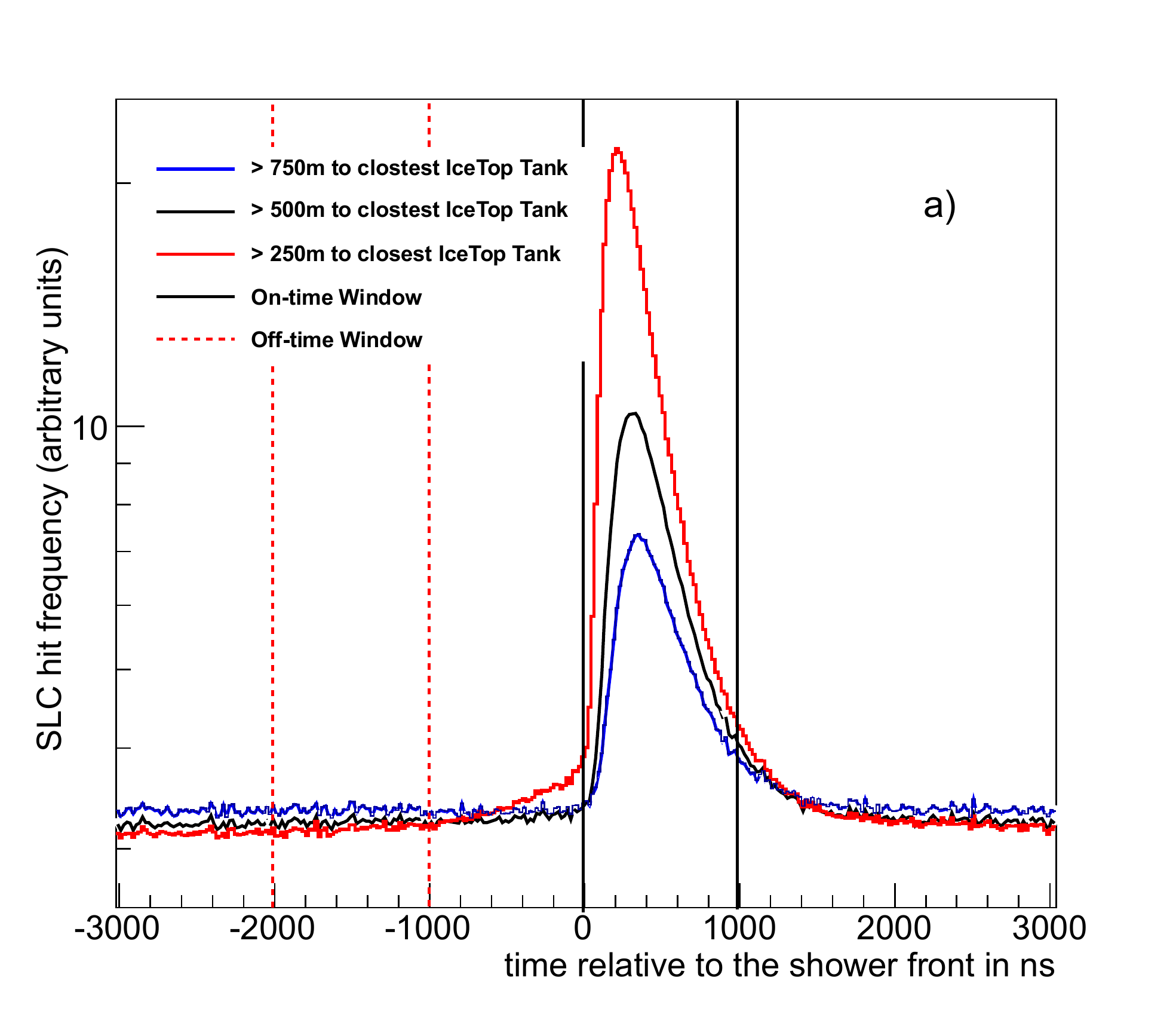}
		\includegraphics[width=0.49\textwidth]{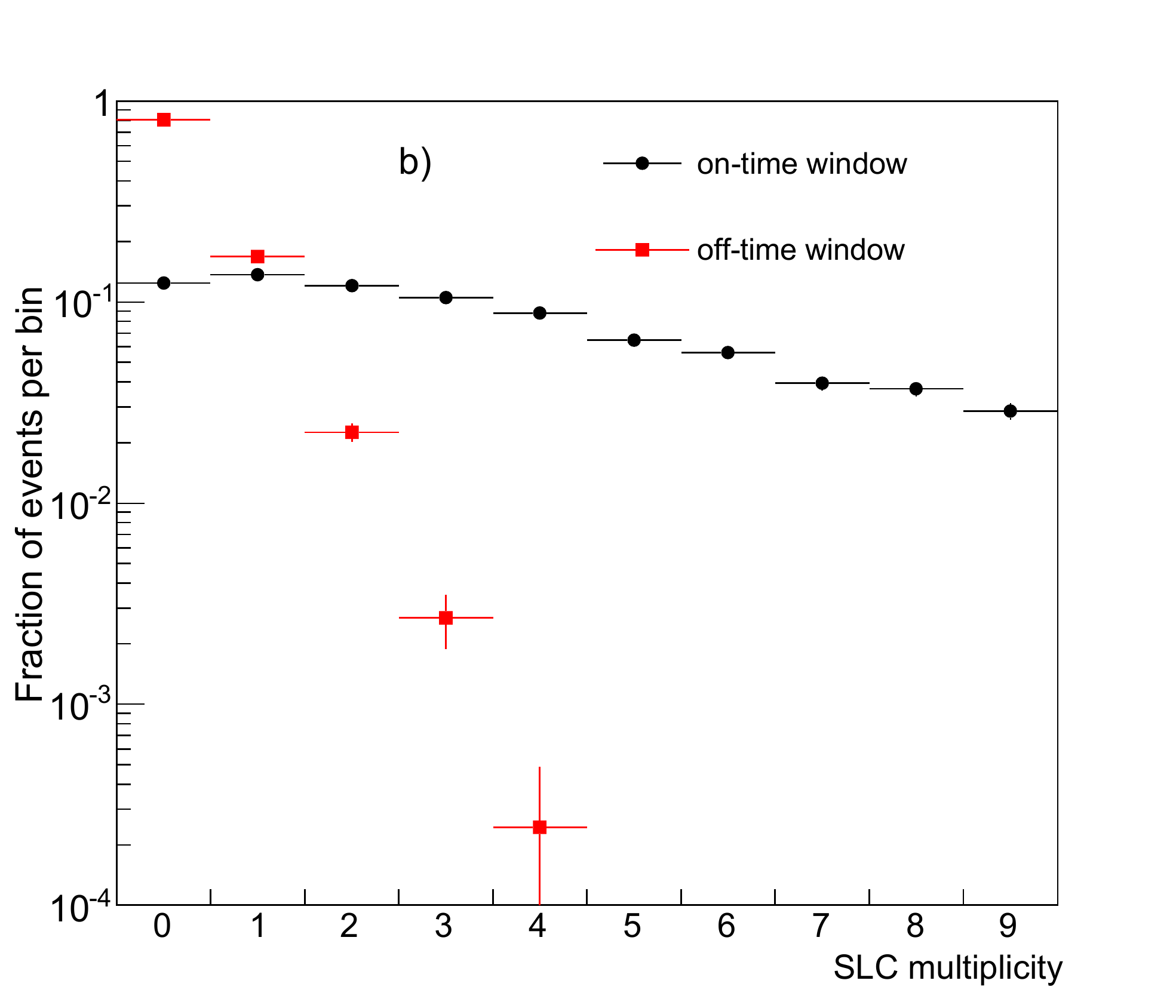}
	\caption{IceTop veto against cosmic ray background in the search for extremely high-energetic neutrinos studied using data from one month running in the IC79/IT73 configuration. 
	 a) The plot shows the time distribution of any SLC hit in IceTop  relative to the shower front  expected if the event was a cosmic ray event (see details in text).  The different distributions are for in-ice tracks which would pass the nearest IceTop tank at the distances given in the legend. The vertical solid black  lines define the on-time window (cosmic ray signal), the dashed red lines the off-time window (background from random IceTop hits). b) The plot shows for the $>500$\,m curve in a) the distribution of the number of hits in the on-time window (black) and the off-time window (red).}
	\label{fig:EHE_veto}
\end{figure}

  In \cite{icrc2011_0778} a study is presented which explores the possibilities of using the IceTop array to improve background rejection while keeping a large fraction of the neutrino signal.  In a limited zenith range around the vertical this is obviously possible with high efficiency. However, by exploiting the IceTop SLC hits (single tank hits without additional conditions, see Section \ref{subsubsec:SLC}) air showers can also be efficiently rejected at larger zenith angles where the shower core does not hit IceTop.

  The plots in Fig.\ \ref{fig:EHE_veto} include data corresponding to 33.4 days from runs distributed over the whole year 2010 (IC79/IT73 configuration). Events were selected as neutrino candidates if they had  more than $10^4$ photons detected in IceCube, corresponding to a neutrino energy of more than \unit{10^{17}}{eV}. Under the assumption that the candidate is a cosmic ray event, a hypothetical shower front can be constructed from the in-ice track and its movement over the IceTop array can be followed.   The plot in Fig.\ \ref{fig:EHE_veto}\,a shows the time distribution of any SLC hit in IceTop  relative to the expected shower front.  The different distributions are for in-ice tracks which would pass the nearest IceTop tank at different distances of the shower core to the nearest IceTop tank. With the condition that the track passes through the in-ice detector the distances of the shower axes from the nearest IceTop tank correspond to zenith angle ranges: \unit{250}{m}, \unit{500}{m} and \unit{750}{m}   correspond to approximately 10-30\degree ,  30-50\degree\ and 25-60\degree , respectively. The vertical solid black  lines define the on-time window containing the cosmic ray signal, the dashed red lines the off-time window which yields the background of random hits in IceTop. The plot in Fig.\ \ref{fig:EHE_veto}\,b shows for the $>500$\,m curve in a) the distribution of the number of hits in the on-time window (black) and the off-time window (red).
  
  In the off-time window the  hit multiplicity distribution drops very quickly, while the distribution of coincident hits is much flatter. The off-time hit distribution can be well described by a Poisson distribution with a mean of about 0.25/$\mu$s. That means requiring for a neutrino event to have less than 1 or 2 SLC hits yields neutrino signal reductions by 22\% and 2.6\%, respectively. The veto efficiency depends strongly on the required amount of light detected in IceCube and on the the distance of the extrapolated track to the nearest IceTop tank. From Fig.\ \ref{fig:EHE_veto}\,b, where more than $10^4$ photons in IceCube and a closest approach of more than \unit{500}{m} are required, one can read off a veto efficiency of about 75\% for less than 2 SLC hits. Requiring $10^5$ photons in IceCube,  the same criteria result in a veto efficiency well above 98\%. Exploiting the IceTop veto the cuts on neutrino candidates can be widened gaining some 10\% overall acceptance.


\subsection{Calibration of IceCube with IceTop} \label{subsec:perf_calib}

Events reconstructed with IceTop that are also seen
in the deep detector can be used to verify event reconstruction and the consistency of the relative alignment of both detector components.  
Examples of verification of timing
and direction with IceTop are given in  \cite{achterberg06}.

Showers that trigger IceTop produce usually  bundles of several (at 1 PeV) or many muons in the deep detector.
In contrast, much of the atmospheric muon background in deep
IceCube consists of single muons. Most of these events are from cosmic-rays with primary energies below \unit{10}{TeV}.
By selecting a sample of coincident events in which
both tanks at one and only one IceTop station are hit, it is possible to discriminate against high-energy events and find a sample enriched in single muons. Such a sample can be selected by the filter InIceSMT\_IceTopCoin  (Table \ref{tab:filter_scheme} in Section \ref{subsubsec:ITfilters}).
Coincidences involving only an interior IceTop station provide a
sample in which about 75\% are single muons in the deep detector \cite{Bai:2007zzm}. 

\begin{figure}
	\centering
		\includegraphics[width=0.40\textwidth]{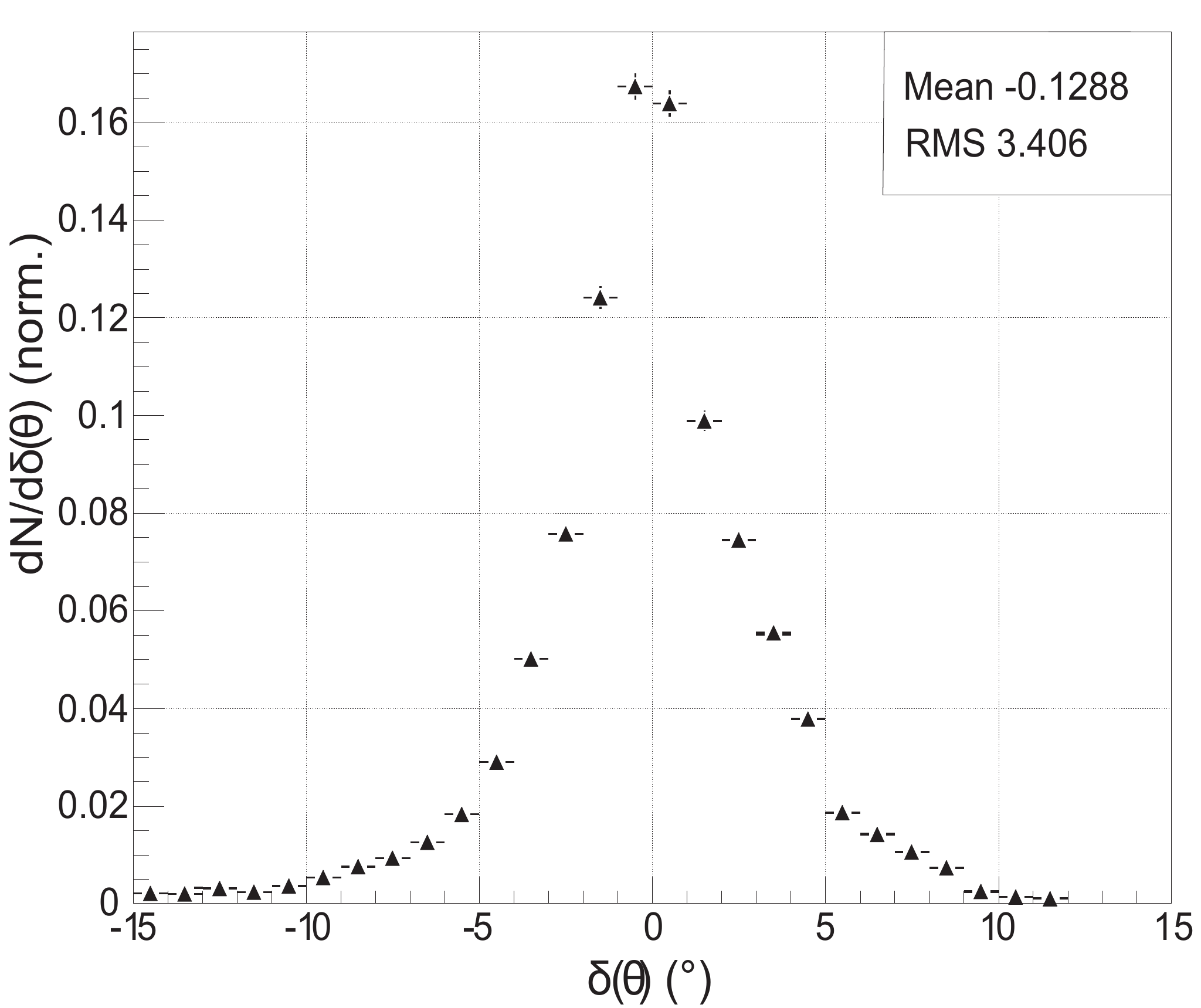}
	\caption{The difference between the zenith angle defined
by the line connecting triggered IceTop station with the
the center-of-gravity of triggered in-ice DOMs and the track reconstructed by the in-ice detector. In IceTop it was required that only one station has triggered and that the signal is made up by more than 400 photoelectrons.}
	\label{fig:InIce-IT_direction_resolution}
\end{figure}

 An analysis of such coincident events using data taken in 2006 with the IC9/IT16 configuration is presented in \cite{Bai:2007zzm}.  Figure \ref{fig:InIce-IT_direction_resolution} shows the distribution of the differences  between the zenith angle defined by the line from the hit station to the center-of-gravity of hits in the deep detector and the direction obtained from the muon reconstruction algorithm in the deep detector.
The estimated angular accuracy of the connecting line is about 3\degree . The observed rms of the zenith difference distribution of 3.4\degree\ means that the in-ice reconstruction of muons has an accuracy of better than 2\degree . The mean of the distribution being close to 0  confirms the accuracy of positioning of both detectors as quoted in Section \ref{subsec:survey}.


\subsection{Heliospheric physics with IceTop}  \label{subsec:perf_helio}

The IceTop tanks detect secondary particles from multi-GeV cosmic rays with a counting rate exceeding 1 kHz per detector.  The detector response to secondary cosmic rays  is discussed in \cite{Clem:2007zz}. 

A certain class of transient events, like solar flares or gamma ray bursts, can be detected with very good time resolution via a common rate increase in all IceTop tanks. The observation of the Dec 13, 2006 Sun flare event \cite{Abbasi08} demonstrates these abilities of IceTop. Following this observation the detector readout has been setup such that counting rates could be obtained at different thresholds (Section \ref{subsubsec:DiscrThresholds}) 
 allowing the unfolding of cosmic ray spectra during a flare (see Fig.\ \ref{fig:icrc0921_fig02}). 

\section[Summary and Outlook]{Summary and Outlook}
\label{sec:outlook}
%
%
%

The IceTop air shower array, the surface component of the IceCube Neutrino Observatory, is operating in its final configuration since the beginning of 2011. In all aspects it has reached design performance. With an in-fill array the energy threshold is even lower than in the original design, leading to an energy coverage from about \unit{100}{TeV} to \unit{1}{EeV}. The detector is primarily designed to study the mass  composition of primary cosmic rays by exploiting the correlation between the shower energy measured in IceTop and the energy deposited by muons in the deep ice. 

First results on the energy spectrum and mass composition have been obtained with the still incomplete detector. We look forward to composition results from the complete detector. These results should lead to a substantial improvement of our understanding of cosmic rays in an energy range where the transition from galactic to extra-galactic origin is expected.

In addition, the IceTop array has found various other applications. Counting rates of individual tanks as a function of different thresholds allow investigation of transient events from the sun and maybe other astrophysical objects. Using single tank hits IceTop can be very efficiently used for vetoing cosmic rays in the search for high-energy neutrinos. IceTop is also employed for direction and time calibration of the in-ice detector.

For the future an extension with radio antennas is being studied to enhance the air shower detection capabilities \cite{icrc2011_1102}.  With the measurement of the shower development in air a radio detector would add complementary observables  sensitive to the composition of cosmic rays.

\section*{Acknowledgments}
\label{sec:acknowledgments}
\addcontentsline{toc}{section}{Acknowledgments}

We acknowledge the support from the following agencies: U.S. National Science Foundation-Office of Polar Programs, U.S. National Science Foundation-Physics Division, University of Wisconsin Alumni Research Foundation, the Grid Laboratory Of Wisconsin (GLOW) grid infrastructure at the University of Wisconsin - Madison, the Open Science Grid (OSG) grid infrastructure; U.S. Department of Energy, and National Energy Research Scientific Computing Center, the Louisiana Optical Network Initiative (LONI) grid computing resources; National Science and Engineering Research Council of Canada; Swedish Research Council, Swedish Polar Research Secretariat, Swedish National Infrastructure for Computing (SNIC), and Knut and Alice Wallenberg Foundation, Sweden; German Ministry for Education and Research (BMBF), Deutsche Forschungsgemeinschaft (DFG), Research Department of Plasmas with Complex Interactions (Bochum), Germany; Fund for Scientific Research (FNRS-FWO), FWO Odysseus programme, Flanders Institute to encourage scientific and technological research in industry (IWT), Belgian Federal Science Policy Office (Belspo); University of Oxford, United Kingdom; Marsden Fund, New Zealand; Australian Research Council; Japan Society for Promotion of Science (JSPS); the Swiss National Science Foundation (SNSF), Switzerland.




\section*{References}
\addcontentsline{toc}{section}{References}

\bibliography{IceTop-Detector}

\end{document}